\documentclass[%
 reprint,
 amsmath,amssymb,
 aps,
rmp,
floatfix
]{revtex4-2}

\usepackage{graphicx}
\usepackage{dcolumn}
\usepackage{bm}
\usepackage{color}
\usepackage{xcolor}
\usepackage[normalem]{ulem}
\usepackage{hyperref}

\newcommand\apjl{The Astrophysical Journal Letters}
\newcommand\mnras{Monthly Notices of the Royal Astronomical Society}
\newcommand\aap{Astronomy \& Astrophysics}
\begin{document}

\title{Neutron stars and the dense matter equation of state: from microscopic theory to macroscopic observations}

\author{Katerina Chatziioannou}%
\email{kchatziioannou@caltech.edu} 
\affiliation{Department of Physics, California Institute of Technology, Pasadena, California 91125, USA}

\author{H.~Thankful Cromartie}%
\email{thankful.cromartie@nanograv.org}
\affiliation{National Research Council Postdoctoral Associate, National Academy of Sciences, Washington, DC 20001, USA; resident at Naval Research Laboratory, Washington, DC 20375, USA
}

\author{Stefano Gandolfi}%
\email{stefano@lanl.gov}
\affiliation{Theoretical Division, Los Alamos National Laboratory, Los Alamos, NM 87545, USA}

\author{David Radice}%
\email{david.radice@psu.edu}
\affiliation{Institute for Gravitation \& the Cosmos, The Pennsylvania State University, University Park, PA 16802, USA}
\affiliation{Department of Physics, The Pennsylvania State University, University Park, PA 16802, USA}
\affiliation{Department of Astronomy \& Astrophysics, The Pennsylvania State University, University Park, PA 16802, USA}

\author{Andrew W. Steiner}%
\email{awsteiner@utk.edu}
\affiliation{Department of Physics and Astronomy, University of Tennessee Knoxville}
\affiliation{Physics Division, Oak Ridge National Laboratory}

\author{Ingo Tews}%
\email{itews@lanl.gov}
\affiliation{Theoretical Division, Los Alamos National Laboratory, Los Alamos, NM 87545, USA}

\author{Anna L. Watts}%
\email{A.L.Watts@uva.nl}
\affiliation{Anton Pannekoek Institute for Astronomy, University of Amsterdam, Science Park 904, 1090GE Amsterdam, the Netherlands}

\date{\today}

\begin{abstract}
The past years have witnessed tremendous progress in understanding the properties of neutron stars and of the dense matter in their cores, made possible by electromagnetic observations of neutron stars and the detection of gravitational waves from their mergers.  
These observations provided novel constraints on neutron-star structure, that is intimately related to the properties of dense neutron-rich matter described by the nuclear equation of state.  
Nevertheless, constraining the equation of state over the wide range of densities probed by astrophysical observations is still challenging, as the physics involved is very broad and the system spans many orders of magnitude in densities.
Here, we review theoretical approaches to calculate and model the neutron-star equation of state in various regimes of densities, and discuss the related consequent properties of neutron stars.  
We describe how the equation of state can be calculated from nuclear interactions that are constrained and benchmarked by nuclear experiments.  
We review neutron-star observations, with particular emphasis on information provided by gravitational-wave signals and electromagnetic observations.
Finally, we discuss future challenges and opportunities in the field.

\end{abstract}

\maketitle
\tableofcontents

\section{Introduction}

Neutron stars are the most compact material objects in the universe. 
They represent unique systems of matter under extreme conditions of pressure and density, exhibiting a variety of fascinating properties that can be inferred from astronomical observations.
The existence of neutron stars has been predicted immediately after the discovery of the neutron.
In 1934, Baade and Zwicky stated ``With all reserve we advance the view that a super-nova represents the transition of an ordinary star into a \emph{neutron star}, consisting mainly of neutrons.''~\cite{Baade:1934}.
Neutron stars have typical radii around 10\,km but masses that can reach 2 times the mass of the sun.
These extreme conditions lead to unique properties of neutron stars: densities in the center of neutron stars can reach up to 6 to 8 times the nuclear saturation density~\citep{Koehn:2024set}, which is the density in the center of heavy atomic nuclei.
These densities can currently not be realized in experiments on Earth. 
Theoretical predictions of the properties of this matter can therefore be verified only with astronomical observations.

\begin{figure*}[h!]
\begin{center}
\includegraphics[width=\textwidth]{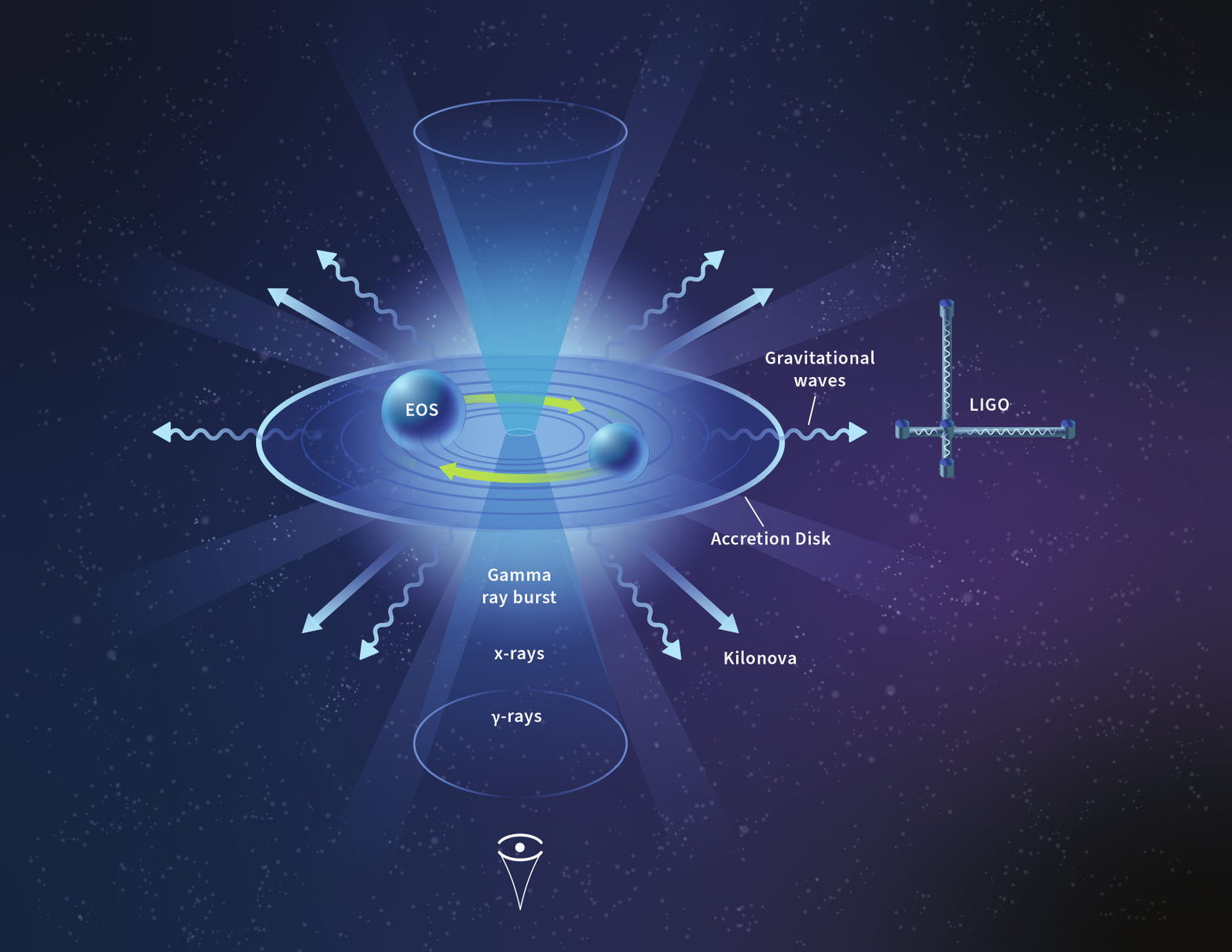}
\caption{Schematic illustration of the merger of two neutron stars with the corresponding emissions that can be observed on Earth.
}
\label{fig:binary}
\end{center}
\end{figure*}

Neutron stars have a layered and physics-rich structure~\cite{Lattimer:2000nx,Lattimer:2004pg}. 
The crust (the outer 1\,km) primarily contains neutron-rich nuclei arranged on a lattice due to the long-range nature of the Coulomb force.
From the outer to the inner part of the crust the properties of these nuclei change drastically. 
At low densities their properties resemble ordinary matter but with increasing density the nuclei become increasingly neutron-rich.
These nuclei cannot be realised in terrestrial experiments, as neutron-rich nuclei are very unstable and their half life is low.
When densities are sufficiently high, neutron begin to drip from these nuclei, forming a ``sea'' of neutrons.
This neutron fluid separates the outer from the inner crust.

The sea of neutrons that surrounds the nuclei in the inner crust may be in a Fermionic superfluid phase that alters the specific heat and the neutrino emission of the star and plays a crucial role in the star cooling.
The superfluid phase is also relevant to sudden observed spin jumps, called glitches, probably related to the angular momentum stored in the rotating superfluid neutrons in the inner crust. 
This superfluid is the most strongly correlated one known in nature. 
Fermionic superfluids have been studied for decades, but the properties of neutron matter in these regimes  are unique. 
The strong pairing consists in strongly correlated (but unbound) pairs of neutrons, that interact at quite large distances. 
The nature of this Fermionic superfluidity has been  confirmed by analyzing neutron star cooling observations.

At even higher densities, the extremely neutron-rich nuclei undergo a number of transitions, manifesting in peculiar shapes and deformations in phases known as nuclear pasta (see for example~\citet{Watanabe:2000rj}). 
In these phases, nuclei are no longer separated; instead protons and neutrons of neighboring nuclei organize themselves in different ``pasta" geometries, such as spaghetti (rods) and lasagna (slabs).
Eventually, the system becomes more and more neutron rich until the homogeneous matter phase, the outer core, is reached.

The outer core of neutron stars consists of matter composed by nucleons, electrons and muons. 
Most nucleons are neutrons with a small fraction of protons, and proton fractions are of the order of 5\% but grow again with density.
The degenerate neutron-rich gas, aided by repulsive interactions among the nucleons, provides pressure against the gravitational pull and keeps the star from collapsing to a black hole.
This region is of particular interest to nuclear physics, as the underlying interactions between nucleons are of great interest and can be tested by comparing predicted properties of dense matter with and data from terrestrial experiments and astrophysical observations of neutron stars. 
In this regime, the nucleons are strongly interacting, and sophisticated models describing the correlations among two and three nucleons are necessary for an accurate description of dense matter but also atomic nuclei.

The inner core of neutron stars is the least understood part of the neutron star. 
Densities reach their maximum, up to multiple times the saturation density, and it is unclear what degrees of freedom describe this region.
There are different hypothetical models.
There could be charged mesons, such as pions or kaons, that eventually form a Bose condensate. 
Other heavier baryons with strangeness, called hyperons, might form from nucleons through weak processes.
Finally,  at higher densities hadrons might not be relevant degrees of freedom and quark matter, in which quarks are deconfined, might appear~\citep{Annala:2019puf}.

\section{History of neutron-star observations}

Neutron stars were discovered in 1967 by Dr. Jocelyn Bell Burnell in the form of radio pulsars \cite{Hewish68} and these still form the majority of the observed neutron star population \cite{atnf,Lorimer08}\footnote{\url{http://www.atnf.csiro.au/research/pulsar/psrcat}}. Many rotation-powered pulsars also emit other wavebands, though most observations are confined to only the lowest (radio $\lesssim$\,3\,GHz) and highest (X-rays and gamma-rays) EM frequencies \cite{fermipulsarcat}. Timing of rotation-powered pulsars in binaries enables precision inferences of neutron star mass that can be used to constrain the dense matter EOS, with the tightest constraints coming from the measurement of pulsars exceeding $2\,M_{\odot}$, the first direct measurement of which was reported in 2010 by \citeauthor{Demorest10} (see Section \ref{sec:radio.masses}). 

The first X-ray telescopes led to the discovery accreting neutron stars, in the form of accreting pulsars \cite{Giacconi71} and thermonuclear bursters \cite{Belian76,Grindlay76}. We now know of a few hundred accreting neutron stars, in both low and high mass X-ray binaries \cite{batsecat,amxps,minbar}. X-ray observations with ROSAT, Chandra, and XMM had sufficient sensitivity to enable the study of cooling of young isolated neutron stars and accreting neutron stars in quiescence \cite{Walter96,Brown98}.  Spectral modeling of all of these sources is now used to infer masses and radii (Section \ref{sec:xrayspec}). Meanwhile the discovery of thermal X-ray pulsations from rotation-powered pulsars would eventually open up new techniques for constraining mass and radius \cite{Becker93,Pavlov97} (Section \ref{sec:ppm}). 

Gravitational waves were first discovered indirectly~\cite{Hulse:1975} before the development of interferometric detector, see~\cite{Pitkin:2011}, that enabled the first direct detection~\cite{Abbott:2016}.  The first BNS merger followed shortly after~\cite{Abbott:2017}.

\section{The equation of state from nuclear theory}
\label{sec:theory}

In this section, we will give an overview of the nuclear EOS in terms of its general properties, such as the symmetry energy and its slope~\citep{Lattimer:2012xj}.
We will discuss other macroscopic properties that are constrained by nuclear experiments. 
We will then give an overview of the various density regimes explored in neutron stars. 
We will discuss low-density constraints on the EOS, the complicated "crustal" regime, where very deformed nuclei are surrounded by a superfluid neutron gas, and the EOS at higher densities, where heavier particles might appear. 
In particular, we will focus on discussing theoretical approaches to calculate the EOS from microscopic nuclear interactions that include two- and three-body forces that also reproduce several properties of nuclei.
Finally, we will discuss additional degrees of freedom, such as strangeness, and their impact on the EOS, such as the appearance of phase transitions.

\subsection{Gross properties of the equation of state at nuclear densities}

As a first approximation, matter in the core of neutron stars, from about one half up to about twice nuclear saturation density, can be modeled by considering nuclear matter, i.e., an infinite system of matter formed only by neutrons and protons. 
Properties of nuclear matter are very difficult to extract directly from data because this system cannot be realized in terrestrial experiments. 
However, experiments can provide constraints.
Furthermore, properties of nuclear matter, such as its EOS, can be extracted from theoretical calculations.
Here, we begin our discussion of the EOS by introducing a generic parameterization which provides a good approximation to models more closely connected to quantum chromodynamics (QCD) and allows us to define the basic physical quantities of interest.

We define the number density of neutrons as $n_n$, the number density of protons as $n_p$, the total baryon density $n=n_n+n_p$, and the asymmetry $\beta\equiv (n_n-n_p)/n$.
A useful expansion of the EOS is given in terms of the asymmetry, 
\begin{align}
\frac{E}{N}(n,\beta)=\frac{E}{N}(n,0)+S(n)\beta^2+\dots \,,
\label{eq:eosgeneral}
\end{align}
where $E(n,0)/N$ is the energy of symmetric nuclear matter, and $S(n)$ is called the nuclear symmetry energy.
It is oftentimes sufficient to approximate this expansion by its quadratic expansion~\citep{Wellenhofer:2015qba,Somasundaram:2020chb}.
From this general definition, one can compute the energy density of nuclear matter
\begin{align} 
\varepsilon(n,\beta)=n\frac{E}{N}(n,\beta) \,,
\label{eq:enedensgeneral}
\end{align}
the neutron and proton chemical potentials,
\begin{align}
\mu_i=\frac{\partial \varepsilon}{\partial n_i} \,,
\end{align}
and the pressure 
\begin{align}
P=n^2\frac{\partial(E/N)}{\partial n}=
-\varepsilon + \mu_n n_n + \mu_p n_p\,.
\end{align}
These expressions do not contain the terms for the rest mass energy density, which we define to be $m_n n_n + m_p n_p$. 
An important density in nuclear systems is the nuclear saturation density, $n_0\approx 0.16\,\mathrm{fm}^{-3}$, that is related to the size of atomic nuclei. 
The nuclear saturation density corresponds to a mass density of $\approx 2.7\cdot 10^{14}$\,g/cm$^3$.
The general expressions for the EOS can be conveniently expanded around
$\chi=\beta=0$,
\begin{align}
\frac{E}{N}(n,\beta)\approx-B+\frac{K}{2}{\chi}^2 
+\left(S+L\chi+\frac{K_{\mathrm{sym}}}{2}{\chi}^2\right)\beta^2 \,,
\end{align}
where $\chi=(n-n_0)/3n_0$.
This form uses the fact that symmetric nuclear matter saturates at density $n_0$, implying that the pressure vanishes, $P(n_0)=0$. 
More sophisticated EOS also include higher-order terms proportional to $\chi^3$ and $\beta^4$ and beyond~\citep{Margueron:2017eqc,Margueron:2017lup}.
Just below the saturation density, isospin-symmetric matter with $n_n=n_p$ ($\beta=0$) is unstable as the pressure is negative. 
This is the mathematical manifestation of the physical reality that nuclear matter prefers to separate into two phases at low density: a dense phase, given by atomic nuclei, and the vacuum.
Such non-homogeneous matter appears in the crust of neutron stars and will be described separately below.
Most many-body methods permit one to calculate the EOS of homogeneous matter (in a model-dependent way) which can be described approximately by the form above, without including the degrees of freedom associated with clustering. 

The $\beta$ and $\chi$ expansions above motivate the definition of several quantities. 
The saturation density, $n_0$, is the density at which the pressure in symmetric matter ($\beta=0$) vanishes.
The binding energy $B$ of symmetric nuclear matter, $B\approx16\,\mathrm{MeV}$, is the energy per particle of symmetric nuclear matter at saturation density $n_0$, which is intimately connected to the binding energy of nuclei.
The nuclear incompressibility (also called compressibility) is defined for $\beta=0$ as
\begin{align}
K= \left. \frac{\partial^2 (E/A)}{\partial \chi^2}\right|_{n_0}
=9n_0^2\left. \frac{\partial^2 (E/A)}{\partial n^2}\right|_{n_0}
=9\left. \frac{\partial P}{\partial n}\right|_{n_0} \,.
\end{align}
The incompressibility can be experimentally constrained by measuring the peak energy of the giant monopole resonance in nuclei~\citep{Youngblood:1999zza} or from heavy-ion collisions~\citep{LeFevre:2015paj}. 
Recent models give $K=230 \pm 20$\,MeV~\citep{Huth:2020ozf}.

The symmetry energy $S(n)$ is an especially important quantity.
At saturation, the parameter $S=S(n_0)$ is related to several nuclear properties.
Furthermore, the symmetry energy determines the pressure of neutron-rich matter and hence, it plays an important role for the structure of neutron stars. 
The symmetry energy $S(n)$ can be defined in two ways.
First, it is given by difference between the EOS of pure neutron matter and symmetric nuclear matter,
\begin{equation}
S(n)=\frac{E}{N}(n,1)-\frac{E}{N}(n,0) \,.
\label{eq:symmene2}
\end{equation}
A second definition is via a second derivative of the energy per particle with respect to the asymmetry,
\begin{equation}
S_2(n)=\frac{1}{2}\frac{\partial^2(E/A)}{\partial \beta^2}\Big|_{n_0} \,.
\label{eq:symmene1}
\end{equation}
If terms beyond $\beta^2$ in Eq.~\eqref{eq:eosgeneral} can be neglected, both definitions agree.
As this approximation is usually good, both definitions lead to comparable results.
From the symmetry energy, the slope parameter $L$, can be defined by 
\begin{equation}
L(n)=3n_0\frac{\partial S}{\partial n}\Big|_{n_0}
=\frac{3}{n_0}P(n_0,\beta=1) \,.
\end{equation}
Note that the latter expression directly relates the value of $L$  to the pressure of pure neutron matter at saturation density.
The symmetry energy is also related to the neutron and proton chemical potentials. 
Using
\begin{equation}
\left(\frac{\partial \varepsilon}{\partial \beta}\right)_{n} = 
\frac{\partial \varepsilon}{\partial n_n} 
\left(\frac{\partial n_n}{\partial \beta}\right)_{n} +
\frac{\partial \varepsilon}{\partial n_p} 
\left(\frac{\partial n_p}{\partial \beta}\right)_{n} 
=\frac{n}{2} \left(\mu_n - \mu_p \right)\,,
\end{equation}
we obtain
\begin{equation}
S(n)=\frac{1}{4}\frac{\partial}{\partial \beta}
\left(\mu_n - \mu_p\right) \,.
\end{equation}
Similarly, the slope of the symmetry energy can be rewritten as 
\begin{equation}
L(n)=\frac{3n}{4}\frac{\partial}{\partial n}
\frac{\partial}{\partial \beta}\left(\mu_n - \mu_p\right) \,.
\end{equation}
Alternatively, using 
\begin{equation}
\left(\frac{\partial \varepsilon}{\partial n}\right)_{\beta} = 
\frac{\partial \varepsilon}{\partial n_n} 
\left(\frac{\partial n_n}{\partial n}\right)_{\beta} +
\frac{\partial \varepsilon}{\partial n_p} 
\left(\frac{\partial n_p}{\partial n}\right)_{\beta} 
= \frac{1}{2} \left(\mu_n + \mu_p \right),
\end{equation}
$L$ can be rewritten
\begin{align}
L(n,\beta) &= 3 n \left[\frac{-1}{2 n^2} 
    \frac{\partial^2 \varepsilon}{\partial \beta^2} + 
    \frac{1}{4 n} \frac{\partial^2}{\partial \beta^2} 
    \left(\mu_n + \mu_p\right)\right]  \nonumber\\
  &= \frac{3}{4}\frac{\partial^2}{\partial \beta^2} 
  \left(\mu_n + \mu_p\right) - 3 S(n) \, .
\end{align}

\subsection{Low-density neutron matter and the unitary Fermi gas}
\label{sec:2b}

At very low densities, i.e. $n\lesssim 10^{-2}n_0$, properties of homogeneous neutron matter are very similar to those of ultracold Fermi gases of atomic nuclei, despite the differences between those two systems and the details of the inter-particle interaction~\cite{Gandolfi:2015}.
This similarity arises because the neutron-neutron interaction is peculiar: its scattering length is very large and the effective range is very small.
Under these conditions, several properties of the system are universal, meaning that they do not depend on the explicit details of the interactions among the particles.
Then, systems that are very different in nature have similar properties. 
The great advantage of using ultracold atomic gases is that they can be experimentally controlled very precisely and then be used to ``simulate'' neutron matter.
This provides valuable information because the latter cannot be directly realized and/or observed in experiments.

These systems are unique because they exhibit the largest superfluid character in Fermionic systems.
Other examples of Fermionic superfluids include superconductors, atomic liquid $^3He$, and high-$T_C$ superconductors, and their pairing gap, i.e. the energy to excite the system, is of the order of $10^{-4}-10^{-2}\,E_{FG}$, wehere $E_{FG}$ is the Fermi gas energy.
Ultracold atomic gases and low density neutron matter have a pairing gap of the order of $E_{FG}$, and this aspect has dramatic consequences on their dynamical properties. 
The superfluid gap of neutron matter has important consequences to the cooling and other properties of neutron stars~\citep{Page:2010aw}.

Ultracold Fermi gases have been extensively studied both experimentally and theoretically (for a review see for example~\cite{Giorgini:2008}). 
Of particular relevance to understand neutron matter is a gas made by two-species of same-mass Fermions in two states: one with spin up and one with spin-down. 
Because of the universality of these systems, which properties being independent of the interactions, one can conveniently describe these systems by a model where particles interact in S-wave and the physics of the many-body system is mainly governed by the two-body scattering length $a$. 
Since the effective range $r_e$ of the interaction is very small compared to the interparticle separation $r_0$, the Fermi momentum $k_F$ of the system is the only relevant scale, i.e. the system is in a dilute regime where $r_e\ll r_0\ll a$.
Then, 
in the dilute limit the system's properties are only dependent by the natural (unitless) quantity $a k_F$. 
The energy and other quantities, like the pressure and temperature, can be conveniently measured in units of $k_F$ or of the Fermi gas energy $E_{FG}$.
Of particular interest is the extreme limit where $a\rightarrow\infty$ and $r_e\rightarrow 0$.
This is called the unitary limit. 
In this case, the energy is often indicated through the so-called Bertsch parameter $\xi=E/E_{FG}$, and its value is independent to the detail of the interaction between particles~\citep{}.
 
In the weakly interacting limit of $a k_F\ll 1$, the EOS of ultracold Fermi gases can be described by a perturbative expansion~\cite{Lenz:1929,Lee:1957}
\begin{equation}
  \frac{E}{E_{\mathrm{FG}}} = 1 + \frac{10}{9 \pi} a k_F + \frac{4}{21
    \pi^2} \left( 11- 2 \ln 2 \right) \left( a k_F \right)^2 +\dots \,.
\end{equation}
When the strength of the interaction increases, the system becomes highly correlated, and non-perturbative methods have to be used to solve for the EOS.
Among other techniques, Quantum Monte Carlo (QMC) methods have been employed to investigate properties of these systems, from the weakly to the strongly interacting regime~\cite{Carlson:2003,Bulgac:2006,Carlson:2011,Forbes:2011,Gandolfi:2011}.
For small and negative values of $a k_F$ the two-body system is unbound, the system is weakly interacting, and atoms in different spin states form Cooper pairs (BCS). 
This is the regime where these atomic gases resemble low-density neutron matter.

\begin{figure}
\begin{center}
\includegraphics[width=0.5\textwidth]{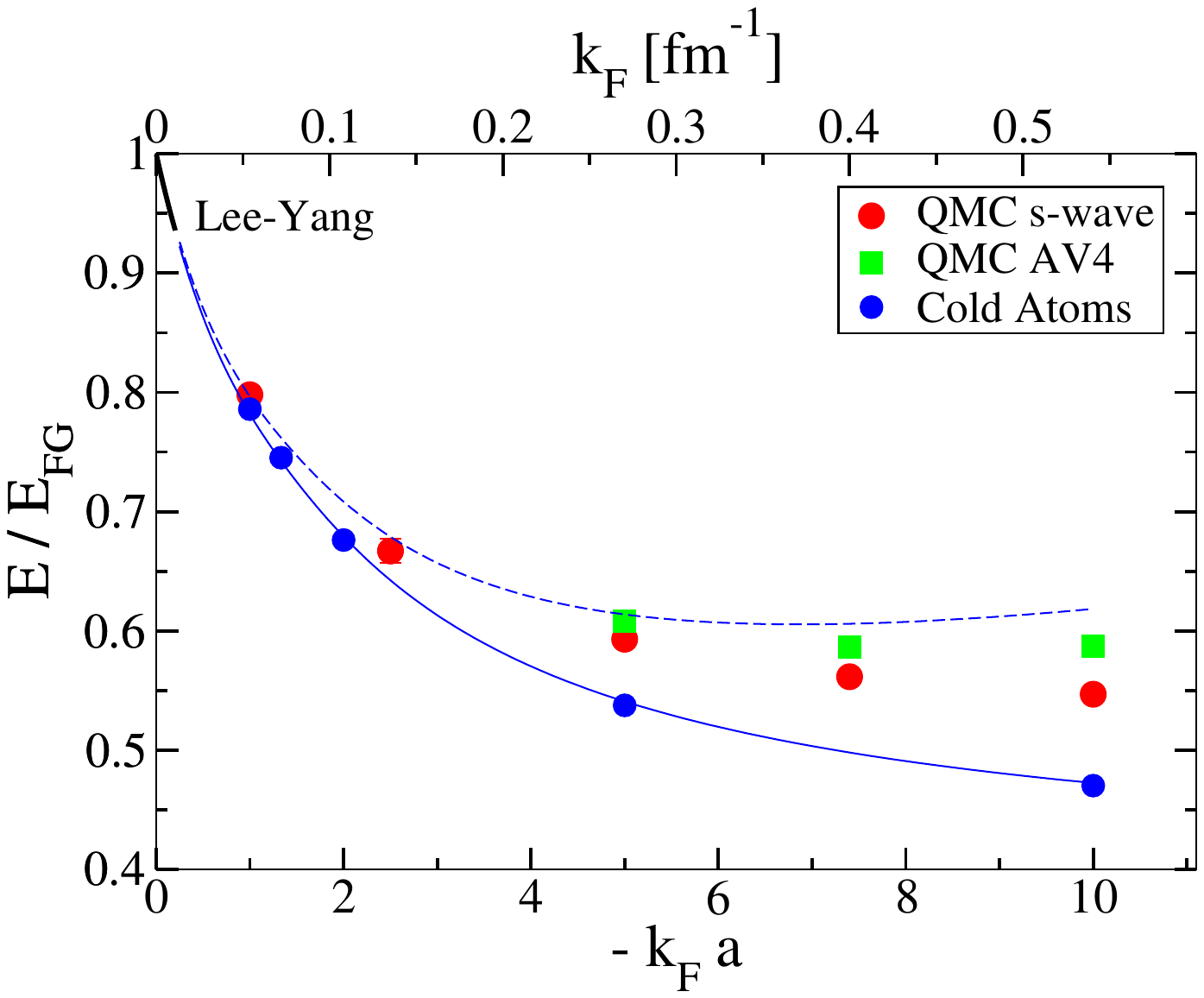}
\end{center}
\caption{The EOS of cold atoms (blue lower line) compared to neutron matter. 
The dashed blue line correspond to cold atoms in the limit of zero effective range, while the dashed line correspond to an interaction with the same effective range of neutron-neutron interaction. The red circles have been obtained by considering the S-wave part of a realistic neutron-neutron interaction, and the green squares also include the effect of P-wave parts~\cite{Gezerlis:2010,Carlson:2012}.}
\label{fig:cold_neut}
\end{figure}

At very low momenta (densities), the neutron-neutron interaction is dominated by the S-wave component and other partial waves are negligible, see for example~\citet{Gandolfi:2015}. 
Given that the neutron-neutron effective range extracted from scattering data is $r_e\approx 2.7$\,fm and the scattering length is $a\approx-18.5$\,fm, there is a density regime where $r_e< r_0< a$, and the properties of neutron matter can be compared to ultracold atomic Fermi gases. 
The energy of cold atoms as a function of $a k_F$ has been used to extrapolate the EOS of neutron matter~\cite{Carlson:2012}. 
In Fig.~\ref{fig:cold_neut} we show the EOS of cold atoms (solid blue line) compared to neutron matter (red and green symbols). 
For cold atoms the interaction is short-range, and the results have been obtained in the limit of $r_e\rightarrow 0$~\cite{Forbes:2012}. 
The EOS of neutron matter has been calculated by considering the S-wave part of a realistic neutron-neutron interaction~\cite{Gezerlis:2008} (red symbols), and also by including the contribution of P-wave~\cite{Gezerlis:2010} (green symbols). 
As we see, the properties of homogeneous neutron matter can be quite well extrapolated from cold atoms calculations in the dilute regime ($|a k_F|$ not too large). 
This has been  very important to benchmark theoretical calculations with experimental constraints.

A similar benchmark has also been done for the superfluid pairing gap, and theoretical calculations are in a very good agreement 
with experimental measurements~\cite{Carlson:2005}.

\subsection{Crust equation of state}

The neutron-star crust is the outer region of the neutron star which lies between the ``ocean'' and the ``core'' and is composed of a Coulomb lattice of nuclei. The ocean is the microscopically thin region at the surface which is sufficiently hot and tenuous to remain in a fluid phase because the ambient temperature is larger than the melting temperature of the crystal. (Note that we will also use the term atmosphere below, which refers to the region where the temperature drops precipitously and sets the emergent X-ray spectrum. The precise ordering of these regions depends on the neutron star's temperature and accretion state.)
The ocean (see, e.g.~\citet{Page:2013hxa}) does not contain sufficient mass to impact the global neutron-star structure. 
At the lowest densities, the crust is composed of atoms in a crystal and the electron wave functions are localized near their respective nuclei. 
At a mass density near $\rho \sim 6 \times 10^{-7}\,\mathrm{MeV}/\mathrm{fm}^3$ $(10^{6}\,\mathrm{g}/\mathrm{cm}^3)$, the density becomes large enough that the electrons become degenerate and form a nearly uniform background. 
This region is referred to as the ``outer crust'' (see, e.g., \citet{Oyamatsu07} and \citet{Piekarewicz08}). 
In isolated neutron stars, it becomes energetically favorable for some of the neutrons to leave their nuclei at a mass density of $\rho \sim 0.2\,\mathrm{MeV}/\mathrm{fm^3}$ $(4 \times 10^{11}~\mathrm{g}/\mathrm{cm}^3)$ (this threshold density may increase by about a factor of two in accreting neutron stars (see, e.g.,~\citet{Steiner:2012bq}). 
This begins the ``inner crust'', which survives up to a mass density of about $\rho \sim 60\,\mathrm{MeV}/\mathrm{fm}^3$ $(10^{14}\,\mathrm{g}/\mathrm{cm}^3$), or about half the nuclear saturation density. 

In cold isolated neutron stars which have achieved nuclear statistical equilibrium, the EOS is determined by the mass of the equilibrium nucleus, the electron EOS, and small corrections due to the Coulomb lattice. 
The temperature is small enough to render the electrons as degenerate but not small enough to make the nuclei degenerate. 
The energy density of matter, $\varepsilon$ can thus be written as a function of the baryon density, $n$ (in the case $T=0$), as
\begin{equation}
\varepsilon(n) = n_{(Z,N)} m_{(Z,N)} + \varepsilon(n_e) + \varepsilon_{\mathrm{Lattice}} \, .
\end{equation}
where $n_{(Z,N)}$ is the number density and $m_{(Z,N)}$ is the total mass of the nucleus with proton number Z and neutron number N. 
The last quantity, $\varepsilon_{\mathrm{Lattice}}$, is the lattice binding energy associated with a one-component BCC Coulomb lattice. (Note that the nucleon rest masses are sometimes not included in the EOS, but we choose to include them here because it simplifies the discussion of beta equilibrium). 
The ideal gas contribution to the EOS from the nondegenerate nuclei is small compared to the rest mass contribution, so we include only the latter. 
By charge neutrality, the electron density is 
fixed to $n_e = Z n_{(Z,N)}$. 
The equilibrium nucleus is determined by minimizing $Z$ and $N$ at any fixed baryon density, $n = (Z + N) n_{(Z,N)}$. 
Because the nuclear binding energy per nucleon is minimized near $^{56}\mathrm{Fe}$, this nucleus is expected to dominate the composition of the lowest densities in the outer crust in cold isolated neutron stars. 

The neutron and proton number densities can be written as $n_n = N n_{(Z,N)}$ and
$n_p = Z n_{(Z,N)}$, respectively. 
One can compute the neutron and proton chemical potentials, $\mu_i = (\partial \varepsilon)/(\partial n_i)$. 
As the density increases, the electron chemical potential rises more quickly than that of the nucleons because the nuclei are nondegenerate and the electrons are degenerate. 
This fact, when combined with the beta-equilibrium condition, $\mu_n = \mu_p + \mu_e$, (see also the description of beta equilibrium in Section~\ref{sec:2e} below) means that nuclei must become more neutron-rich as the density increases. 
At the largest densities in the outer crust, the relevant nuclei are near the neutron drip line. There is thus still a small uncertainty in the EOS due to the experimental uncertainties in the nuclear masses at the highest densities in the outer crust, near $^{118}\mathrm{Kr}$ (which has $N=82$). 
The EOS variation in the outer crust for isolated neutron stars is not important for the global structure of the neutron star. 

In accreting neutron stars, material in the outer crust is not necessarily in nuclear statistical equilibrium, and thus the EOS is also determined by the unknown nuclear composition. 
Stellar evolution suggests that most of the material accreted onto the neutron-star surface comes from the outer layers of white dwarfs or main-sequence neutron stars. 
In either case, this matter likely consists of light nuclei, from hydrogen to oxygen. 
This accreted matter can become unstable to a thermonuclear explosion, and this is the origin of X-ray bursts or superbursts. 
X-ray bursts and superbursts lead to the creation of nuclei up to the $A\sim 107$ region~\cite{Schatz2001}, thus average nuclear mass number ranges from $1$ near the surface to $\sim 100$ at the transition between the outer and inner crust~\cite{Haensel:2003ti}. 
As with isolated neutron stars, the EOS variation in the outer crust does not impact the global structure of the neutron star. 
However, it can be important for the interpretation and analysis of X-ray bursts and superbursts.

The neutron star inner crust is one of the most complicated many-body quantum systems in the universe. There are two basic types of physics input: (i) the EOS of low-density neutron matter (described in Section~\ref{sec:2b}) and (ii) the structure of neutron-rich nuclei beyond the neutron drip line. 
In addition, the competition between the Coulomb and nuclear surface energy leads to strongly deformed nuclear shapes, often called the ``nuclear pasta``~\cite{Ravenhall83}. 
It is important to note that this deformation is qualitatively different from the deformation experienced by laboratory nuclei, which is strongly connected to nuclear shell effects. 
The nuclear pasta would exist in the neutron star inner crust even if nuclear shell effects could be ignored. 
There is a measurable amount of mass in the inner crust, on the order of 0.1 solar masses, thus the uncertainty in the EOS can potentially have an impact on the global neutron star structure. 

Theoretical models of the nuclei in the inner crust typically take two forms: either phenomenological models based on either liquid drop (or droplet) models or mean-field Hartree (or Hartree-Fock) calculations~\cite{Bender03} (pioneered by \citet{Negele73}). 
The uncertainties in the EOS of the inner crust thus originate in (i) our imperfect knowledge of the nucleon-nucleon interaction~\cite{Tews:2016ofv} and (ii) our imperfect ability to perform quantum many-body calculations of nuclei. 
These nuclear structure calculations can be combined with the Wigner-Seitz~\cite{Wigner33} approximation for the Coulomb lattice, to form a model of the inner crust. 
The basic idea is that each nucleus resides at the center of a nearly spherical volume (a ``Wigner-Seitz (WS) cell'') and that these cells are then arranged into a lattice.
The radius of a spherical WS cell is fixed by two constraints: (i) the constraint that the total electric charge inside the cell is zero, and (ii) the constraint that all baryons exist in WS cells (i.e. there are no baryons ``outside''). 
In an accreting star, where the crust may contain an ensemble of nuclear species at every density, each nuclear species has its own type of Wigner-Seitz cell with a distinct radius.   

The Wigner-Seitz approximation breaks down at the highest densities in the crust because the nuclear size and the WS cell size are nearly equal. 
This means that the calculation of the Coulomb energy is nontrivial. Molecular dynamics methods have been used to tackle this issue (see, e.g.,~\citet{Horowitz:2004yf}), but a complete exploration of the crust with all of the concomitant uncertainties is not yet possible. 

Finally, the neutrons which are outside nuclei are expected to be superfluid, with a pairing energy on the order of a MeV~\citep{Gandolfi:2015,Gandolfi:2022}. 
Unlike in condensed matter systems, where the lattice phonons are critical in order to provide an attractive interaction between the participating fermions, the long-range interaction between neutrons is attractive (coming from one-pion exchange). 
The neutron Cooper pairs at this density are likely to be arranged in a singlet configuration, where the spins of the neutrons are oppositely aligned. 
This superfluid has a smaller impact on the EOS, but is observable in the cooling of transiently accreting neutron stars~\cite{Brown09}.

\subsection{Microscopic theory for the equation of state at nuclear densities}

\subsubsection{Nucleon-nucleon interactions and three-body forces}

At densities $n\gtrsim 10^{-2}n_0$, neutron matter cannot be described by s-wave interactions only, and higher partial waves have to be included in the interaction, see 
Fig.~\ref{fig:cold_neut}.
From this figure, we see the the equation of state of neutron matter obtained with s-wave interactions only and when perturbatively including p-wave contributions start to differ at higher densities. 

To describe nucleon-nucleon scattering in various partial waves, 
nucleon-nucleon interactions are usually dependent on the relative spin and isospin state of the nucleons.
A large amount of empirical information about the nucleon-nucleon scattering problem has been accumulated. 
In 1993, the Nijmegen group analyzed all nucleon-nucleon scattering data below 350\,MeV published in physics journals between 1955 and 1992~\cite{Stoks:1993}. 
Older nucleon-nucleon interaction models that fit the Nijmegen database with a $\chi^2/N_{data}\sim$1 are called ``phenomenological'' and fit experimental data very accurately. 
Examples of these interactions are the Nijmegen models~\cite{Stoks:1994} (Nijm93, Nijm I, Nijm II and Reid-93), Argonne models~\cite{Wiringa:1995,Wiringa:2002} and the CD-Bonn potential~\cite{Machleidt:2001}.

Although two-nucleon interactions can describe nucleon-nucleon scattering data and the properties of the deuteron very accurately, they fail to describe the binding energies of nuclei with A$\ge$2~\cite{Pieper:2001},
(see for example \citet{Hammer:2013}).
Furthermore, several calculations suggest that by considering nuclear Hamiltonians that include only two-body interactions the equation of state of symmetric nuclear matter saturates at too high densities with too low energies; see, for example, results for the  Argonne AV18 interactions~\cite{Akmal:1998}, older variational calculations~\cite{Pandharipande:1979}, Brueckner
calculations~\cite{Day:1985,Baldo:2012}, and other recent calculations using Hamiltonians from chiral effective field theory (EFT)~\cite{Hebeler:2011,Carbone:2013}.
The missing ingredient in these calculations are three-nucleon forces. 

\begin{figure}
\begin{center}
\includegraphics[width=0.5\textwidth]{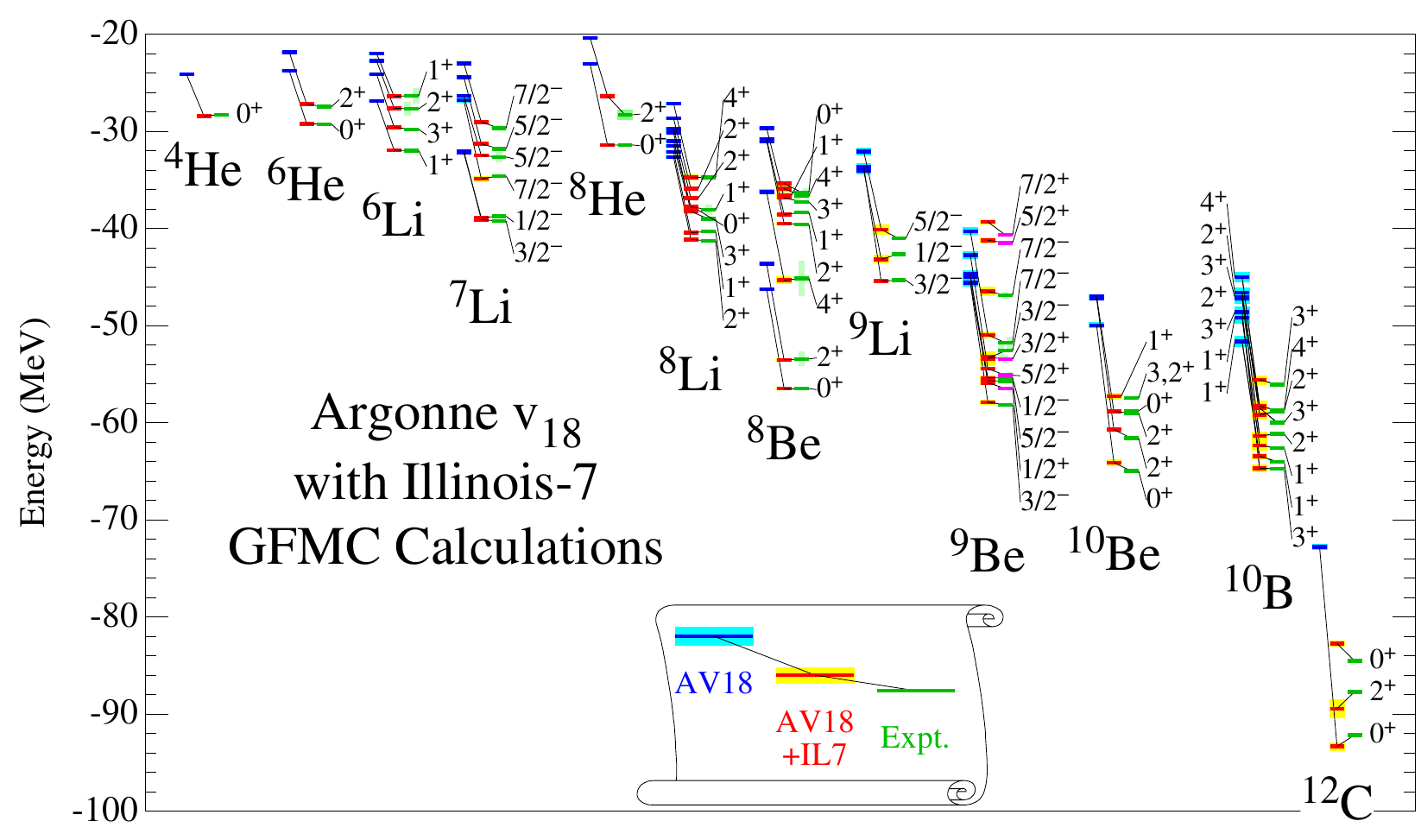}
\end{center}
\caption{The energy of several ground- and excited-state of light nuclei 
compared to experiments. For each level, the green symbol at the right represent the 
experimental measurement, the blue symbol at the left is the result by only including
two body interactions in the nuclear Hamiltonian, and the red symbol in the middle
is the prediction by also including three-body forces~\cite{Carlson:2015}.}
\label{fig:lightnuclei}
\end{figure}

While nucleon-nucleon scattering data have been measured very
precisely, unfortunately there are no experiments to constrain solely three-body forces that reach similar accuracies. 
Typically, three-body interactions are constrained by fitting calculations to reproduce properties of light nuclei,
i.e. binding energies, radii, or other matrix elements of tritium or helium nuclei.
Phenomenological interactions, such as the Argonne AV18 and AV8$'$ potentials, have been combined with three-body forces to describe properties of light nuclei. 
In particular, the Urbana IX (UIX) three-body interaction has been constrained to reproduce the saturation of nuclear matter and the ground state of
$^4$He~\cite{Pudliner:1995}. 
However, the UIX slightly underbinds light
nuclei~\cite{Pudliner:1997}, and the Illinois models of three-nucleon
forces have been constructed later to reproduce better properties of
p-shell nuclei~\cite{Pieper:2001}. They give an excellent description of
several properties of light nuclei, including ground- and excited-states,
matrix elements, scattering, and response functions~\cite{Carlson:2015}. 
As an example, we show in Fig.~\ref{fig:lightnuclei} the energy spectrum of several ground- and excited-states of nuclei up to $^{12}C$. 
In this figure, the Argonne AV18 two-body interaction and the Illinois 7 three-body interaction were used, but similar results have been obtained using different Hamiltonians.
Predictions including only two-body interactions systematically underbind all the nuclei, while the 
inclusion of three-body forces add additional repulsion necessary for agreement with experimental data. 

Several calculations of the equation of state of homogeneous nuclear and neutron matter using the UIX three-nucleon interaction have been performed.
It is worth mentioning that the corresponding EOS predicted the existence of two solar masses neutron stars well before their discovery~\cite{Wiringa:1988,Akmal:1998}. 
In contrast, even though the Illinois three-nucleon interactions describe properties of nuclei very accurately, they produce very soft EOS of neutron-rich matter that do not reproduce even qualitatively the observed maximum mass of neutron 
stars~\cite{Sarsa:2003,Maris:2013}.

The above-mentioned phenomenological approaches do not provide a systematic scheme of improving two-nucleon interactions and they also do not provide a means to derive consistent three-nucleon forces.
A solution to these shortcomings was provided by the modern approach to nuclear forces within the framework of chiral EFT~\cite{Epelbaum:2009,Machleidt:2011zz}.
In chiral EFT, the interactions between nucleons and pions are systematically
organized in powers of momenta $Q$ over a breakdown scale $\Lambda_b$, $Q/\Lambda_b$.
Here, $Q$ are the typical momenta of nucleons of the order of the pion mass, $140$~MeV, and $\Lambda_b$ is the breakdown scale, i.e., the scale where the chiral EFT expansion is expected to break down.
The breakdown scale has been estimated to be of the order of $500-600$~MeV~\cite{Drischler:2020hwi}.
One great advantage
of the chiral EFT framework is that interactions can be systematically improved. 
Another advantage is that all the couplings between nucleons (or Delta degrees of freedom) and pions are fully determined by pion-nucleon scattering data~\cite{Hoferichter:2015hva}. 
The parameters associated with nucleon 
contact terms, that encode missing high-momentum physics, are obtained by fitting to nucleon-nucleon scattering data.
For more details on these interactons, see for example Refs.~\cite{Epelbaum:2009,Machleidt:2011zz,Piarulli:2019cqu,Tews:2020hgp}.

The breakdown of chiral EFT signalizes that a systematic description of nuclear forces in terms of nucleons and pions is valid only at low energies.
Hence, their applicability to study very dense neutron matter in the core of neutron stars is limited. 
This also applies to phenomenological forces which are, however, typically fit to scattering data up to higher momenta than chiral EFT interactions.
With a very simple qualitative analysis, the momentum of two neutrons is related to the laboratory energy in a nucleon-nucleon collision with $k\approx\sqrt{E_{\rm lab}m/2}$, where $E_{\rm lab}$ is the energy of the nucleon-nucleon scattering in the laboratory frame.
By equating the momentum $k$ with the Fermi momentum we can estimate the density at which these momenta are important,  $\rho\approx(E_{\rm lab}m/2)^{3/2}/3\pi^2$.
For a lab energy of $500$~(600)~MeV, this leads to a density estimate of three (four) times the nuclear saturation density.

Several families of two-body forces have been derived within the chiral EFT framework~\cite{Entem:2003,Epelbaum:2004fk,Gezerlis:2013,Gezerlis:2014,Piarulli:2014bda,Ekstrom:2015,Reinert:2017usi,Entem:2017gor,Somasundaram:2023sup}.
For calculations of neutron matter, potentials at N$^2$LO~\cite{Lynn:2015jua,Tews:2018kmu} and next-to-next-to-next-to-leading-order (N$^3$LO)~\cite{Hebeler:2009iv,Tews:2012fj,Drischler:2017wtt,Keller:2022crb}  have been employed. 
The most sophisticated chiral EFT interactions 
have been developed at  next-to-next-to-next-to-next-to-leading order(N$^4$LO)~\cite{Entem:2017gor,Reinert:2017usi}, and they reproduce nucleon -nucleon scattering data up to high energies very well.  
Finally, interactions with explicit Delta degrees of freedom in the
chiral expansion have been explored~\cite{Piarulli:2014bda,Jiang:2020the}. 

Within chiral EFT, three-body forces are naturally predicted to appear at
N$^2$LO~\cite{vanKolck:1994} and come with two low-energy couplings, $c_D$ and $c_E$. 
Given the fact that two-body interactions are constructed from nucleon-nucleon scattering data, the two extra parameters entering in the chiral three-body forces are typically fitted to properties of light nuclei with A=3,4. 
Often, the two parameters are constrained by reproducing the experimental binding energies and/or charge radii of these systems~\cite{Navratil:2007}.  
Similarly, $c_D$ can be constrained using the Gamow-Teller matrix elements related to the $\beta$-decay of A=3 nuclei~\cite{Marcucci:2012}. 
It also  has been suggested to constrain both the two- and three-body interactions at N$^2$LO byreproducing selected observable of scattering data and many-body systems
simultaneously~\cite{Ekstrom:2013,Ekstrom:2015}.
The resulting interaction reproduced the properties of medium-mass nuclei and the equation of
state of symmetric nuclear matter very well, but has a worse description of scattering data and neutron-rich matter dense matter.
Three-body terms at N$^3$LO do not have new parameters~\cite{Bernard:2007sp,Bernard:2011zr}, but,
because of their very complicated structure involving many diagrams,
they have been employed so far only in calculations of nuclear
matter~\cite{Tews:2012fj,Kruger:2013,Drischler:2017wtt}.

The main advantages of interactions obtained within chiral EFT are that i) they can be systematically improved and, more importantly,
ii) it is possible to estimate the theoretical uncertainty due to the  truncation of the series expansion~\cite{Epelbaum:2015,Drischler:2020hwi}.
The easiest way to estimate these uncertainties was provided by \citet{Epelbaum:2015}. 
Using this approach, the 
uncertainty associated to an observable $X$ at, e.g., $N^2LO$ is given by
\begin{align}
\Delta X^{N^2LO}&={\rm max}(Q^4|X^{LO}|,Q^2|X^{NLO} \\
&-X^{LO}|,Q|X^{N^2LO}-X^{NLO}|) \,,
\end{align}
where order-by-order calculations of $X$ iare used, and the scale $Q$ is given by
\begin{align}
Q={\rm max}\left(\frac{p}{\Lambda_b},\frac{m_\pi}{\Lambda_b}\right) \,.
\end{align}
For homogeneous matter, it is reasonable to take $p=k_F$, and $\Lambda_b\sim 500 MeV$ as breakdown scale.
More sophisticated approaches using Gaussian processes have also been introduced~\cite{Drischler:2020hwi}, but they assume uncertainties to be Gaussian. 
However, both approached usually give very similar uncertainty estimates~\cite{Keller:2022crb}.

\subsubsection{Microscopic many-body calculations}

The EOS of nuclear matter is given by the solution of the Schroedinger equation 
for many-body systems.  
The nuclear Hamiltonian is in general non-perturbative, and hence, advanced  numerical methods are required in order to calculate, for example, the dense-matter EOS. 
Several works have been devoted to calculate the EOS of nuclear and neutron matter by starting from nuclear Hamiltonians that include two- and three-body interactions. 
The general form of the Hamiltonian is
\begin{align}
H=-\frac{\hbar^2}{2m}\sum_i\nabla_i^2+\sum_{i<j}v_{ij}+\sum_{i<j<k}V_{ijk}+\dots \,,
\label{eq:hamiltonian}
\end{align}
which includes the kinetic energy operator, a two-nucleon interaction
$v_{ij}$, and a three-nucleon interaction $V_{ijk}$ discussed above.
This Hamiltonian is non-relativistic, and nucleons are considered as point-like particles.

The various methods to calculate the EOS can be essentially divided in two different groups.  
The first group contains several methods that modify the Hamiltonian in such a way that the many-body problem is easier to solve. 
Examples of these methods include the well known Brueckner-Hartree-Fock (BHF)~\cite{Vidana:1999jm}, and perturbative calculations~\cite{Drischler:2021kxf}. 
The BHF method essentially recasts the Hamiltonian into a one-body operator, and the residual part is evaluated by solving the Brueckner-Bethe-Goldstone (BBG) equation.
The accuracy of the results obtained within the BBG approach depends on the expansion level adopted;
the perturbative expansion of BBG is not convergent, but the cluster diagrams can be grouped with the number of independent hole lines~\cite{Pandharipande:1979}. 
Several attempts to study the convergence of hole-line expansion has been explored, but they are very difficult to implement~\cite{Song:1998}, and systematic uncertainties cannot be
quantified.  
Methods based on perturbation theory require the use of
soft Hamiltonians, i.e., Hamiltonians without a strong repulsion at short distances~\cite{Hoppe:2017lok,Drischler:2021kxf}.
This is generally achieved using techniques like the $V_{lowk}$~\cite{Schwenk:2001hg,Schwenk:2002fq,Bogner:2003wn} or Similarity Renormalization Group (SRG)~\cite{Bogner:2006pc,Hebeler:2015hla}, where the latter approach has been dominant in the past years.
With softenened interactions, the convergence in the perturbative order is much faster, but residual induced SRG three- and many-body forces are difficult to be included.
Other methods in this group, that have been used to investigate properties of neutron matter, include the Self Consistent Green's Function (SCGF)~\cite{Carbone:2013eqa,Carbone:2014mja} and Coupled-Cluster (CC)~\cite{Baardsen:2013vwa,Hagen:2013yba,Ekstrom:2017koy} approaches.
These also rely on the use of soft interactions in order to achieve a good convergence.
Using these methods the inclusion of three-body interactions is not straightforward. 
A density-dependent reduction or a normal ordering approximation are used to reduce the three-body force to a density-dependent two body interaction, that can be naturally included in addition to the original nucleon-nucleon interaction~\cite{Carbone:2013eqa,Hebeler:2009iv}.

The second group of many-body methods make use of correlated wave functions where short-range two- and many-body correlations are included explicitly. 
This makes the wave function much more accurate,
in particular when the system is strongly correlated.
Such methods are generally more difficult to implement, because in most of the cases high-dimensional integrals have to be solved, which requires invoking numerical techniques based on Monte Carlo integrations~\cite{Carlson:2015}. 
Most of the results presented here have been obtained using the Auxiliary Field Diffusion Monte Carlo method, that 
provided to be a very accurate method to calculate properties of nuclei
and the nuclear and neutron matter EOS~\cite{Schmidt:1999,Gandolfi:2014b,Lonardoni:2018prl,Lonardoni:2018prc,Lonardoni:2020}

A totally different approach to calculating the EOS of dense matter is based on effective interactions that are
constructed in the framework of the mean-field Hartree-Fock approximation.
Since the original work by Skyrme in the 1950s~\cite{Skyrme:1959zz}, Skyrme interactions became  very popular because they have an easy analytical form, and their energy density functionals provide a good description of several properties of nuclei and infinite nuclear matter.
They also serve as basis for the metamodeling approach of \citet{Margueron:2017eqc} and \citet{Margueron:2017lup}.
Skyrme interactions are basically contact interactions between nucleons, and they describe correlations at very low momenta. 
A great advantage of the Skyrme interactions is that they can be written in terms of a density functional, and an explicit analytical expression for nuclear matter can be derived.
Given their relatively simple form, they do not (and are not intended to) accurately describe nucleon-nucleon phase shifts and properties of light nuclei. 
However, since the calculations of nuclear states within the Hartree-Fock approximation is relatively simple, the Skyrme models are very common to investigate properties of medium and heavy nuclei.

The EOS of nuclear and neutron matter around saturation density has been extensively studied using different methods and Hamiltonians.
As discussed before, the general properties of the EOS of symmetric nuclear matter can be inferred by different experiments. 
In particular, the saturation point, the energy, and the compressibility are well constrained.
However, the situation for pure neutron matter, that is more relevant for the physics of neutron stars, is much more complicated.
As already discussed, the pure neutron matter EOS is related to the symmetric matter EOS by the symmetry energy. 
On the other side, calculations based on realistic nuclear Hamiltonians, that include precise information on 
two- and three-nucleon interactions, have been very useful to predict the 
EOS of pure neutron matter.

\begin{figure}
\begin{center}
\includegraphics[width=0.5\textwidth]{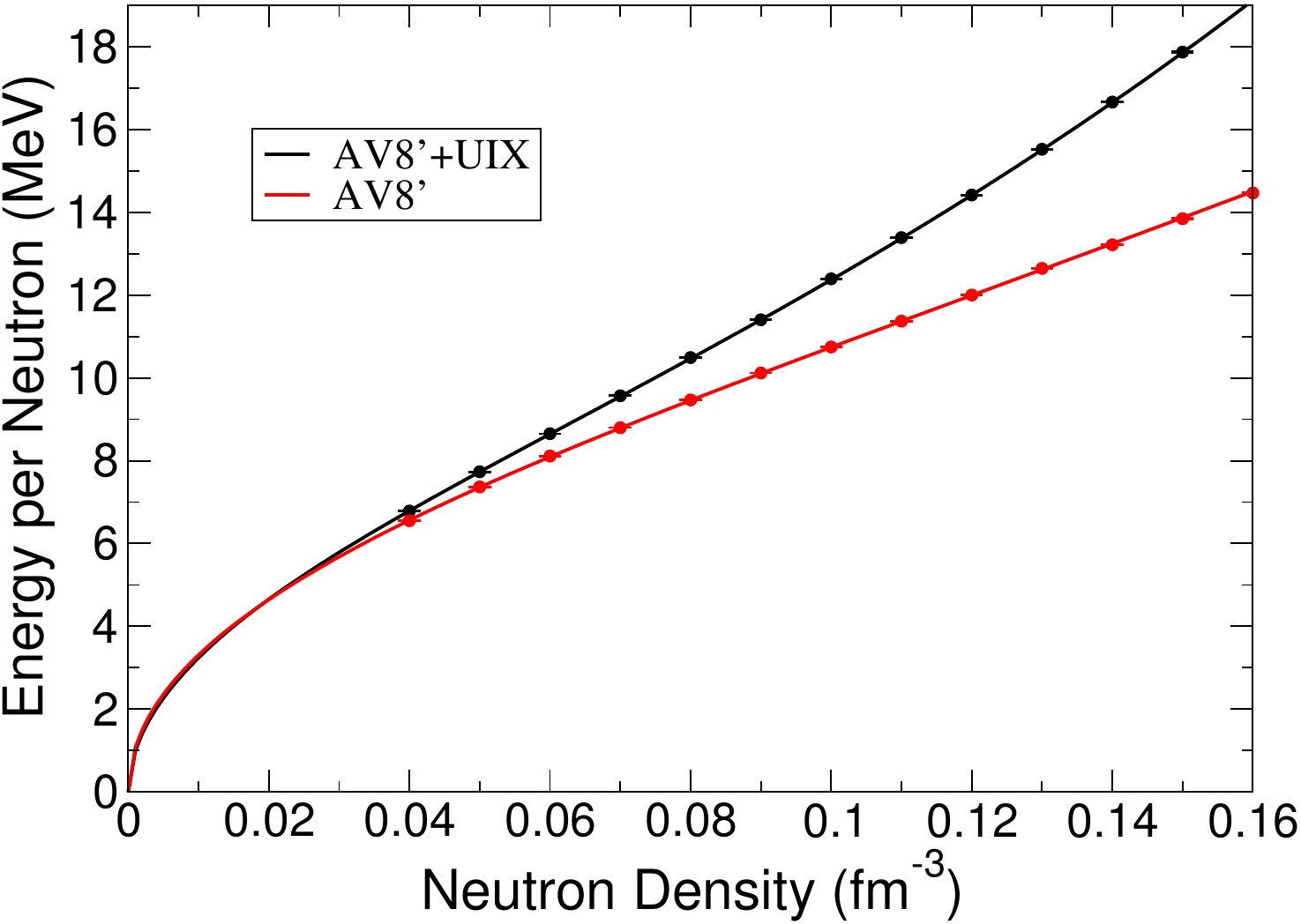}
\end{center}
\caption{The EOS of pure neutron matter calculated using the Argonne AV8' two-body 
interaction (lower red symbols and line), and by also including the Urbana 
IX three-body interaction (upper black symbols and line)~\cite{Gandolfi:2014}.}
\label{fig:pnm1}
\end{figure}

We first would like to focus to the role of three-body interactions. 
They are typically repulsive in pure neutron matter, and this fact is dramatically important for the EOS, and hence, neutron stars. 
In Fig.~\ref{fig:pnm1}, we show as an example the EOS of pure neutron matter calculated with and without a three-body forces. 
The two-body interaction is the Argonne AV8', and the three-body interaction is the Urbana IX model.
This nuclear Hamiltonian gives a good, although not perfect, description of light nuclei, and it has been extensively used in nuclear and neutron matter calculations.
However, the key point here is the role of three-neutron interactions. 
The three-neutron forces are overall repulsive, and their role becomes more and more important as the density increases.

\begin{figure}
\begin{center}
\includegraphics[width=0.5\textwidth]{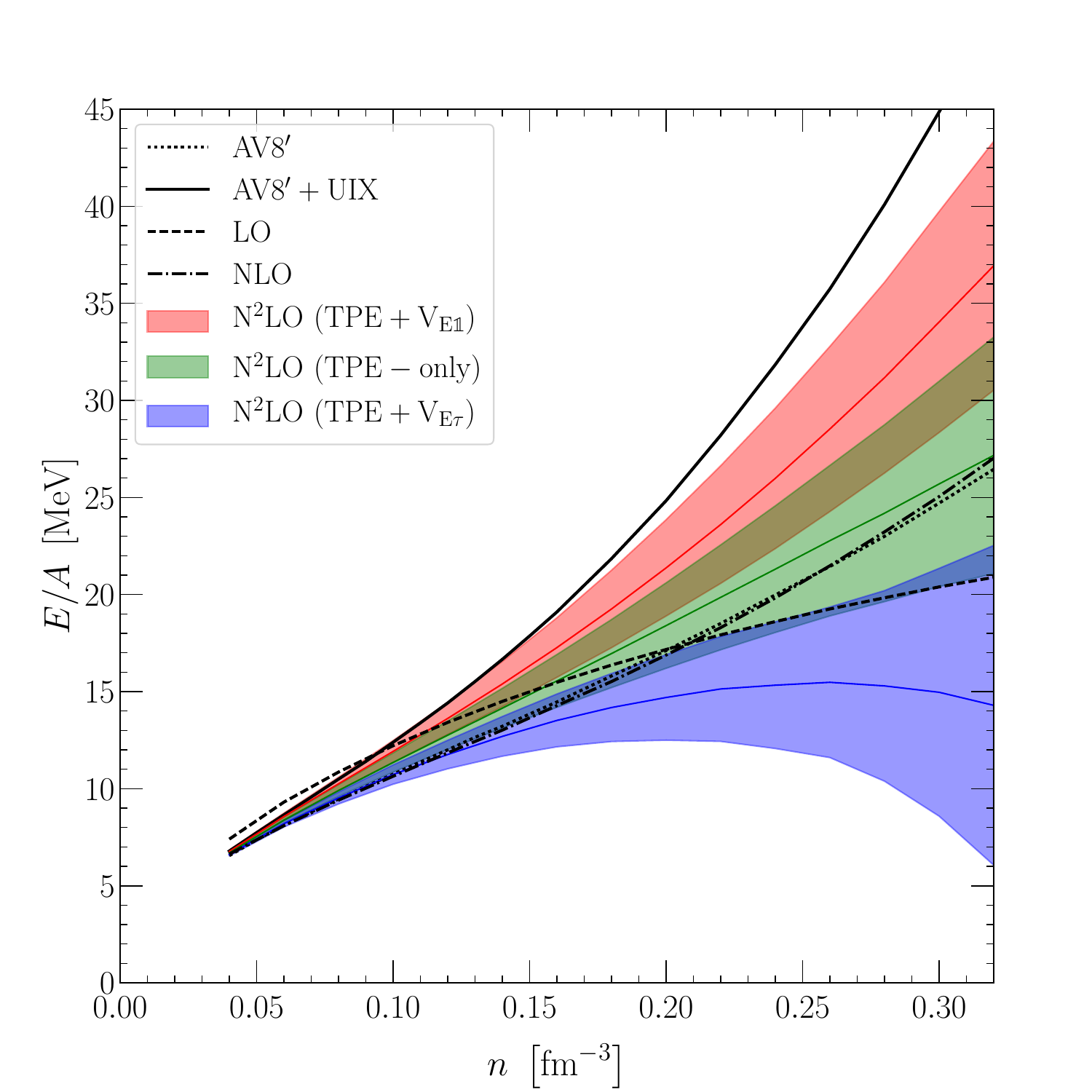}
\end{center}
\caption{The EOS of pure neutron matter calculated using the chiral Hamiltonians.
The figure is adapted from Ref.~\cite{Tews:2018}, and the description of the various results
is provided in the text.}
\label{fig:pnm2}
\end{figure}

More recent results for the EOS of pure neutron matter have been obtained
using chiral EFT Hamiltonians previously described. 
In this case
it is possible to estimate the systematic error due to the truncation in the
chiral expansion as previously discussed. 
Results are shown in Fig.~\ref{fig:pnm2}  at various orders in the EFT expansion, and compared with the results of Fig.~\ref{fig:pnm1}. 
The uncertainties are shown only at N$^2$LO.
The three-different bands (colors) are obtained from taking into account regulator artifacts for local chiral EFT interactions, see for example~\cite{Lynn:2015jua,Dyhdalo:2016ygz,Huth:2017wzw}. 
Although the three choices should be ``equivalent'' within EFT, this is broken by finite-range regulators and the results for the EOS differ.
In particular, the blue band is not physical as neutron batter would become unstable at densities  above saturation.
These regulator artifacts vanish at high regulator scales and show that the two upper bands (green and red) provide a reliable EOS that include  uncertainties.

Generally, the energy of pure neutron matter as a function of the density can be parametrized as
\begin{align}
\frac{E}{N}(n)=a\left(\frac{n}{n_0}\right)^\alpha+
b\left(\frac{n}{n_0}\right)^\beta \,,
\label{eq:pnm}
\end{align}
where $a$, $\alpha$, $b$, and $\beta$ are fitting parameters.
From this form, the value of the symmetry energy and its slope at saturation
density are given by
\begin{align}
&S=a+b+16 \,,
\nonumber \\
&L=3\,(a\alpha+b\beta) \,.
\end{align}
The parametrizations of the EOS showed in Fig.~\ref{fig:pnm1} are reported 
in Table~\ref{tab:fit}, together with the corresponding symmetry energy
and slope.

\begin{table*}[]
\centering
\begin{tabular}{@{} lcccccc @{}}
\hline
Hamiltonian & $S$ &$L$ & $a$    &  $\alpha$ & $b$ & $\beta$ \\
               & (MeV)  & (MeV) & (MeV)    &        & (MeV)  &  \\
\hline
AV8'                        & 30.5 & 31.3 & 12.7 & 0.49  & 1.78 & 2.26 \\
AV8'+UIX                    & 35.1 & 63.6 & 13.4 & 0.514 & 5.62 & 2.436\\
N$^2$LO (TPE+V$_{E1}$)      & 33.3 & 46.4 & 13.2 & 0.503 & 4.07 & 2.17 \\
N$^2$LO (TPE-only)          & 31.6 & 33.8 & 11.0 & 0.426 & 4.59 & 1.44 \\
\hline
\end{tabular}
\caption{Fitting parameters for the neutron matter EOS defined in Eq.~\ref{eq:pnm} using
different nuclear Hamiltonians.
For chiral interactions, the fit are obtained from the central values (solid lines) of
each band, and we omitted the unphysical results.}
\label{tab:fit}
\end{table*}

Within Skyrme models, the equation of state expressed as the energy per particle

\begin{align}
\frac{E}{N}&=\frac{3\hbar^2}{10m}\left(\frac{3\pi^2}{2}\right)^{2/3}n^{2/3}H_{5/3}
+\frac{t_0}{8}n[2(x_0+2) 
\nonumber \\
&-(2x_0+1)H_2]+\frac{1}{48}\sum_{i=1}^{3}t_{3i}n^{\sigma_{i}+1}[2(x_{3i}+2)
\nonumber \\
&-(2x_{3i}+1)H_2]+\frac{3}{40}\left(\frac{3\pi^2}{2}\right)^{2/3}n^{5/3}\left(aH_{5/3}+bH_{8/3}\,,
\right)
\label{eq:eosskyrme}
\end{align}
with
\begin{align}
&a=t_1(x_1+2)+t_2(x_2+2) \,, \nonumber \\
&b=\frac{1}{2}\left[t_2(2x_2+1)-t_1(2x_1+1)\right] \,, \nonumber \\
&H_n(y)=2^{n-1}[y^n+(1-y)^n]\,,
\end{align}
where $y=n_p/n$ is the proton fraction.
All the various parameters entering in the Skyrme functional are typically fit to reproduce selected properties of atomic nuclei and the properties of the known quantities of nuclear and neutron matter, like binding energy and charge radii of nuclei, saturation density and energy and nuclear compressibility of symmetric nuclear matter, 
symmetry energy, etc.
From the above expression, it is possible to derive the formula for the pressure,
compressibility, symmetry energy, effective mass, and others.

\begin{table*}[]
\centering
\begin{tabular}{@{} lcccccccccccc @{}}
\hline
Skyrme & $t_0$ & $t_1$ & $t_2$ & $t_{3}$ & $x_0$ & $x_1$ & $x_2$ & $x_{3}$ &
$\alpha$ & Reference \\
\hline
NRAPR & -2719.7 & 417.64 & -66.687 & 15042 & 0.16154 & -0.047986 & 0.027170 & 
0.13611 & 0.14416 & \citet{Steiner:2004fi} \\
SV-min & -2112.2 & 295.78 & 142.27 & 13989 & 0.24389 & -1.4349 & -2.6259 & 
0.25807 & 0.25537 & \citet{klupfel:2009} \\
UNEDF2 & -1735.5 & 262.81 & 1183.31 & 12293 & 0.17225 & -3.7669 & -1.3835 &
0.052864 & 0.35146 & \citet{Kortelainen14} \\
\end{tabular}
\caption{Parameters of the Skyrme equation of state for selected models.
The unit for $t_0$ is MeV fm$^3$, $t_1$ and $t_2$ are in MeV fm$^5$, the $t_{3i}$ are 
in MeV fm$^{3(1+\alpha)}$, and the other parameters are dimensionless.
}
\label{tab:skyrme}
\end{table*}

\subsection{The equation of state in $\beta$ equilibrium}
\label{sec:2e}

We have so far mainly discussed the EOS of pure neutron matter around saturation density.
However, as the chemical potential of neutrons in neutron stars is sufficiently high, it is energetically more favorable to have a finite fraction of protons and negatively charged leptons. 
Several neutron-star properties, like the mass and radius, can be calculated in a good approximation directly starting from the pure neutron matter EOS, but such an approximation misses the effects of the neutron-star crust and the remaining protons. 
Therefore, a more realistic EOS should include these effects which can be estimated from the PNM EOS and information on the saturation pint of symmetric nuclear matter.
Here, we describe how to include the effects of protons in the EOS
by starting from a few assumptions.
We will then discuss how to calculate neutron-star properties in the next chapter.

By starting from Eqs.~\eqref{eq:eosgeneral} and~\eqref{eq:enedensgeneral} we can solve 
for $\beta(\rho)$ by imposing $\beta$-equilibrium between neutrons, protons, 
electrons, and muons. This can be done by imposing the conservation of the total chemical potential.
Let us rewrite Eq.~\eqref{eq:eosgeneral} in terms of the proton fraction
\begin{align}
x=\frac{n_p}{n} \,,
\end{align}
and consider the total energy density of the system:
\begin{align}
\epsilon&=n\,[E(n,x)+m_n(1-x)+m_p\,x] \,,
\end{align}
where $m_n$ and $m_p$ are the neutron and proton rest mass.
It is easy to show that the difference between neutron and proton chemical potentials is given by
\begin{equation}
\mu_n-\mu_p=4(1-2x)S(n) \,,
\end{equation}
where $S(n)$ can be calculated from the EOS as in Eqs.~\eqref{eq:symmene2}
and~\eqref{eq:symmene1}.
Beta decay requires:
\begin{equation}
\mu_n-\mu_p=\mu_e=\mu_\mu\,. \\
\end{equation}
A good approximation is to consider electrons and muons as non-interacting particles.
This is reasonable as their fraction is small compared to the number of neutrons that screen the Coulomb interaction between them. 
Note that this is not true for protons, as they interact with neutrons via  strong interactions.

The electron chemical potential for relativistic and degenerate electrons is given by
\begin{equation}
\mu_e=(m_e^2+\hbar^2 k_F^2)^{1/2}=[m_e^2+\hbar^2 (3\pi^2 n x_e)^{2/3}]^{1/2} \,,
\end{equation}
where $x_e$ is the electron fraction defined in a similar way as the proton fraction, and $n$ is the total density of the system.
Since the mass of muons is much larger, they can be treated as a non-relativistic free Fermi gas. 
Their chemical potential is given by
\begin{equation}
\mu_\mu=[m_\mu^2+\hbar^2 (3\pi^2 n x_\mu)^{2/3}]^{1/2} \,.
\end{equation}

All the fractions $x$, $x_e$ and $x_\mu$ can be calculated by imposing charge neutrality and using the relations for the chemical potentials as above:
\begin{equation}
x=x_e+x_\mu \,.
\end{equation}

\begin{figure}
\includegraphics[width=0.5\textwidth]{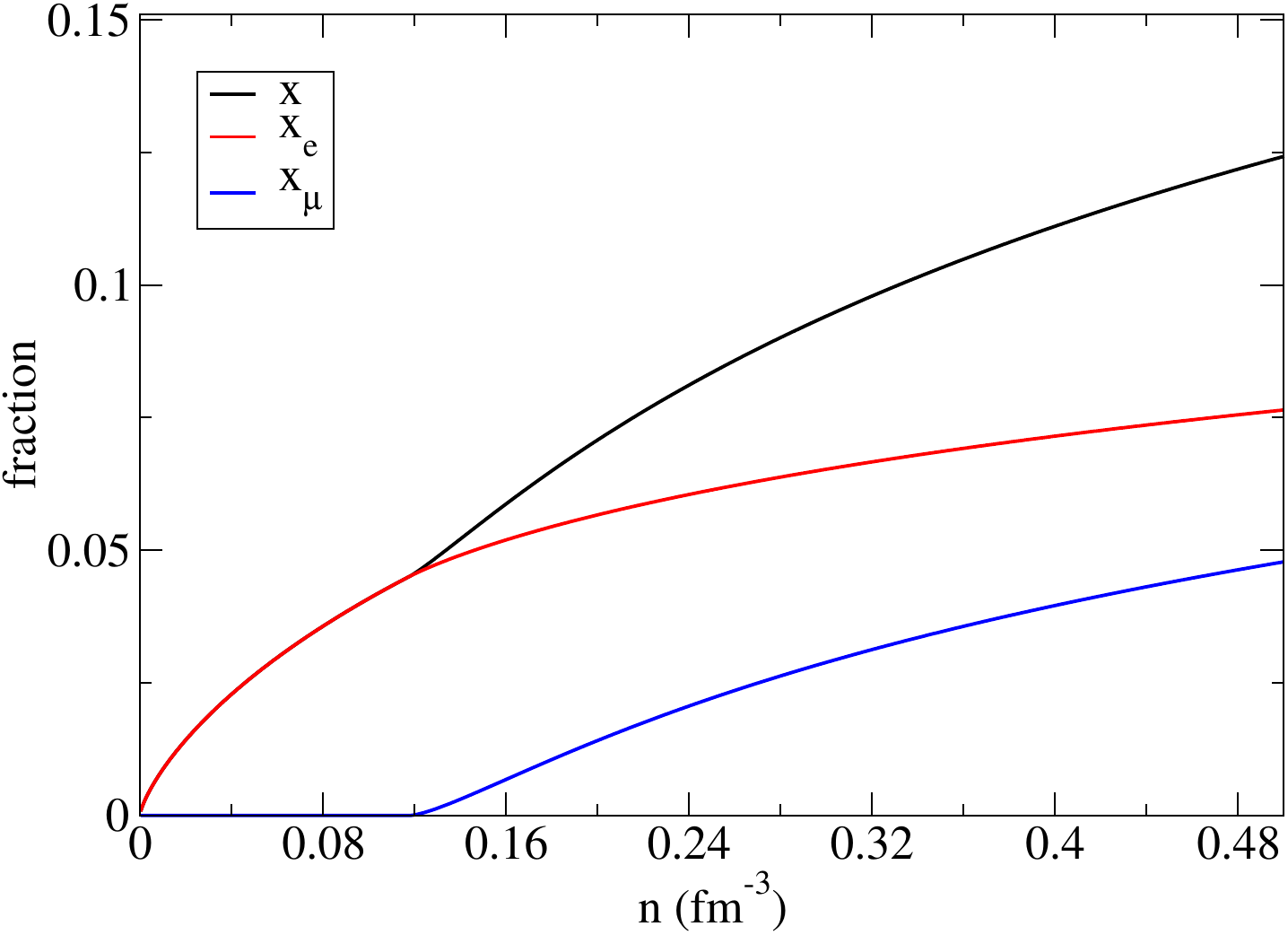}
\caption{Proton and lepton fractions obtained from
the AFDMC EOS defined in Table~\ref{tab:fit} using the AV8'+UIX 
parametrization. The figure is adapted from Ref.~\cite{Gandolfi:2019}.}
\label{fig:fraction}
\end{figure}

\begin{figure}
\includegraphics[width=0.5\textwidth]{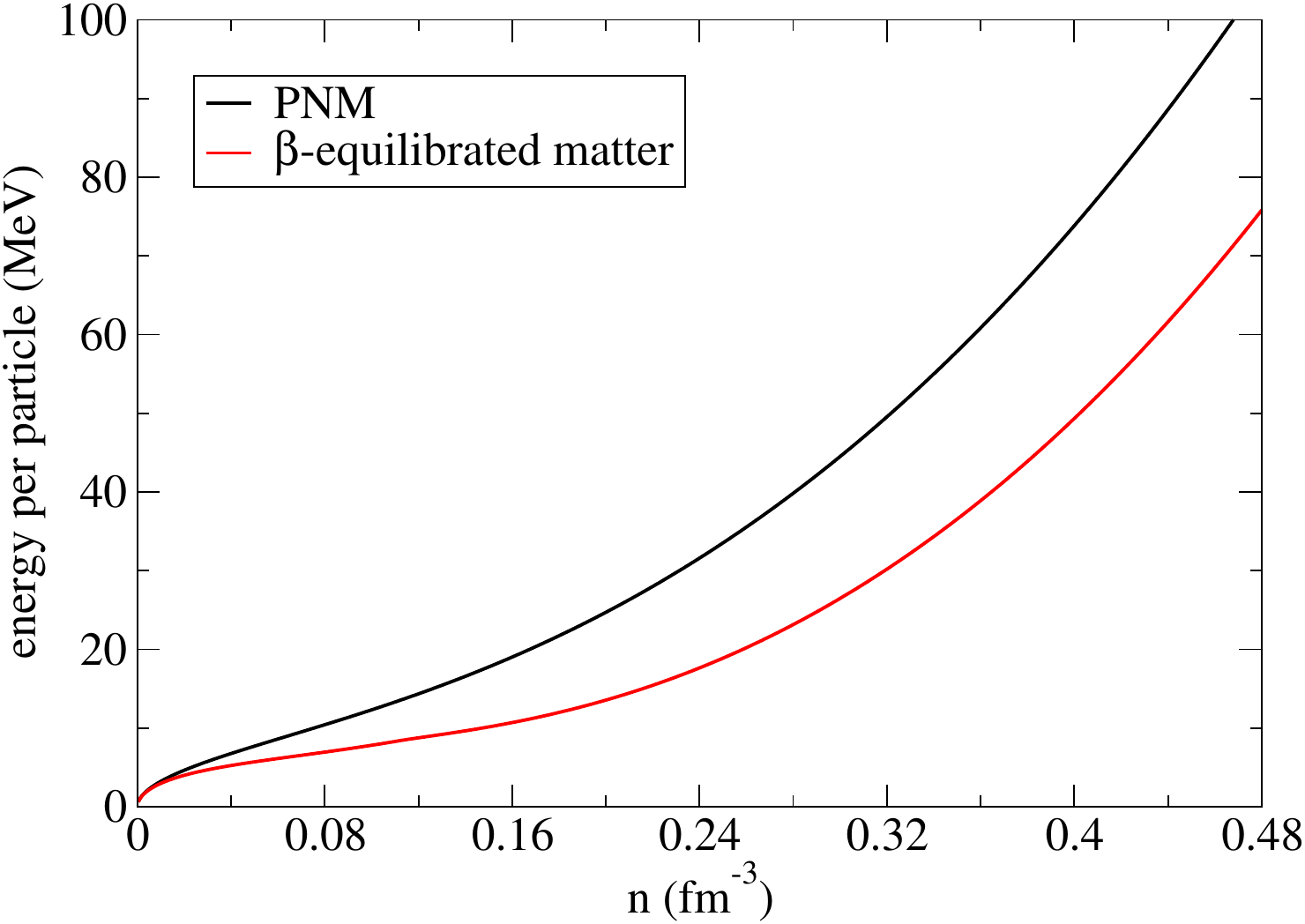}
\caption{The EOS of pure neutron matter compared to the $\beta$-equilibrated
using the AV8'+UIX nuclear Hamiltonian.}
\label{fig:eosbeta}
\end{figure}

As an example, we show the proton and lepton fractions for the EOS in Table~\ref{tab:fit} using the AV8'+UIX parametrization of  Fig.~\ref{fig:fraction}.
Protons and electrons  have finite fractions starting at very low densities because the symmetry energy $S(n)$ is finite at any density.
The muons, instead, appear  at larger densities, around $n\approx 0.12$ fm$^{-3}$, because of their larger mass compared to the electrons.

The EOS for pure neutron matter is compared to the $\beta$-equilibrated one 
in Fig.~\ref{fig:eosbeta}. When including protons and leptons, the EOS is softened.
The EOS of homogeneous matter in $\beta$-equilibrium is a valid  model for high densities in the outer crust of neutron stars, where nuclei are not present anymore.

\subsection{Matter at high densities: quarks, hyperons, pions, and other degrees of freedom}

At higher densities, degrees of freedom other than nucleons and electrons may appear, and a reasonable first step is to presume that matter may consist of degrees of freedom which we have observed in the laboratory, e.g. exotic mesons and baryons. 
Because QCD is asymptotically free, quarks may also become deconfined. 
Finally, the relevant degrees of freedom in neutron stars may be qualitatively new degrees of freedom which we have not yet otherwise observed. 

We expect baryon number to be approximately conserved and electric charge to be exactly conserved.
Thus, the degrees of freedom can be organized according to the chemical potential for baryon number $\mu_B$ and charge $\mu_Q$
\begin{equation}
    \mu_i = B_i \mu_B + Q_i \mu_Q\,,
    \label{eq:chem_pot}
\end{equation}
where $B_i$ is the baryon number of particle $i$ and $Q_i$ is the charge of particle $i$ (keeping in mind that the chemical potentials of each particle include the contribution from the rest mass). Except in dynamical systems like core-collapse supernovae and neutron star mergers, we expect strangeness not to be conserved.

Most of the work in dense matter is done in the mean-field approximation. 
Additionally, the dispersion relations (the relationship between energy and momentum) of the relevant degrees of freedom are often assumed to be similar to their noninteracting counterparts, except that the mass and chemical potential are modified by the mean field. 
In the mean-field approximation at $T=0$, exotic fermions appear if their chemical potential is greater than their mass (keeping in mind that both the chemical potential and mass may be modified by the dense medium). 
The maximum baryon chemical potential in neutron star interiors is typically less than about $2\,\mathrm{GeV}$, and thus, the quark chemical potential is less than $700\,\mathrm{MeV}$, which rules out tau leptons and hadrons containing charm, bottom, or top quarks (but strange quarks may occur). 
At $T=0$, bosons appear when they form a Bose condensate, i.e. when their chemical potential is equal to their mass.

Models of hyperons in dense matter typically take two forms: extensions of relativistic mean-field theories~\cite{Horowitz81} to include hyperon degrees of freedom (see e.g. Ref.~\cite{Reinhard:1989zi}) and applications of Brueckner-Hartree-Fock theory~\cite{Vidana:1999jm}. 
These models can be constrained by the increasing amount of data on hypernuclei (see quantum Monte Carlo results in~\citet{Lonardoni:2013,Lonardoni:2014}). 
For example, the binding energy of the $\Lambda$ hyperon in isospin-symmetric nuclear matter is approximately $-28\,\mathrm{MeV}$. 
There is also an increasing amount of information on hyperon-nucleon interactions from Lattice QCD~\cite{Beane:2006gf}. 
Adding degrees of freedom to a system of degenerate fermions tends to lower the pressure, thus hyperons tend to lower the maximum mass of a cold neutron star~\cite{Glendenning:1991es}.
Hyperonic models sometimes need to be supplanted with quark models at higher densities in order to increase the pressure enough to explain the existence of two solar mass neutron stars. 

An early model of Bose condensation based on a chirally-symmetric Lagrangian comes from Ref.~\cite{Kaplan86}. 
At lower densities, pions can be described by EFTs similar to those without explicit pion degrees of freedom described above.
By Eq.~\eqref{eq:chem_pot} above, negatively-charged pions and muons have the same chemical potential. 
Because pions and muons also have similar masses one naively expects that pions and muons appear in similar conditions in dense matter. 
However, strong interactions involving pions strongly modify the in-medium values of the mass and the chemical potential.

Descriptions of deconfined quark matter typically take one of four forms: (i) purely phenomenological models similar to the ``MIT Bag'' model, (ii) models related to the Nambu--Jona-Lasinio model where quarks are described with a non-renormalizable Lagrangian with chiral symmetry~\cite{Klevansky92,Hatsuda:1994pi}, (iii) high-density effective theories~\cite{Hong:1998tn}, and (iv) and calculations from perturbative QCD (see, e.g.~\citet{Komoltsev:2021jzg,Gorda:2022jvk,Komoltsev:2023zor}. 
Also, the relevant degrees of freedom in the core of neutron stars may be different than those encountered in vacuum, for a recent example see Ref.~\cite{Jeong:2019lhv}.

In lieu of these more complex models of dense matter which are specialized to a particular class of degrees of freedom, more generic phenomenological models have been used. 
These more generic models can express all possible equations of state in a form which is easy to evaluate. 
These kinds of models are often useful in the analysis of neutron-star observational data to obtain EOS of $T=0$ matter in beta-equilibrium.
In these models, above some density $n_t$, that is usually chosen as $n_0<n_t<2n_0$, the EOS is
phenomenologically modeled.

One posible approach is  using polytropes, i.e.
\begin{equation}
P_i(\epsilon)=K_i\epsilon^{1+1/n_i} \,,
\end{equation}
where $P$ if the pressure now as a function of the energy density $\epsilon$,
and $K_i$ and $n_i$ are parameters that can change the soft-stiffness of the EOS.
Several polytropes can be used, in such a way the EOS can have more 
freedom, and also include eventual phase transitions. The different polytropes
are used to model the EOS in different density regimes (see, e.g., \citet{Read09}) 
For example:
\begin{align}
P(\epsilon)=
\begin{cases}
K_1\epsilon^{1+1/n_1} \quad  n_t<n<n_{t1}   \\
K_2\epsilon^{1+1/n_2} \quad  n_{t1}<n<n_{t2} \\
K_3\epsilon^{1+1/n_3} \quad  n_{t2}<n<n_{t3} \,.
\end{cases}
\end{align}
In this case, given the transition densities $n_t$, $n_{t1}$, $n_{t2}$ and $n_{t3}$,
the parameters must be chosen in order to ensure the pressure is monotonically increasing across a finite number of jumps in energy density (which correspond to phase transitions).

\begin{figure}[t]
\centering
\includegraphics[width=0.5\textwidth]{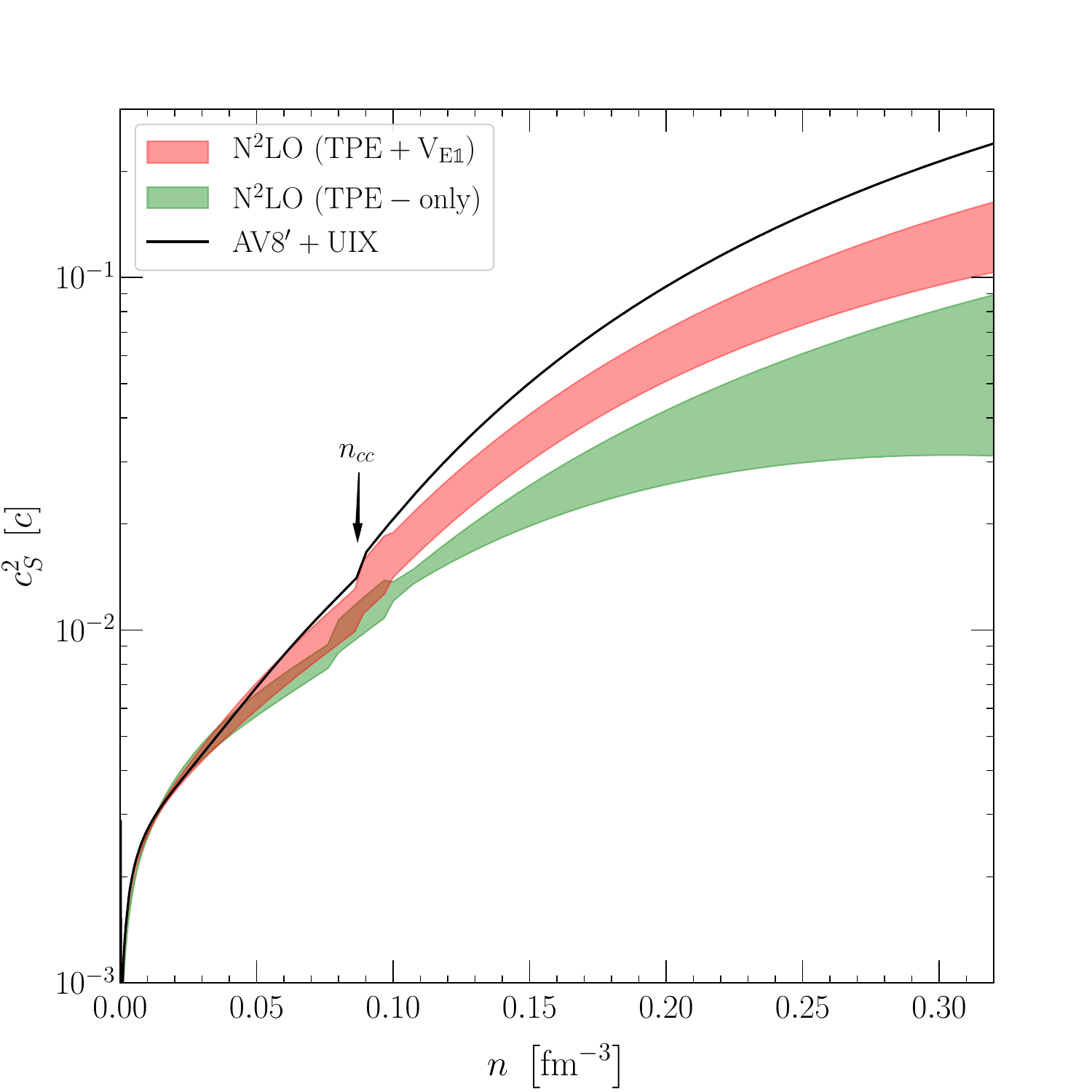}
\caption{Speed of sound as a function of density for NS matter based
on selected EOS previously presented in Fig.~\ref{fig:pnm2}. In this case an EOS for the crust has
been included for densities lower than $n_{cc}\approx n_0/2$, The Figure is taken from Ref.~\cite{Tews:2018}.}
\label{fig:cs}
\end{figure}

\begin{figure}[t]
\centering
\includegraphics[width=0.5\textwidth]{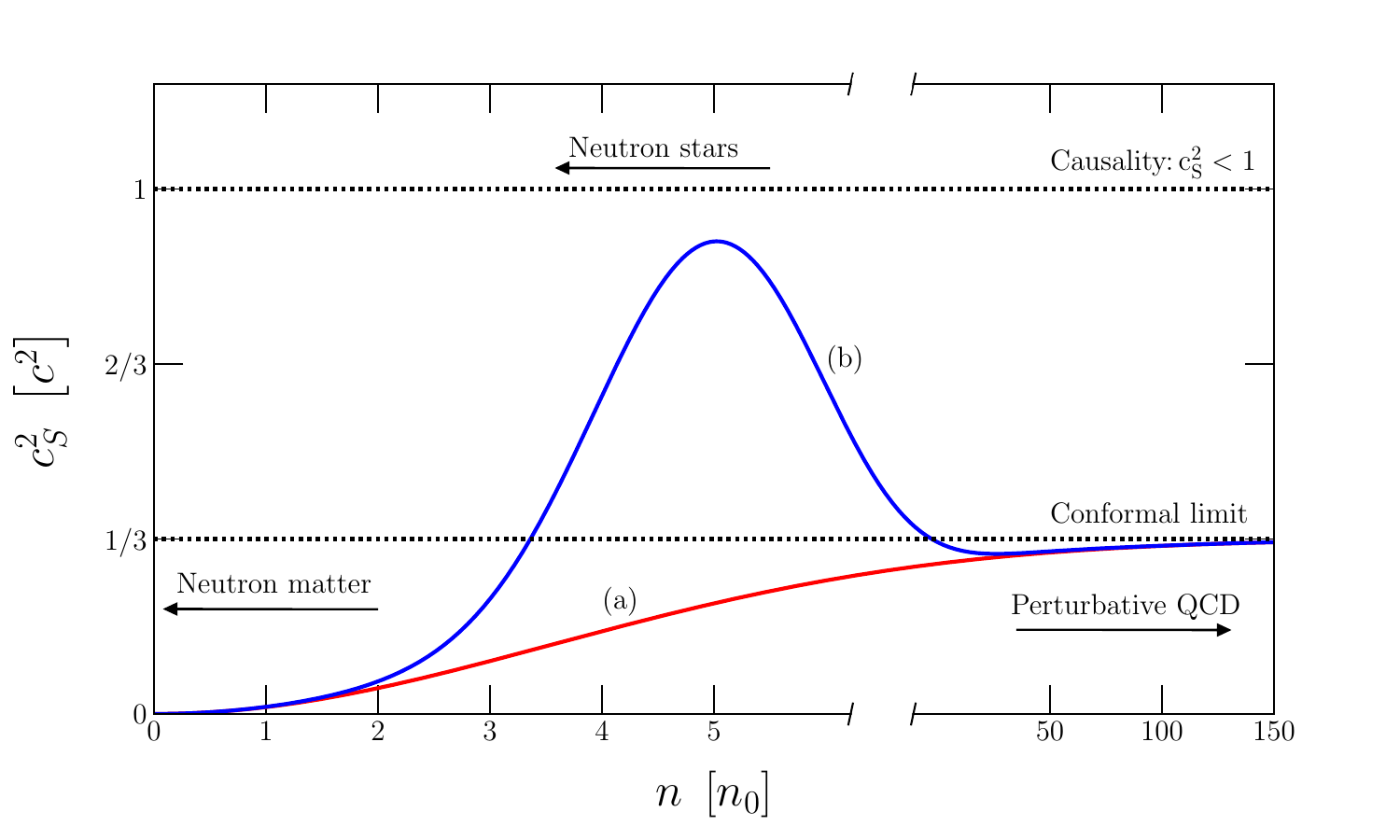}
\caption{Two possible scenarios for the evolution of the speed of sound in dense matter, see Ref.~\cite{Tews:2018} for details.}
\label{fig:cs2}
\end{figure}

Another parametrization for the high-density EOS is in terms of the speed of sound $c_s$ (see, e.g., \citet{Tews:2018kmu} or \citet{Greif:2018njt}).
The speed of sound can be obtained in different ways, for example:
\begin{align}
c_s^2=\frac{\partial P(\epsilon)}{\partial\epsilon} \,.
\end{align}
Causality imposes the constraint that $c_s$ cannot exceed the speed of light, mechanical stability imposes a non-negative speed of sound, and at a sufficiently high density $c_s/c\rightarrow 1/3$ since quarks at high momenta are asymptotically free. 
In the regime where the EOS is dominated by neutron matter (as discuss in the above sections),
the value of $c_s$ can be directly obtained from the calculated EOS; the speed of sound can be obtained from selected Hamiltonians presented in the previous sections and  is 
shown in Fig.~\ref{fig:cs}. 
We find that $c_s$ grows with the density for different Hamiltonians.
At higher densities, $c_S$ can have different possible behaviours.
In Fig.~\ref{fig:cs2} we show schematically two possible curves for $c_S$, one exceeding the 
limit of $1/3$ at intermediate densities and another one where this limit is valid. 
Exceeding the limit of $1/3$ implies the existance of strongly correlated matter in the core of neutron stars.
It is possible to parametrize the behavior of $c_S$ with some generic model.

\section{Constraints from nuclear experiments}

We will now give an overview of experimental constraints to the EOS. 
We will describe how nuclear observables, like binding energies, radii and others, are used to construct nuclear interactions. 
Then we will focus on experimental data that is commonly used to constrain the EOS, in particular on measurements of the neutron-skin thickness and its connection to macroscopic properties like symmetry energy and its derivatives.
This section will also describe the connection between the neutron-star EOS and heavy-ion collisions. 
We expect that, in the near future, gravitational wave observations will provide a significant amount of information about the equation of state. 
However, heavy-ion collisions provide two important constraints.
First, they potentially constrain neutron-star matter between the nuclear saturation density (where the momenta are sufficiently small that chiral effective theory provides an accurate description of neutron star matter) and higher densities which will be constrained by future GW observations. 
Second, they potentially constrain the in-medium nucleon single-particle potentials, information which is difficult to access from GW observations.

\subsection{Nuclear structure experiments}

As described in the previous section, the nuclear EOS can be calculated by starting from nuclear Hamiltonians
that include two- three- and many-body interactions in general.
These interactions contain free parameters, that need to 
be constrained by available experiments.

Nuclear Hamiltonians have been derived in a phenomenological
way, or based on chiral effective field theory. Although several parameters can be extracted from nuclear experiment,
for example $\pi$-nucleon couplings, many other parameters have to be (indirectly) constrained by other type of experiments that need to solve the nuclear two- and many-body problem, for example the binding energy and other properties of light nuclei.
The parameters entering the two-body interactions are commonly fit in order to reproduce nucleon-nucleon scattering
data, see for example~\cite{Wiringa:1995,Epelbaum:2004fk,Entem:2003}.

The fitting of free parameters entering in the three-body interaction need a different approach. The most precise measured
nuclear observables include the binding energy (or nuclear masses) and charge radii, provided by experimental 
electron scattering off nuclei that are very well under control. 
Typically, three-body interactions are constrained
by calculating properties of very light nuclei, where exact calculations of the nuclear ground state are available.
Observables that are commonly chosen include the binding energy and/or charge radii of light nuclei like $^3$H, $^3$He, $^4$He, and neutron-$\alpha$
scattering, see for example Ref.~\cite{Lynn:2016}.
Other observables of such light systems can also be used, for example
neutron-deuteron scattering.
Free parameters entering the nuclear Hamiltonians have
also been constrained to reproduce the saturation density and energy of symmetric nuclear matter and/or properties of 
medium nuclei~\cite{Ekstrom:2015}.
However, this latter approach relies on many-body calculations of such systems that are not as 
accurate as calculations of few-body systems.
Within chiral effective field theory, also nuclear $beta$-decays can be included in the fit.
This is because the electro-weak operators that describe the interactions between nucleons with an external lepton and neutrino are 
related to three-body interactions~\cite{Baroni:2018}.

Density functionals and/or models based on the mean-field approximation, like Skyrme interactions, are typically constrained by considering only binding energies and charge radii and distributions of medium- heavy-nuclei.

\subsection{Neutron-skin thickness}

Heavy atomic nuclei typically have an excess of neutrons over protons, with proton fractions of the order of 40\%.
The radius of protons in the nucleus, as measured by the point proton radius $R_p$, is then usually smaller than the radius of the neutrons measured by the point neutron radius $R_n$.
The neutron-skin thickness is defined as their difference, $R_{\rm skin}=R_n-R_p$~\cite{Thiel_2019}.
Qualitatively, the neutron-skin thickness is proportional to the pressure among neutrons, as a larger pressure leads to the neutrons occupying a larger volume.
Therefore, the neutron-skin thickness is an important observable to learn about the behavior of the neutron-matter EOS.
Usually, the neutron-skin thickness is used to constrain the slope of the symmetry energy $L$ introduced before, because $R_{\rm skin}$ and $L$ have been found to be  correlated~\cite{horowitz:2001,Typel:2001,carriere:2003,klupfel:2009,Vinas:2013hua,Reinhard:2016mdi, Mondal2016}.
A larger skin implies a larger value of $L$ and hence, a larger pressure of pure neutron matter at saturation density.

In the past years, several experiments aimed at measuring the neutron-skin thickness to constrain the EOS of neutron matter. 
These include measurements of the neutron-skin thickness in $^{208}$Pb via the dipole polarizability, see, e.g. Refs.~\citep{Birkhan_2017, Tamii_2011, Roca-Maza_2015}, or via parity violating electron scattering by the PREX-II experiment~\cite{PREX2}.
Analogously to the latter, the skin has also been measured in $^{48}$Ca by the CREX experiment~\cite{Adhikari_2022}.

\begin{figure}
    \centering
    \includegraphics[width = \linewidth]{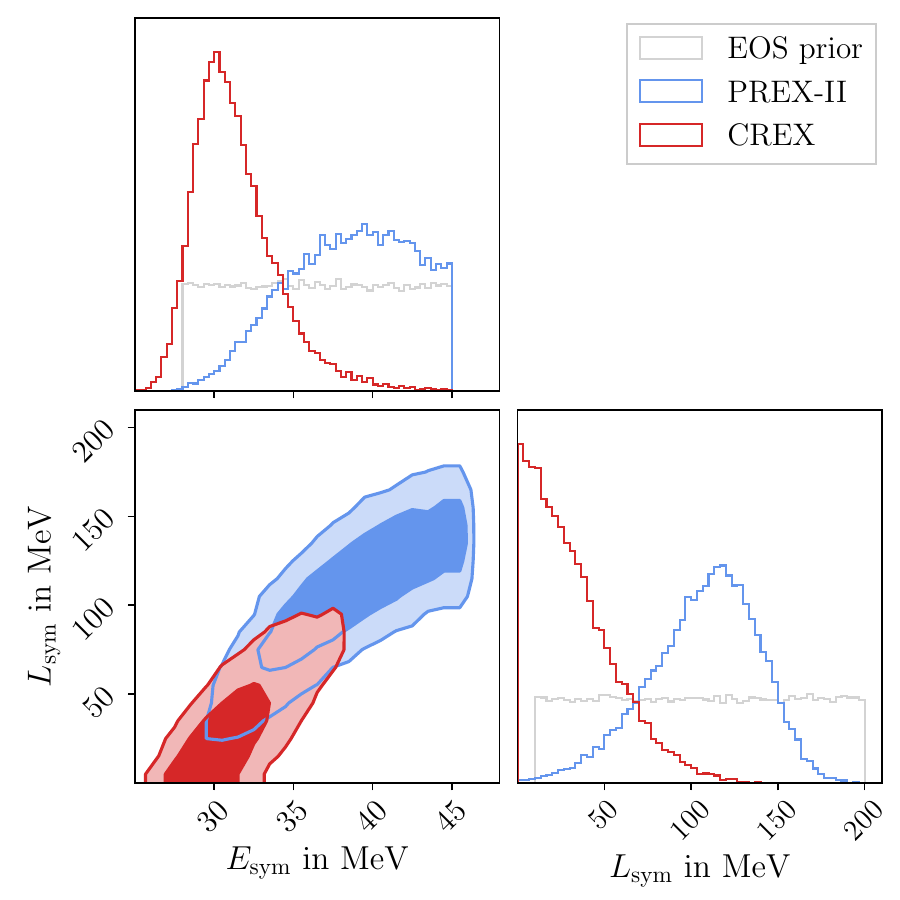}
    \caption{Posteriors of the symmetry energy $E_{\rm sym}$ and its slope $L_{\rm sym}$ at saturation density inferred from the PREX-II (blue) and CREX (red) experiments from Ref.~\cite{Koehn:2024set}. 
    The contours denote the 68\% and 95\% credibility intervals. 
    The prior is shown in gray.
    Figure taken from Ref.~\cite{Koehn:2024set}.
}
    \label{fig:ENP}
\end{figure}

The impact of these measurements on the EOS were recently analyzed in Refs.~\citep{Reed_2021,Essick_2021,Essick:2021ezp,Koehn:2024set}.
Figure~\ref{fig:ENP} shows the resulting posterior for the symmetry energy and its slope at $n_0$ extracted from the CREX and PREX-II campaigns in Ref.~\cite{Koehn:2024set}.
It can be seen that both experiments give posteriors that have only modest overlap~\citep{Miyatsu_2023,Reed_2023}. 
This might point at the need for improved theoretical modeling to reduce systematic uncertainties in both measurements.
In addition to these experiments, Ref.~\citep{Essick:2021ezp} also analyzed the impact of measurements of the dipole polarizability of $^{208}$Pb, and found these to be in good agreement with predictions from chiral EFT and CREX~\cite{Tamii:2011pv, RocaMaza:2013, RocaMaza:2015}.

\subsection{Heavy-ion collisions}

Intermediate-energy heavy-ion collisions (HICs) have the promise to provide novel constraints on the EOS of dense matter because they probe matter above the nuclear saturation density\footnote{Higher energy heavy-ion collisions achieve higher temperatures but achieve lower densities than intermediate energy collisions.}. 
However, this is not without complications: during the time which high-densities are obtained, the temperatures are still large and matter is not in equilibrium. 
In addition, at these energies, nucleons are still often clustered into light nuclei. 
Finally, the particles which are observed in HICs(nuclei, nucleons, pions, etc.) are not necessarily connected in a model-independent way to the hot and dense matter achieved in the collision itself. 
Thus, complicated modeling is required to convert information obtained from intermediate-energy HIC observables to quantitative constraints on the nature of hot and dense matter in thermodynamic equilibrium.

There are several methods for analyzing intermediate-energy HIC data, the most popular are based on the Boltzmann-Uehling-Uhlenbeck (BUU) equation~\cite{Danielewicz91,Li93} and quantum molecular dynamics (QMD)~\cite{Zhang06}. 
The BUU equation describes a semi-classical formalism that describes the space-time evolution of the particle distribution functions in terms of collision integrals, which encode the information about reactions between the associated degrees of freedom. 
In QMD, nucleons are represented by Gaussian wavepackets which interact with each other and with the background mean field. 
QMD is particularly adept at treating clusters of nucleons, which are more difficult to describe in the BUU formalism. 

Despite these challenges, HICs provide vital information for hot and dense matter, see for example~\cite{Li:2008gp,Tsang:2008fd,Huth:2021bsp,Agnieszka:2024}. 
First, they are the only constraint on the nature of isospin-symmetric matter at high densities (pioneered by \citet{Danielewicz:2002pu})
since neutron stars are neutron-rich. 
Second, they have demonstrated that the simple noninteracting dispersion relation for nucleons is not likely appropriate for all densities and temperatures~\cite{Aichelin87}. 
This latter point may be particularly important for the proper modeling of neutron-star mergers. 
A particularly important frontier on the theory side is the systematic calibration of various numerical methods used to analyze HIC data~\cite{Xu16}

\section{From microphysics to neutron star structure}
\label{sec:TOVtides}

This section reviews how the microphysical description on dense nuclear matter impacts the macroscopic properties of neutron stars.
We review the general-relativistic equations of stellar structure, the Tolman, Oppenheimer, Volkoff (TOV) equations, in Sec.~\ref{sec:TOV} and how they directly determine the mass and radius. 
Tidal properties of neutron stars and the tidal deformability parameter are discussed in Sec.~\ref{sec:tidaldeformability}.

\begin{figure}
\centering
\includegraphics[width=0.99\linewidth]{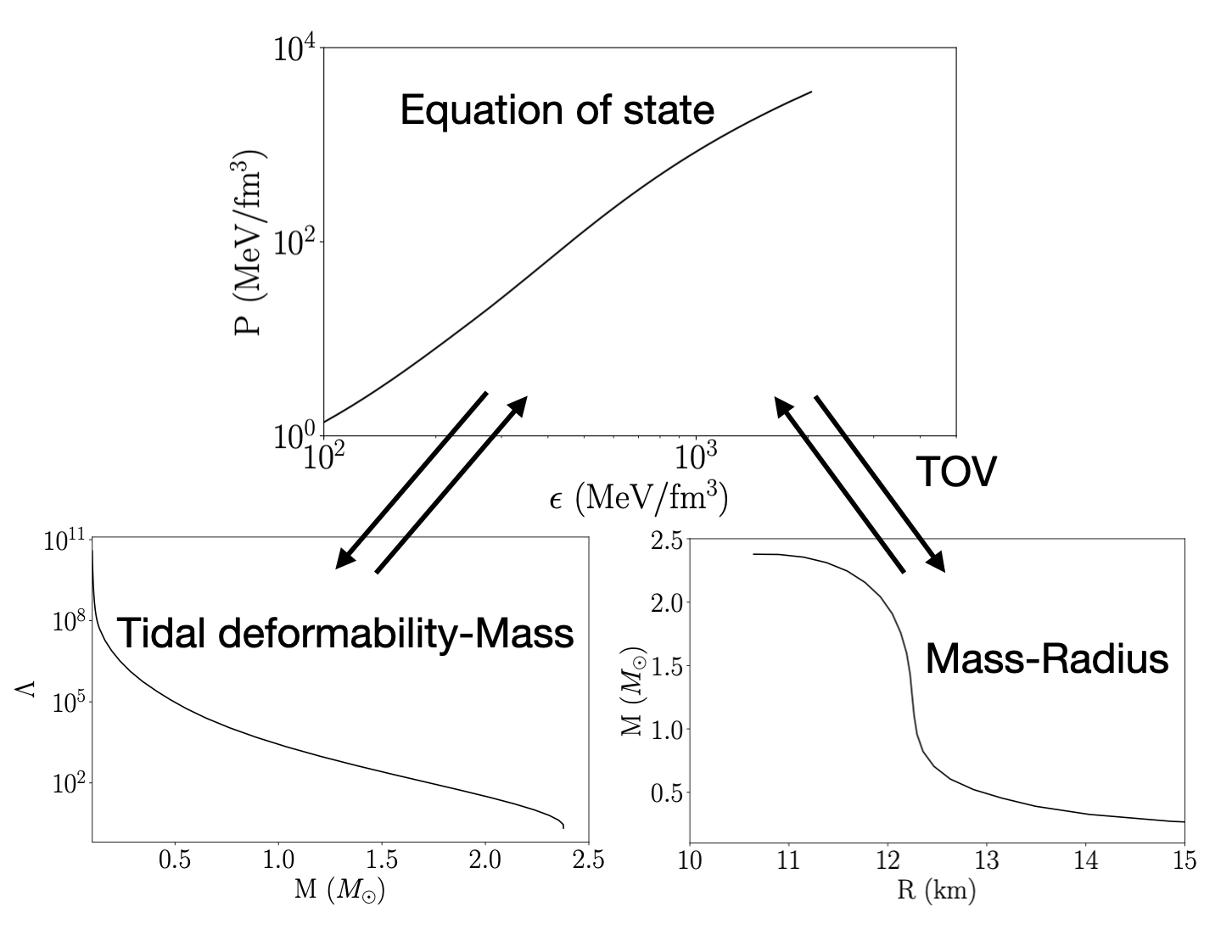}
\caption{From microscopic to macroscopic neutron-star properties. 
A nuclear EOS for the pressure $P$ as a function of the energy density $\epsilon$ (top right) completes the equations of stellar structure, Eqs.~\eqref{eq:mass} and~\eqref{eq:TOV}, and leads to a unique curve on the mass $M$ and radius $R$ plane (middle left) for static, spherically symmetric neutron stars.
Perturbed neutron stars are deformed away from equilibrium. 
The $\ell=2$ perturbation is characterized by the dimensionless tidal deformability $\Lambda$ as a function of the mass (bottom left), Eq.~\eqref{eq:lambda}, which again depends on the EOS.}
\label{fig:TOVseq}
\end{figure}

\subsection{The Tolman-Oppenheimer-Volkoff equation}
\label{sec:TOV}

Neutron stars are the most compact material objects we know of, with compactness $C\equiv M/R\sim 0.2$. 
This suggests that general relativistic effects meaningfully impact their macroscopic appearance.
Luckily, spherically symmetric solutions in General Relativity are quite simple: Birkhoff's theorem states that vacuum spherically symmetric solutions are also static and asymptotically flat, which implies that the spacetime outside a nonrotating neutron star is equal to the well-known Schwarzschild spacetime. 
The interior solution must match to the Schwarzschild one on the star surface and satisfy the non-vacuum Einstein Equations.

The most simple interacting fluid is the perfect fluid, with no heat conduction or viscosity, that can be fully characterized by its isotropic pressure $P$ and its energy density $\rho$. 
These conditions lead to a diagonal stress-energy tensor
\begin{equation}
T_{\mu \nu}=\left(\rho+P\right) u_\mu u_\nu+P g_{\mu \nu}\,,
\label{eq:stress-energytensor}
\end{equation}
where $u_\mu$ is the fluid 4-velocity and $g_{\mu \nu}$ is the spacetime metric. The time-time component of the Einstein Equations is 
\begin{equation}
\frac{d m(r)}{d r}=4 \pi r^2 \rho(r)\,,
\label{eq:mass}
\end{equation}
where $r$ is the radial coordinate and $m(r)$ is, so far, a function that parametrizes the time-time component of the metric. Matching to the Schwarzschild solution allows us to identify $m(R)=M$, the neutron star mass as measured by a distant observer.\footnote{This is not the density integrated over the star volume, the difference being  the (negative) gravitational binding energy.} Then, $m(r)$ is the enclosed mass and local flatness at the star center imposes $m(r=0)=0$.

Momentum-energy conservation, $T^{\mu \nu}_{\quad;\nu}=0$ and the r-r component of the Einstein Equations yield the TOV equation~\cite{Tolman:1939,Oppenheimer:1939}, the general relativistic equation of hydrostatic equilibrium between outward pressure and inward gravitational attraction
\begin{equation}
\frac{d P}{d r}=- \left[\varepsilon(r)+P(r)\right]\left[\frac{m(r)}{r^2}+4 \pi r P(r)\right]\left[1-\frac{2 m(r)}{r }\right]^{-1}\,.
\label{eq:TOV}
\end{equation}
This is the special and general relativistic generalization to the Newtonian equation of hydrostatic equilibrium, and to which it reduces for $P(r)\ll\rho(r)$ and $m(r)\ll r$.
In each square bracket, the second term provides corrections beyond Newtonian physics: the first two are special relativistic, while the third is a general relativistic correction. 
The last term also justifies use of $C=M/R$ as a measure of how important general relativistic corrections are.
Notably, all three corrections increase the effect of gravitational attraction.

Finally, the pressure $P$ appears both on the left-hand and the right-hand side of Eq.~\eqref{eq:TOV}. The former is pressure's usual role: it provides an outward force stabilizing the star. 
The latter is novel. 
In General Relativity all forms of energy gravitate and therefore pressure contributes to the gravitational pull as well. 
This suggests that, regardless of how much pressure nuclear physics can create, there is a limit to how massive neutron stars can become before the pressure derivative diverges and the star collapses.

Equations~\eqref{eq:mass} and~\eqref{eq:TOV} form a system of two equations for three unknowns: $m(r), P(r),$ and $\varepsilon(r)$. The system is closed by a third external equation, the neutron star EOS
\begin{equation}
    P=P(\varepsilon)\,.
    \label{eq:eos}
\end{equation}
This system of three equations for three unknowns defines a sequence of neutron stars as a function of mass $M$ or, equivalently, central density $\rho_c$ or pressure $P_c=P(\rho_c)$, c.f.,~\cite{Silbar:2003wm}.
For a given central pressure, Eqs.~\eqref{eq:mass} and~\eqref{eq:TOV} can be integrated outward with boundary conditions $m(r=0)=0$ and $P(r=0)=P_c$. 
The star's surface is reached when $P(r=R)=0$ at which point $M=m(r=R)$ is the star's mass. 
Repeating this process for different values of $P_c$ ``converts" an equation of state $P(\rho)$ to a stellar sequence $M(R)$ as shown in Fig.~\ref{fig:TOVseq}.
The mapping is one-to-one and invertible~\cite{Lindblom:1992}.

\subsection{Tidal deformability}
\label{sec:tidaldeformability}

The TOV equations describe the equilibrium configuration of a static, spherically symmetric neutron star.
Neutron stars in binaries will, however, be perturbed from their spherically symmetric equilibrium due to the gravitational impact of their binary companion.
An external gravitational field that is spatially inhomogeneous will result in a varying spacetime curvature throughout the star's $\sim25\,$km-wide body. 
This tidal interaction is conceptually similar to Newtonian tidal forces~\cite{Poisson_Will_2014}, though now the phenomenon needs to be described in terms of quantities derived from the spacetime metric, rather than forces.

In a binary, the external tidal field created by the companion varies in the orbital timescale given by Kepler's law
\begin{equation}
\tau_{\rm orb} \sim \frac{1}{\omega_{\rm orb}}\sim \sqrt{\frac{r^3}{M_T}}\,,
\end{equation}
where $\omega_{\rm orb}$ is the orbital frequency, $r$ is the binary separation and $M_T$ is the binary total mass.
Responding to this, the neutron star adjusts to a new non-sperical equilibrium configuration at a timescale characteristic of its internal oscillation modes, the fundamental of which is dominant 
\begin{equation}
    \tau_{\rm resp} \sim \frac{1}{f_{\rm 2}}\sim \sqrt{\frac{R^3}{M}}\,,
\end{equation}
where $f_{\rm 2}$ is the fundamental mode frequency, $R$ is the perturbed neutron star radius and $M$ is its mass. 
For sufficiently widely separated binaries,
\begin{equation}
\frac{\tau_{\rm resp}}{\tau_{\rm orb}}\sim \left(\frac{R}{r}\right)^{3/2}\ll 1\,,    
\end{equation}
implying that the perturbation is static and the neutron star ``instantaneously" adjusts the location of the induced tidal bulge to face toward and away from its companion.
Physically, the companion tidal field acts as a non-resonant driving force on the star's fundamental modes.

In General Relativity, the tidal deformation manifests in the asymptotic behavior of the spacetime metric at large distances from the star.
In the asymptotic Cartesian rest-frame of the neutron star whose origin coincides with its center of mass, the time-time component of the metric $g_{t t}$ can be expressed via a multipolar expansion~\cite{Thorne1998,Thorne1980}
\begin{align}
\frac{1-g_{t t}}{2}  =&-\frac{M}{r}-\frac{3 Q_{i j}}{2 r^3}\left(n^i n^j-\frac{1}{3} \delta^{i j}\right)+{\cal{O}}\left(r^{-4}\right) \nonumber\\
& +\frac{1}{2} \mathcal{E}_{i j}r^2 n^i n^j+{\cal{O}}\left(r^3\right)\,,
\label{eq:metric_multipolar}
\end{align}
where $r$ is the distance from the neutron star, $n_i$ are unit vectors pointing along the coordinates, and $\delta^{i j}$ is the Kronecker delta.
The first line in Eq.~\eqref{eq:metric_multipolar} represents a multipolar expansion of the field sourced by the neutron star.
The $M/r$ term represents the standard monopolar term. 
The $\sim 1/r^2$ term vanishes identically, since the neutron star is placed at the origin of the coordinate frame.
The $\sim 1/r^3$ term represents the quadrupolar correction, with the coefficient $Q_{ij}$ defining the tidally-induced quadrupole moment. 
Higher-multipolar corrections are represented by ${\cal{O}}\left(r^{-4}\right)$.
The second line in Eq.~\eqref{eq:metric_multipolar} is the contribution of the external field, with the $\sim r^2$ term representing the quadrupolar tidal field defined as $\mathcal{E}_{i j}$. Higher-order corrections are included in ${\cal{O}}\left(r^3\right)$.
In Newtonian gravity, $Q_{ij}$ would be given by integrating density perturbations over the star while $\mathcal{E}_{i j}$ would be given by spatial derivatives of the companion's field~\cite{Poisson_Will_2014}. In General Relativity, both quantities are defined via Eq.~\eqref{eq:metric_multipolar}.

Though their definitions and calculations differ, in both Newtonian and Relativistic gravity and under static perturbations the tidally-induced quadrupole moment is proportional to the perturbing field, with the constant of proportionality defining the tidal deformability
\begin{equation}
\lambda \equiv-\frac{Q_{i j}}{\mathcal{E}_{i j}}\,.
\end{equation}
Given an external perturbation $\mathcal{E}_{i j}$, the tidal deformability $\lambda$ quantifies how much the star is deformed. 
Intuitively, a bigger (in size) star should be more deformable, therefore $\lambda$ should increase with the neutron star radius $R$.
Moreover, dimensional analysis based on Eq.~\eqref{eq:metric_multipolar} suggests that $Q_{i j}\sim \mathrm{[L]}^3$, while $\mathcal{E}_{i j}\sim \mathrm{[L]}^{-2}$, suggesting that $\lambda\sim\mathrm{[L]}^{5}$. 
Indeed, $\lambda$ is related to the (dimensionless) tidal love number $k_2$ as
\begin{equation}
\lambda=\frac{2}{3} k_2 R^5\,.
\end{equation}
In the context of gravitational-wave observations, Sec.~\ref{sec:nsmergers}, tidal effects are imprinted on the signal through the dimensionless tidal deformability 
\begin{equation}
\Lambda \equiv \frac{\lambda}{M^5}=\frac{2}{3} k_2 \frac{R^5}{M^5}=\frac{2}{3} k_2 C^{-5}.
\end{equation}

For a neutron star of a given mass, both $k_2$ and $R$ depend on the neutron star equation of state, and thus offer complementary information.
Typical values range $k_2\sim 0.05-0.1$~\cite{Hinderer:2009ca}, with a $10-20\%$ reduction compared to Newtonian gravity~\cite{Hinderer:2007mb}, while $\Lambda$ spans about one order of magnitude. 
For a given equation of state, $\Lambda$ is  a steep function of the mass. For $m\gtrsim 1\,M_{\odot}$, $k_2\sim M^{-1}$~\cite{Zhao:2018nyf}, bringing the total dependence to $\Lambda\sim M^{-6}$. 
Indeed $\Lambda\sim{\cal{O}}(10^4)$ for $M=1\,M_{\odot}$ and drops to $\Lambda\sim{\cal{O}}(10)$ for $m=2\,M_{\odot}$, c.f., Fig.~\ref{fig:TOVseq}, suggesting that the measurability of the tidal deformability is a sensitive function of the binary mass. The tidal deformability of a black hole is zero~\cite{Binnington:2009bb,Chia:2020yla}.

Calculating $\lambda$ for a given neutron star amounts to introducing a static $\ell=2$ perturbation to the Einstein equations and computing the metric asymptotically~\cite{Thorne1967}, extracting the appropriate terms by comparison to Eq.~\eqref{eq:metric_multipolar}, and taking their ratio.
Continuity of the metric and its derivatives across the neutron star surface results in an expression for 
\begin{equation}
    \lambda=\lambda(C,y)\,,
    \label{eq:lambda}
\end{equation}
where $C$ is the neutron star compactness and $y$ expresses a metric quantity and its derivative at $R$ and which can be obtained by numerically integrating a second-order ordinary differential equation from the neutron star center to its surface~\cite{Hinderer:2007mb,GuerraChaves:2019foa}. 

Beyond the leading-order adiabatic quadrupolar tidal deformability, higher-order terms take into account higher-multipolar electric-type and magnetic-type corrections in Eq.~\eqref{eq:metric_multipolar}~\cite{Damour:2009vw,Binnington:2009bb,GuerraChaves:2019foa}, spin-tidal couplings to the neutron star rotation~\cite{Landry:2015zfa,Pani:2015hfa,Pani:2015nua,Landry:2017piv,Abdelsalhin:2018reg}, and dynamical tides~\cite{Hinderer:2016eia,Steinhoff:2016rfi,Schmidt:2019wrl,Pratten:2019sed,Gamba:2022mgx}. 
Though these effects are partially (or fully in some cases) included during analysis of gravitational-wave data, their influence is expected to be subdominant to $\lambda$ for current-generation detectors.


\section{Compact Binary Mergers involving Neutron stars}
\label{sec:nsmergers}

Neutron stars in compact binaries around other neutron stars or black holes can be observed through their gravitational wave signal by the ground-based LIGO~\cite{LIGOScientific:2014pky} and Virgo~\cite{VIRGO:2014yos} detectors and electromagnetic emission. 
To date, the coalescence of two neutron star binaries~\cite{LIGOScientific:2017vwq,LIGOScientific:2020aai}, three neutron star-black hole binaries~\cite{LIGOScientific:2021qlt,LIGOScientific:2024elc} and further sub-$5\,M_{\odot}$ objects of interest~\cite{LIGOScientific:2020zkf,KAGRA:2021vkt} have been observed in gravitational waves. 
These observations have the potential to inform about the nuclear equation of state. 

\begin{figure}
\centering
\includegraphics[width=0.5\textwidth]{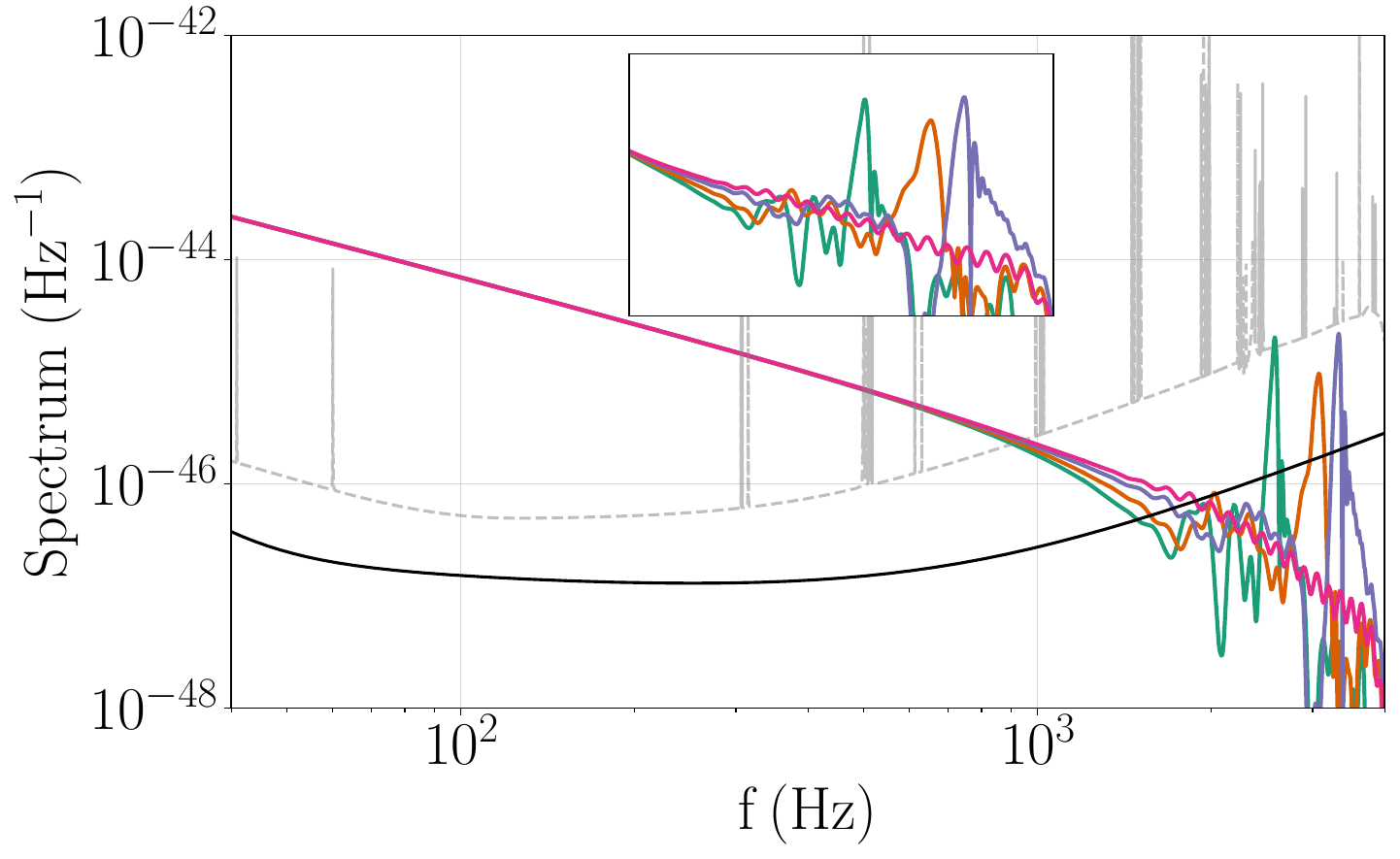}
\caption{Spectrum of neutron star binary signals for different equations of state. Binaries enter the LIGO/Virgo frequency band at a gravitational-wave frequency of $\sim20\,$Hz at $\sim 1000$\,km apart. 
Over the ensuing $2-3$ minutes emit their binding energy in gravitational radiation and inspiral.  
Tidal deformation speeds up the inspiral and imprints the equation of state in the waveform in frequencies $\gtrsim 300\,$Hz.
The neutron stars merge around $1000-1500\,$Hz, after which post-merger emission is dominated by a spectral peak who frequency is a strong function of the equation of state.
Inspiral data are created with {\tt IMRPhenomD\_NRTidalv2}~\cite{Dietrich:2018uni}, while merger and postmerger data correspond to numerical simulations from~\cite{Torres-Rivas:2018svp}.
The GW170817~\cite{LIGOScientific:2019lzm} and design sensitivity~\cite{KAGRA:2013rdx} power spectral density for LIGO Livingston~\cite{LIGOScientific:2014pky} are shown in grey dashed and black black respectively.
The inset focuses on the high-frequency portion of the $4$ signals.}
\label{fig:GWspectrum}
\end{figure}

Compact objects in binaries inspiral around each other, slowly emitting their binding energy in gravitational radiation. 
This long inspiral phase, observed by up to minutes, offers information about the neutron star masses and spins. 
During the late stages of the coalescence and depending on the exact binary parameters, the neutron star gets tidally distorted due to the tidal field of its companion. 
The degree of tidal deformation, quantified through the tidal deformability depends on the equation of state.
After the merger of two neutron stars, the final remnant keeps emitting high-frequency gravitational radiation, whose properties depend on the nature of the remnant, which in turn depends on the binary parameters and the equation of state.
Figure~\ref{fig:GWspectrum} shows the gravitational wave spectrum from neutron star binary coalescences with different equations of state.
In this section we discuss the phenomenology, physics, and observational signatures of compact binary coalescencies in both gravitational and electromagnetic radiation. 
For a more in depth discussion of the qualitative and quantitative features of compact binary mergers, see, e.g., \citet{Baiotti:2016qnr, Shibata:2019wef, Radice:2020ddv, Kyutoku:2021icp}.

\subsection{Compact binary inspirals}

\begin{figure}
\centering
\includegraphics[width=0.45\textwidth]{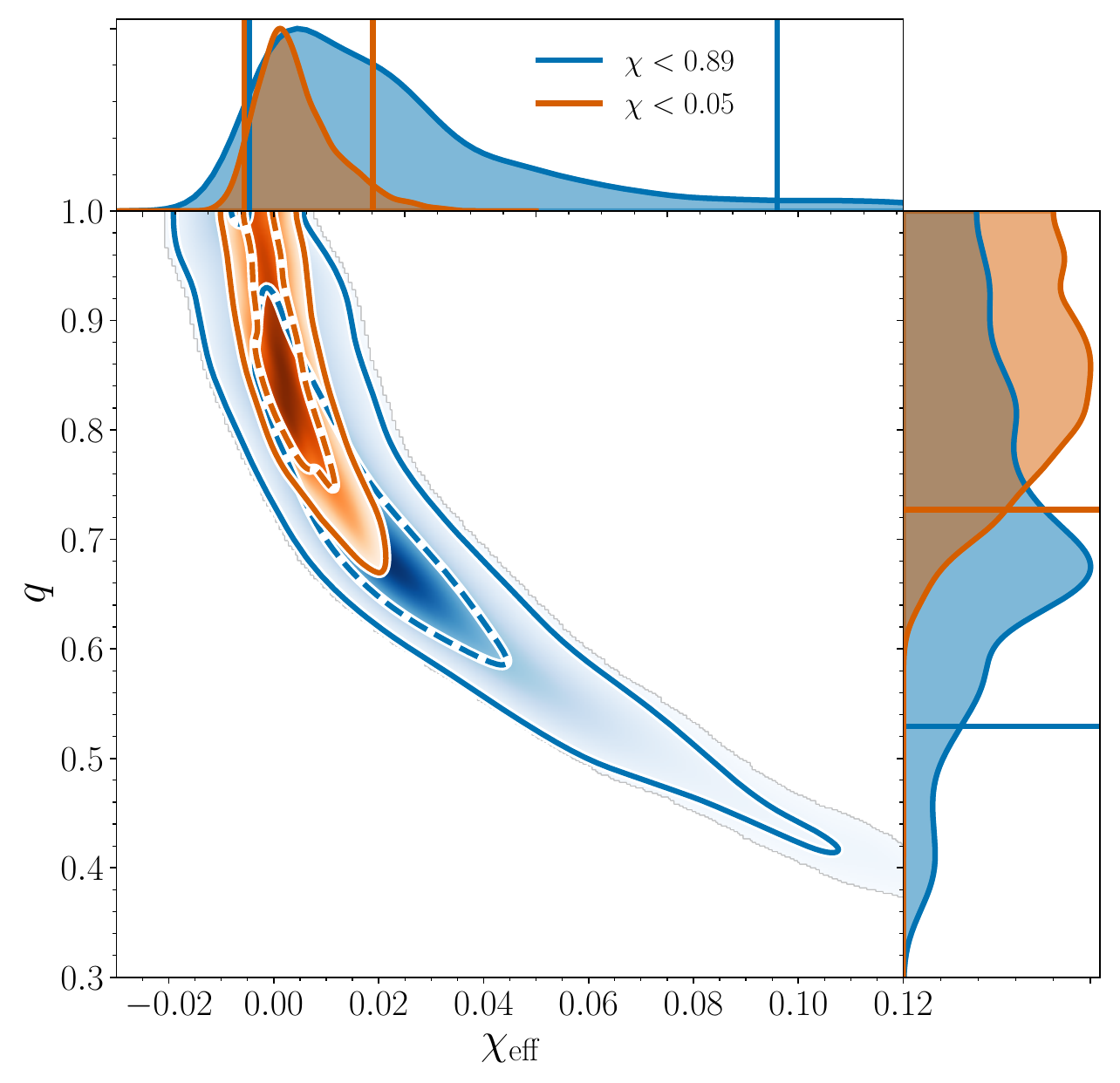}
\caption{Marginalized posterior for the mass ratio and effective spin of GW170817 showcasing the inherent mass-spin correlation of gravitational wave inspiral measurements, reproduced from~\citet{LIGOScientific:2018hze}. 
Assuming GW170817 shares similar spin properties as the Galactic neutron star binaries restricts the spin to $\chi<0.05$ and leads to improved mass ratio and component mass inference.
For reference, a dimensionless spin of $\chi\sim 0.4$ approximately corresponds to a $1\,$ms neutron star.}
\label{fig:GW170817mass-spin}
\end{figure}

Compact binaries consisting of neutron stars and/or black holes emit their binding energy in gravitational-waves and inspiral toward each other until their eventual merger.
Sourced to leading order by the time-varying orbital quadrupole moment~\cite{Peters1963,Peters1964,Blanchet:2013haa}, the rate of emission increases rapidly as the compact objects inspiral and speed up, leading to a runaway process.
Ground-based detectors are sensitive to signals above a gravitational-wave frequency of $10-20\,$Hz, corresponding to the end of the inspiral and merger of compact objects with masses $\lesssim {\cal{O}}(10^2)\,M_{\odot}$.

During the inspiral stage, the compact object velocities $u$ are low compared to the speed of light.
The post-Newtonian (PN) description, therefore, expands all quantities in powers of $u$. 
Terms proportional to $u^{2N}$ compared to the leading order term are labeled as $N$PN corrections.
In the frequency domain and under the stationary phase approximation~\cite{Droz:1999qx}, the gravitational-wave signal $\tilde{h}(f)$ from a compact binary with masses $M_1>M_2$ and spins $\vec{\chi}_1,\vec{\chi}_2$ is
\begin{equation}
\tilde{h}(f)=\mathcal{A} \frac{(\mathcal{M} f)^{5/6} f^{-2}}{D_L} e^{i \Psi(f)}\,,
\end{equation}
where $\mathcal{A}$ is a constant that depends on the binary orientation, $\mathcal{M}$ is the binary chirp mass, $f$ is the frequency, $D_L$ is the luminosity distance to the binary, and $\Psi(f)$ is its phase.
The signal phase can be obtained through energy balance: the loss in binary binding energy is carried away by gravitational waves. 
Then the derivative of the binary orbital frequency is (in the time domain)
\begin{equation}
\frac{d F}{d t}=\frac{d E}{d t} \frac{d F}{d E}\,,
\label{eq:balancechain}
\end{equation}
where $dE/dt$ is the rate of energy emission and can be obtained through the quadrupole formula, and $1/(dE/dF)$ is the binding energy derivative with respect to frequency and corresponds to Kepler's law.

Calculating the gravitational-wave phase for high PN order and transforming to the frequency domain yields~\cite{Blanchet:2013haa,Buonanno:2009zt}
\begin{align}
\Psi(f)  =2 & \pi f t_c+\phi_c-\frac{\pi}{4}  +\frac{3}{128 \eta u^5}\left\{1+f(\eta) u^2\right.\nonumber \\
& \left.+(4 \beta-16 \pi) u^3+[g(\eta)+\sigma] u^4+\ldots\right\} \,.
\label{eq:GWphasePN}
\end{align}
Here $u \equiv(\pi M_T f)^{1 / 3}$, $M_T=M_1+M_2$ is the total mass, $\eta=M_1M_2/M_T^2$ is the symmetric mass ratio, $t_c$ is the time of coalescence, and $\phi_c$ is the phase of coalescence.
The leading order term in Eq.~\eqref{eq:GWphasePN} dominates the overall signal evolution, especially for low-mass compact binaries such as those involving neutron stars.
Its form motivates the definition of the chirp mass $\mathcal{M} \equiv M_T \eta^{3 / 5}$.

The leading order inspiral evolution and Kepler's law allows us to obtain some first estimates of the phenomenology of the signal.
When observed from a frequency $f_0$, the signal inspiral lasts for
\begin{equation}
T_{\rm ins}=\frac{5}{256}\mathcal{M}^{-5 / 3}\left(\pi f_0\right)^{-8 / 3}\,,
\end{equation}
suggesting that it is a strong function of the binary mass. 
For example, a $1.4+1.4\,M_{\odot}$ binary lasts for 1000(157)[53] seconds from 10(20)[30]\,Hz, while a $10+10\,M_{\odot}$ binary lasts for 38(6)[2] seconds from the same frequencies. Therefore the long signal duration is the first indication of the detection of a low-mass, potentially neutron star, binary.

The signal phase and Eq.~\eqref{eq:GWphasePN} prescribe to an excellent approximation which binary properties are measurable and how.
As already mentioned, the leading order term depends on the binary chirp mass $\mathcal{M}$, whose statistical relative measurement accuracy reached ${\cal{O}}(10^{-4})$ for GW170817 at the frame of the detector~\cite{LIGOScientific:2017vwq}.
Uncertainty in the source-frame binary chirp mass is instead dominated by the uncertainty in the distance/redshift used to converted between the two frames~\cite{LIGOScientific:2018hze}. 

A second mass parameter is needed in order to extract the individual component masses. 
The 1PN term in Eq.~\eqref{eq:GWphasePN} introduces the binary symmetric mass ratio $\eta$. 
The equivalent mass ratio $q\equiv M_2/M_1<1$ is measured significantly worse than the chirp mass at $\sim(0.5-1)$ for GW170817~\cite{LIGOScientific:2018hze}. 
Further complicating the picture are the 1.5PN and 2PN terms which introduce spin-orbit $\beta$ and spin-spin $\sigma$ terms that correlate the measurement of the mass ratio and spins.
The latter is typically expressed through the effective spin
\begin{equation}
 \chi_{\rm eff}=\frac{m_1\vec{\chi}_1 \cdot \hat{L}+m_2\vec{\chi}_2 \cdot \hat{L}}{m_1+m_2} \,,  
\end{equation}
where $\hat{L}$ is the direction of the Newtonian orbital angular momentum.
The effective spin is related to $\beta$ in the equal-mass limit and conserved at the 2PN order~\cite{Racine:2008qv}.
The mass-spin correlation means that inference of the binary mass ratio and component spins depends on assumptions about the neutron star spins, as shown in Fig.~\ref{fig:GW170817mass-spin} for GW170817~\cite{LIGOScientific:2018hze}.
Imposing a low spin $\chi<0.05$ motivated by Galactic neutron star binary observations~\cite{Tauris:2017omb} leads to $m_1 \sim(1.36,1.60)\, M_{\odot}$ and $m_2 \sim (1.16,1.36) \,M_{\odot}$ at the 90\% credible level~\cite{LIGOScientific:2018hze}.
An agnostic spin prior that extends to $\chi<0.89$ instead leads to larger uncertainties: $m_1 \sim(1.36,1.89)\, M_{\odot}$ and $m_2 \sim (1.00,1.36) \,M_{\odot}$ at the 90\% credible level.

Though GW170817 is consistent with the Galactic neutron star binary population, the dependence on spin assumptions creates an inherent ambiguity especially for binaries such as GW190425~\cite{LIGOScientific:2020aai}.
At a total mass of $M\sim 3.4\,M_{\odot}$, GW190425 is more massive than GW170817 and Galactic observations whose total masses cluster around $\sim2.7\,M_{\odot}$. 
Similar considerations apply to neutron star-black hole observations regarding the appropriate spin assumptions for the neutron star binary component and their impact on the inferred binary mass ratio.
The expected spins of neutron star binaries or neutron star-black hole binaries are uncertain and it is not clear whether they are representative of the low spins of Galactic binaries.

Absent from the leading-order phase terms is a strong imprint of the neutron-star equation of state.
Modulo resonant excitations of the neutron-star matter, the inspiral signal is dominated by the binary masses and spins.\footnote{Spin induces a quadrupole moment on the neutron star that depends on the equation of state and enters that gravitational wave phase through the 2PN $\sigma$ term~\cite{Poisson:1997ha,Bohe:2015ana}. However, since the term is proportional to the neutron star spin squared and has a small numerical prefactor, it is in general considered to be negligible unless the neutron star has a large spin~\cite{Harry:2018hke,Samajdar:2020xrd}. Nonetheless, it is included in compact binary analyses and inference~\cite{Yagi:2013awa}.}
Therefore the signals emitted by two neutron stars is indistinguishable from the signal emitted by two black holes with the same properties.
Figure~\ref{fig:GWspectrum} shows that the signal spectrum for neutron star binaries is identical up to gravitational-wave frequencies of $\sim 200\,$Hz for different equations of state. A similar conclusion holds for the gravitational-wave phase and neutron star-black hole binaries.

\subsection{Mass observations}

\begin{figure}
\centering
\includegraphics[width=0.5\textwidth]{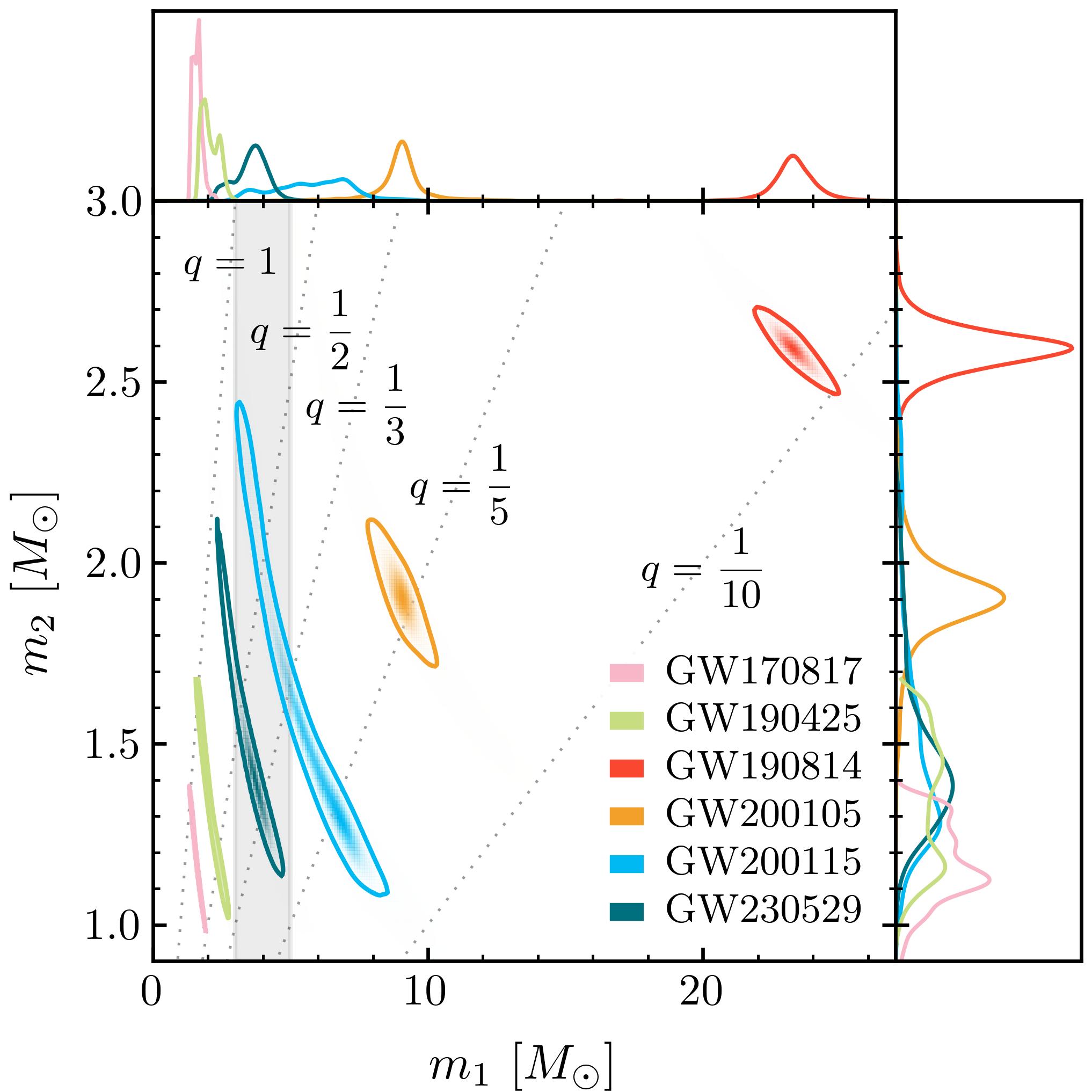}
\caption{
Marginalized mass posteriors for the confident sub-$3\,M_{\odot}$ gravitational wave detections, adapted from from~\cite{LIGOScientific:2024elc}.
Dashed lines denote a constant mass ratio and the shaded region corresponds to $m_1\in[3,5]\,M_{\odot}$, the purported mass gap between neutron stars and black holes.
All analyses allow for large spins and hence represent agnostic results.}
\label{fig:GWMasses}
\end{figure}

Mass measurements of sub$-3\,M_{\odot}$ compact objects in binaries are possible using gravitational wave information.
Figure~\ref{fig:GWMasses} shows the one-and two-dimensional marginalized posteriors for the masses of confidently-detected relevant events.
This plot and the discussion below remains agnostic about the neutron star spins so it is conservative. 
Quoted intervals correspond to 90\% credibility.
\begin{itemize}
    \item At a total mass of $2.77_{-0.05}^{+0.22}\,M_{\odot}$,
 GW170817~\cite{LIGOScientific:2017vwq} is consistent with the Galactic neutron star binary population~\cite{Tauris:2017omb}.
 GW170817 is also the only event on this list with informative tidal constraints.
    \item The second likely neutron star binary, GW190425~\cite{LIGOScientific:2020aai}, has a total mass of $3.4_{-0.1}^{+0.3}\,M_{\odot}$, it is thus more massive than the Galactic neutron star binary.
    Though the components of GW190425 are individually consistent with neutron stars, their combination in a neutron star binary is novel and shows that such massive neutron stars form compact binaries and merge.
    \item The secondary component of GW190814~\cite{LIGOScientific:2020zkf} at $2.59_{-0.09}^{+0.08}\,M_{\odot}$ is the most massive neutron star or the least massive black hole known~\cite{Tan:2020ics,Essick:2020ghc,Dexheimer:2020rlp,Tews:2020ylw,Fattoyev:2020cws}.
    \item GW200105 and GW200115~\cite{LIGOScientific:2021qlt} are the first mixed neutron star-black hole binaries identified. The secondary components at $1.9_{-0.2}^{+0.3}\,M_{\odot}$ and $1.5_{-0.3}^{+0.7}\,M_{\odot}$ respectively, are consistent with Galactic neutron star masses~\cite{Alsing:2017bbc} and nuclear physics expectations.
    \item GW230529~\cite{LIGOScientific:2024elc} originates from the merger of a likely neutron star secondary with a $2.5-4.5\,M_{\odot}$ primary compact object. This mass range corresponds to the observational low-mass gap between neutron stars and black holes, denoted by the shaded vertical band in Fig.~\ref{fig:GWMasses}. Depending on astrophysical assumptions, GW230529 could be a neutron star-black hole binary with a nonspinning $\sim4\,M_{\odot}$ primary, or a heavy neutron star binary with a spin that is negatively aligned with the orbit.
\end{itemize}

The gravitational wave data alone guarantee that the detected objects are very compact, neutron stars or black holes, and primarily constrain the compact object masses.
However, a definitive determination of the nature of the compact objects would require detection of tidal effects, i.e. a lower limit on $\Lambda$.
Since $\Lambda$ rapidly decreases for more massive neutron stars or more asymmetric binaries, lower limits on the tides from $\gtrsim 1.6\,M_{\odot}$ objects are not expected with current or near-future detectors~\cite{Yang:2017gfb,Chen:2020fzm,Brown:2021seh}.
In the absence of a definitive classification from gravitational wave data for $\gtrsim 1.6\,M_{\odot}$ compact objects, external input about the maximum neutron star mass is required.
The most basic classification hinges on the fact that causality limits neutron star masses to $\lesssim3\,M_{\odot}$~\cite{Kalogera:1996ci,Rhoades1974}.

The first option is comparing masses inferred from gravitational wave data to the inferred maximum mass from galactic pulsars $\gtrsim2\,M_{\odot}$~\cite{Alsing:2017bbc,Antoniadis:2016hxz,Farr2020}, though such analyses are based on a heterogeneous set of the observed pulsars with a mass measurements and unquantified selection effects. 
Even then, galactic neutron star masses are not representative of gravitational wave observations.
Galactic neutron star binaries are clustered around $1.4\,M_{\odot}$~\cite{Tauris:2017omb}, while the whole pulsar mass distribution has a sharp peak at $1.4\,M_{\odot}$ and a secondary peak at $\sim1.7\,M_{\odot}$~\cite{Antoniadis:2016hxz}.
The mass distribution of neutron stars observed with gravitational waves is more broad with no evidence of a strong peak at any value~\cite{Chatziioannou:2020msi,Landry:2021hvl,KAGRA:2021duu}.

The second option is comparing masses inferred from gravitational wave data to the maximum TOV mass, which is again $\sim2-2.5\,M_{\odot}$~\cite{Legred:2021hdx,Raaijmakers:2021uju,Miller:2021qha,Koehn:2024set}. 
Inference of the TOV mass is based on a multitude of astronomical data of nuclear experiment and/or calculation that inform the neutron star equation of state at various densities.
Since the maximum TOV mass primarily depends on the highest neutron star densities $\sim5-6\rho_{\rm sat}$, some equation of state model is required to effectively ``translate" constraints across scales introducing a model dependence due to intra-density correlations in the models~\cite{Legred:2022pyp}.
Moreover, such a comparison implicitly assumes that neutron stars up to the maximum mass allowed by nuclear physics are astrophysically produced~\cite{Essick:2020ghc}, which is unclear.

These arguments hinge on neutron stars and black holes having nonoverlapping mass distributions.
If low-mass black holes, of potential primordial origin~\cite{Carr1974}, exist, the overlapping neutron star and black hole populations could be distinguished based on their tidal properties~\cite{Chen:2019aiw}.
Searches for sub-$1\,M_{\odot}$ compact objects have not yielded confident detections~\cite{LIGOScientific:2019kan,LIGOScientific:2022hai,Nitz:2022ltl}, but if such compact objects and binaries exist, the large expected tidal deformability of neutron stars in this mass range would make their classification more straightforward~\cite{Golomb:2024mmt}.

Finally, beyond considerations about tidal inference and nuclear physics input, the whole mass distribution of all objects in compact binaries $\lesssim10\,M_{\odot}$ is not smooth~\cite{Fishbach:2020ryj,Farah:2021qom,KAGRA:2021duu}.
The rate of $\lesssim2\,M_{\odot}$ compact objects is more steep than the rate of $\gtrsim5\,M_{\odot}$ compact objects, with an underdensity between them starting at $\sim2.4\,M_{\odot}$.
Under the assumption of nonoverlapping neutron star and black hole distributions, such a feature could signal the transition between the two populations.
Though its location also roughly corresponds to the neutron star-black hole mass gap at $3-5\,M_{\odot}$, the gravitational wave rate is more consistent with an underdensity than a completely empty gap~\cite{KAGRA:2021duu}.

\subsection{Late inspiral and tidal deformation}
\label{sec:BNSdeformation}

\begin{figure*}
\centering
\includegraphics[width=\textwidth]{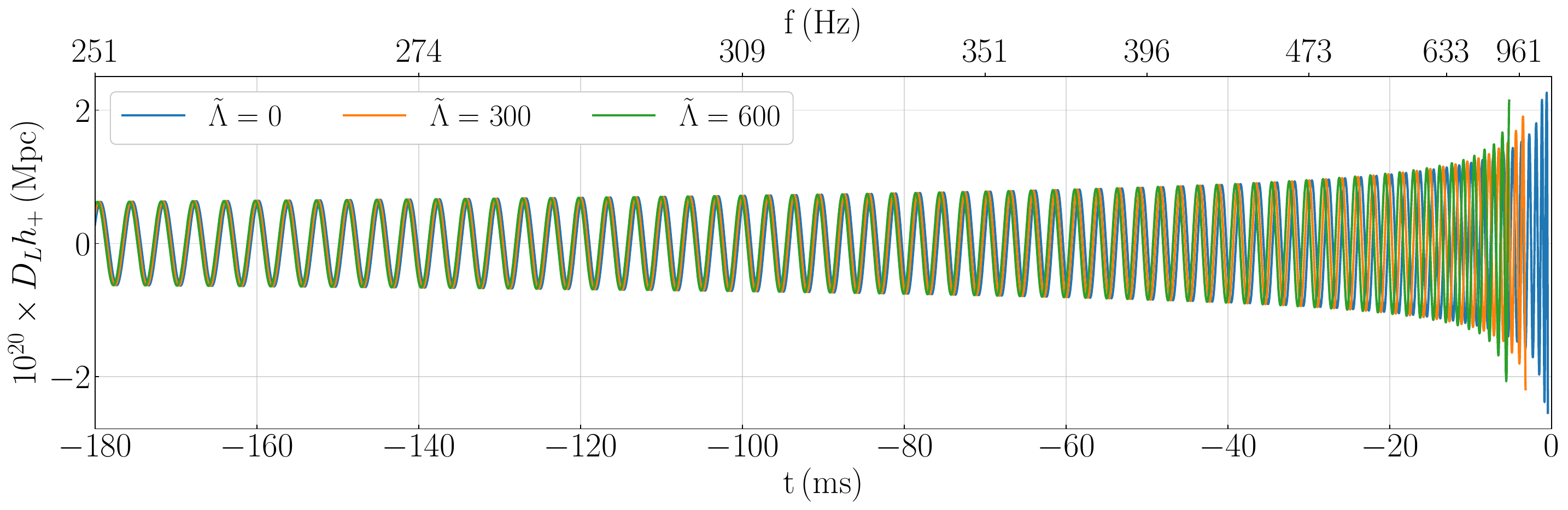}
\includegraphics[width=\textwidth]{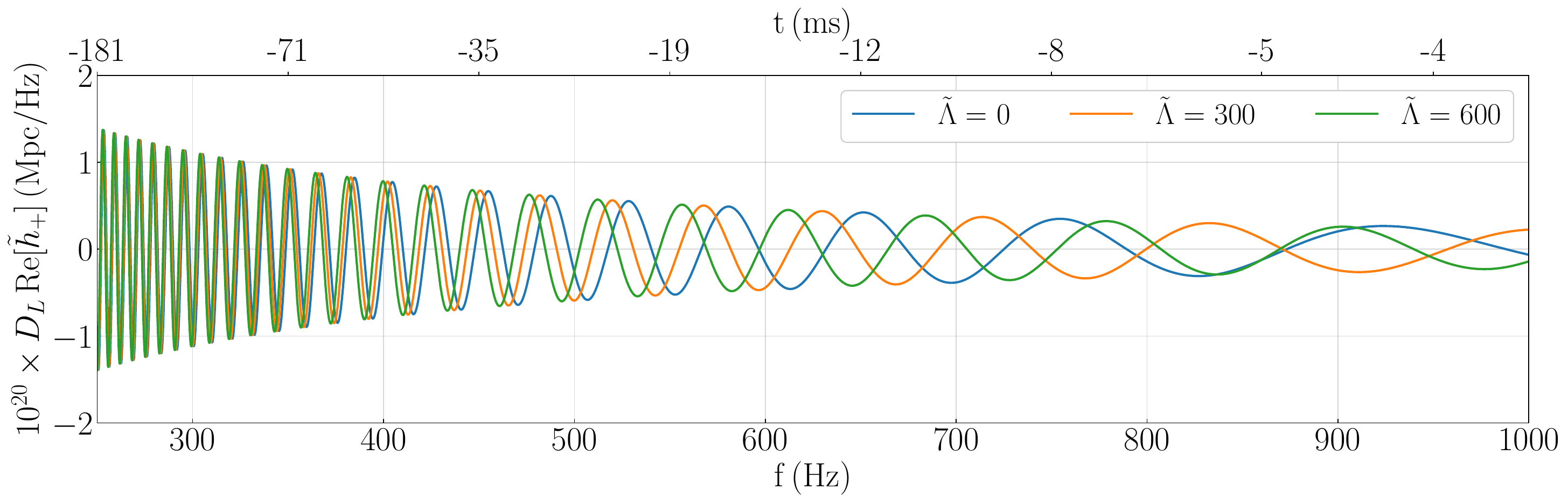}
\caption{Final inspiral stages of a neutron star binary for different values of the tidal deformability. 
We plot the waveform using {\tt IMRPhenomD\_NRTidalv2}~\cite{Dietrich:2018uni} for equal-mass, nonspinning neutron stars up to merger (the peak of the strain in the time domain) in the time (top) and frequency (bottom) domain. The upper x-axis denotes the corresponding frequency (top) and time (bottom) for the $\tilde{\Lambda}=0$ binary. 
Tidal deformation speeds up the binary evolution.
Adapted from~\cite{Chatziioannou:2020pqz}.}
\label{fig:GWBNSwithtides}
\end{figure*}

Once energy loss in gravitational waves brings the compact objects close enough, tidally interactions become important.
A neutron star component will be tidally deformed by the external field created by its companion.
In a neutron star binary this process acts twice, once on each neutron star, while in a neutron star-black hole binary, only the neutron star is deformed.
The leading order deformation is the adiabatic, quadrupole effect described in Sec.~\ref{sec:tidaldeformability}. 
In this case, the deformation is ``coherent" with the binary in the sense that the tidal bulge raised on the deformed neutron star faces toward its companion and the tidally-induced quadrupole moment adds to the dominant orbital quadrupole moment that sources the radiation.
In what follows, we consider the tidal effect on a neutron star with mass $M_1$ and tidal deformability $\lambda_1$ by its companion with mass $M_2$. If the companion is a neutron star as well, its impact is additive with $1\leftrightarrow2$.

Physically, the tidal deformation results in two effects, both of which remove energy from the system and accelerate its inspiral.
Firstly, the energy required to deform an initially spherically symmetric star is removed from the binary binding energy. 
Overall,
\begin{equation}
    E(F)=E_{\rm PP}(F) + E_{\rm Tide}(F)\,,
\end{equation}
where 
\begin{equation}
    E_{\rm PP}(F)=-\frac{1}{2} M_T \eta u^2\left[1+{\cal{O}}(u^{2})\right]\,,
    \label{eq:Etides}
\end{equation}
is the point-particle binding energy~\cite{Buonanno:2009zt,Blanchet:2013haa} with $u=(2 \pi M_T F)^{1/ 3}$.
The tidal contribution is~\cite{Flanagan:2007ix,Hinderer:2009ca}
\begin{equation}
    E_{\rm Tide}(F)\sim Q_{i j}\mathcal{E}_{i j}=-\frac{1}{2} M_T \eta u^2\left[-9\frac{M_2}{M_1} \frac{\lambda_1}{M_T^5} u^{10}+{\cal{O}}(u^{11})\right]\,.
\end{equation}
which lowers (in absolute value) the total energy.
Secondly, gravitational wave emission is sourced by time-varying quadrupole moments. 
Though the orbital quadrupole dominates the emission~\cite{Buonanno:2009zt,Blanchet:2013haa},
\begin{equation}
    \dot{E}(F)=-\frac{32}{5} \eta^2 u^{10}\left[1+{\cal{O}}(u^{2}) \right]\,,
\end{equation}
the tidally-induced quadrupole causes additional coherent energy emission~\cite{Flanagan:2007ix,Hinderer:2009ca}
\begin{align}
\dot{E}(F)\sim \dddot{Q}_{ij}^2=-\frac{32}{5} \eta^2 u^{10}\left[6\frac{M_1+3 M_2}{M_1} \frac{\lambda_1}{M_T^5} u^{10}+{\cal{O}}(u^{11})\right]\,.
\end{align}
such that the total energy emission is enhanced
\begin{equation}
    \dot{E}(F)=\dot{E}_{\rm PP}(F) + \dot{E}_{\rm Tide}(F)\,.
    \label{eq:Edottides}
\end{equation}

The tidally-updated energy, Eq.~\eqref{eq:Etides}, and energy emission, Eq.~\eqref{eq:Edottides} together with Eq.~\eqref{eq:balancechain} result in a 5PN ($u^{10}$ relative to the leading order term) additive tidal phase term in Eq.~\eqref{eq:GWphasePN}:
\begin{equation}
\Psi_{\Lambda}(f)=\frac{3}{128 \eta u^5}\left(-\frac{39}{2} \tilde{\Lambda} u^{10}\right)\,,
\label{eq:Psitides}
\end{equation}
where~\citep{Favata:2013rwa}
\begin{equation}
\tilde{\Lambda} \equiv \frac{16}{3} \frac{\left(M_1+12 M_2\right) M_1^4 \Lambda_1}{\left(M_1+M_2\right)^5}=\frac{16}{3} \frac{\left(M_1+12 M_2\right) }{\left(M_1+M_2\right)^5}\frac{\lambda_1}{M_1}\,.
\end{equation}
is the leading-order and best-measured tidal term in the phase.
Numerically, for a neutron star binary the energy effect in Eq.~\eqref{eq:Etides} is a factor of 2:1 (equal masses) to 3:1 (unequal masses) larger than the energy emission effect in Eq.~\eqref{eq:Edottides}. 
This discussion neglects a number of point-particle and finite-size terms, that are included in modern waveform models~\citep{Vines:2011ud, Bernuzzi:2014owa, Hinderer:2016eia, Hotokezaka:2016bzh, Dietrich:2017aum, Dietrich:2018uni, Dietrich:2019kaq, Steinhoff:2021dsn, Gamba:2020ljo, Gonzalez:2022prs, Abac:2023ujg, Nagar:2018zoe, Thompson:2020nei,Matas:2020wab,Gamba:2023mww}, see~\citet{Dietrich:2020eud} for a review. Additionally, these calculations assume that the external tidal field varies slowly compared to the internal dynamical time of the perturbed neutron star (static tides), and neglect any non-adiabatic or nonlinear effects~\citep{Pratten:2019sed}.

The sign of the phase term in Eq.~\eqref{eq:Psitides} again suggests that tides speed up the binary evolution and the binary accumulates fewer radians before merger.
Figure~\ref{fig:GWBNSwithtides} shows the final stages of the late inspiral of equal-mass, nonspinning neutron star binaries with different values of the tidal deformability in the time and in the frequency domain. 
Tides affect the gravitational wave phase evolution only during the last $\sim100\,$ms before merger and for frequencies above $\sim 300\,$Hz, as expected from a high PN order term.
A larger value of the tidal deformability causes the phase evolution to speed up and the binary to reach merger earlier.
This phase difference is the main tidal observable for neutron star binary signals.
{For a typical binary, with $\tilde\Lambda \sim 500$, the accumulated dephasing with respect to the point particle limit up to the contact frequency is ${\sim}10\,$rad.}

\begin{figure}
    \includegraphics[width=\columnwidth]{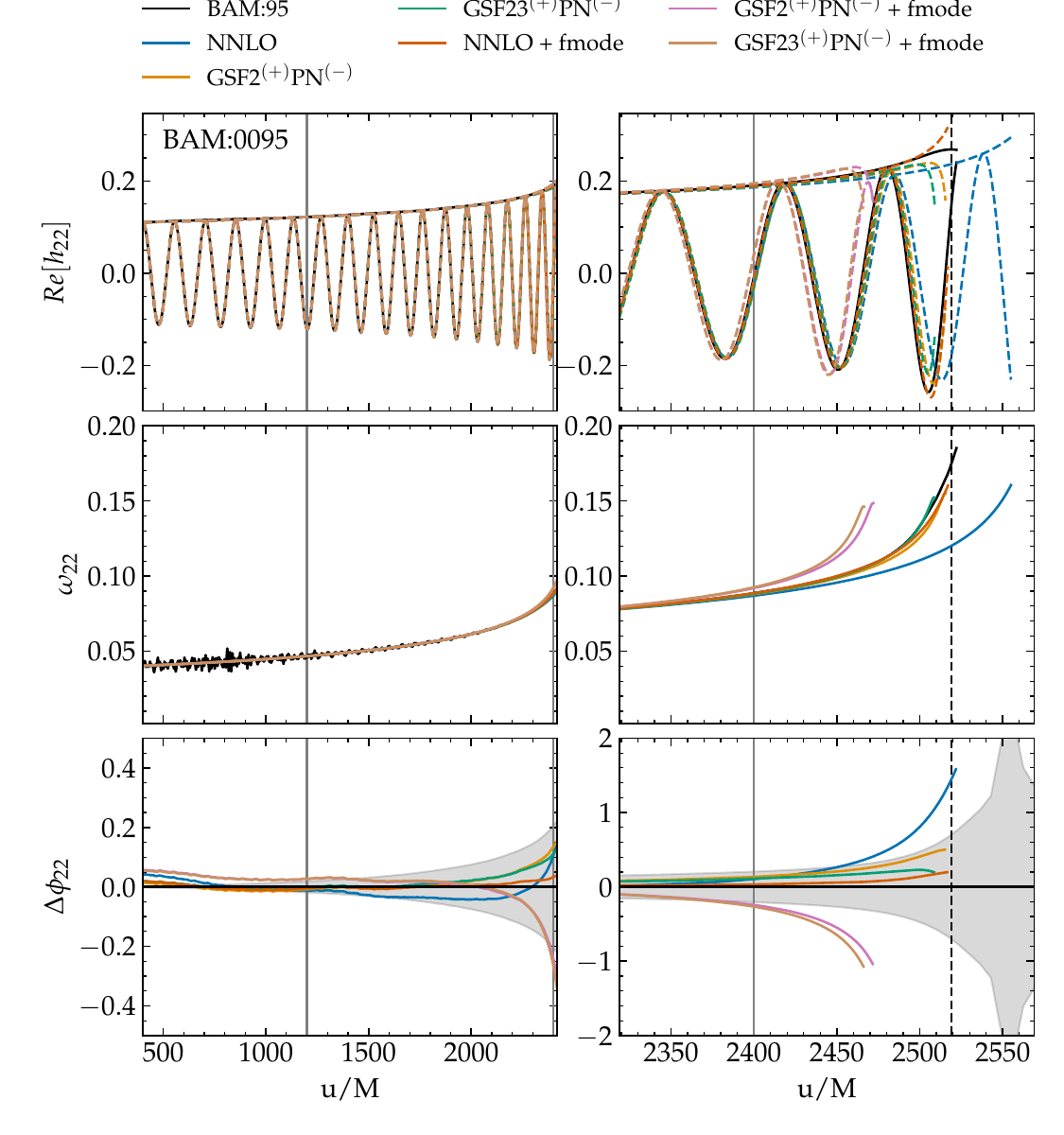}
    \caption{Comparison of different analytical gravitational waveforms within the \texttt{TEOBResum} family with numerical relativity data of inspiralling neutron stars. From Gamba and Bernuzzi~\cite{Gamba:2022mgx}. }
    \label{fig:NR_EOB_comparison}
\end{figure}

In principle, the tidal contribution to the phase can be added to any point-particle post-Newtonian waveform model using Eq.~\eqref{eq:Psitides}. 
In practice, there are conceptual and practical issues. 
In this context, numerical relativity~\cite{Baumgarte:2021skc} simulations play an important role for two reasons. 
First, they are needed to calibrate and validate the baseline point-particle waveform model. 
To this aim, vacuum simulations are used, since they can achieve high precision, with effective dephasing at merger well below one radian~\citep{Nagar:2018zoe, Matas:2020wab, Boyle:2019kee}. 
Second, simulations are needed to understand to what extent matter effects can truly be absorbed into $\tilde\Lambda$~\citep{Steinhoff:2016rfi, Kuan:2023qxo}. 
For example, various recent studies have studied the possible role of dynamical tides in neutron star mergers, but it is unclear that current simulations are sufficiently accurate to reliably extract these effects~\citep{Steinhoff:2021dsn, Gamba:2022mgx}. 
Figure~\ref{fig:NR_EOB_comparison} shows a comparison between various analytical waveforms from the \texttt{TEOBResum} family \citep{Nagar:2018zoe} and numerical relativity data. 
While excellent agreement is found over most of the inspiral, waveforms start to deviate from the numerical prediction in the last gravitational wave cycle.
Unfortunately, this is precisely the phase of the dynamics in which matter effects are the strongest~\citep{Damour:2012yf}. 
The estimated numerical error, shown with a grey band in the figure, is also large at this point.

\begin{figure}
    \centering
    \includegraphics[width=\columnwidth]{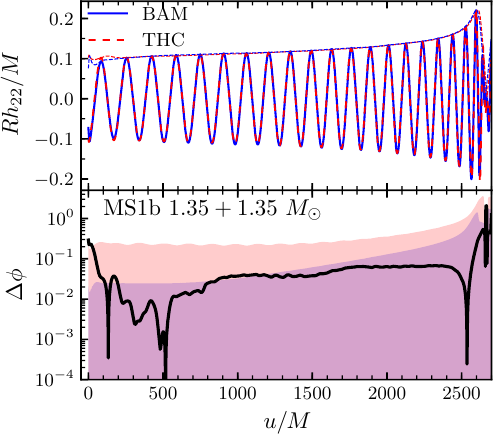}
    \caption{Cross-validation of the \texttt{BAM} and \texttt{THC} numerical relativity codes. From Nagar et al.~\cite{Nagar:2018zoe}. }
    \label{fig:NR_NR_comparison}
\end{figure}

Numerical simulations of tidally interacting neutron stars in binaries employ eccentricity reduced initial data, high-order numerical schemes, and high-resolutions~\citep{Bernuzzi:2012ci, Radice:2013hxh, Bernuzzi:2016pie, Dietrich:2017aum, Kiuchi:2017pte, Doulis:2022vkx}. 
However, they typically employ simplified zero-temperature EOS models. 
This is because, on the one hand, thermal effects are expected to be negligible in the inspiral~\cite{Lai:1993di}. 
On the other hand, smoothness of density and pressure profiles in the stars are necessary for the codes to achieve high accuracy~\citep{Foucart:2019yzo, Knight:2023kqw}. 
For these reasons, matter is described using smooth, barotopic EOSs, i.e., using piecewise polytropic~\citep{Shibata:2005ss, Read:2008iy, OBoyle:2020qvf} or spectral models~\citep{Lindblom:2010bb}, augmented with an ideal-gas prescription to approximately include thermal effects~\citep{Bauswein:2010dn, Raithel:2019gws}. 
Simulations aiming to model the post-merger evolution, or the multi-messenger signals from NS mergers employ more sophisticated equation of state models \citep{Sekiguchi:2011zd}, as discussed below. 
The numerical uncertainty in binary inspiral simulations is typically quantified in terms of phase errors at mergers. 
High-resolution simulations achieve sub-radian accuracy over ${\sim}10$ orbits (20 gravitational wave cycles). 
Figure~\ref{fig:NR_NR_comparison} shows a comparison of \texttt{BAM} and \texttt{THC}, two independently developed and commonly employed numerical relativity codes. 
The two codes are in excellent quantitative agreement. 
Moreover, the independently-estimated uncertainty bands overlap, suggesting that simulations of tidally interacting binaries are reliable.

\begin{figure}
\centering
\includegraphics[width=0.45\textwidth]{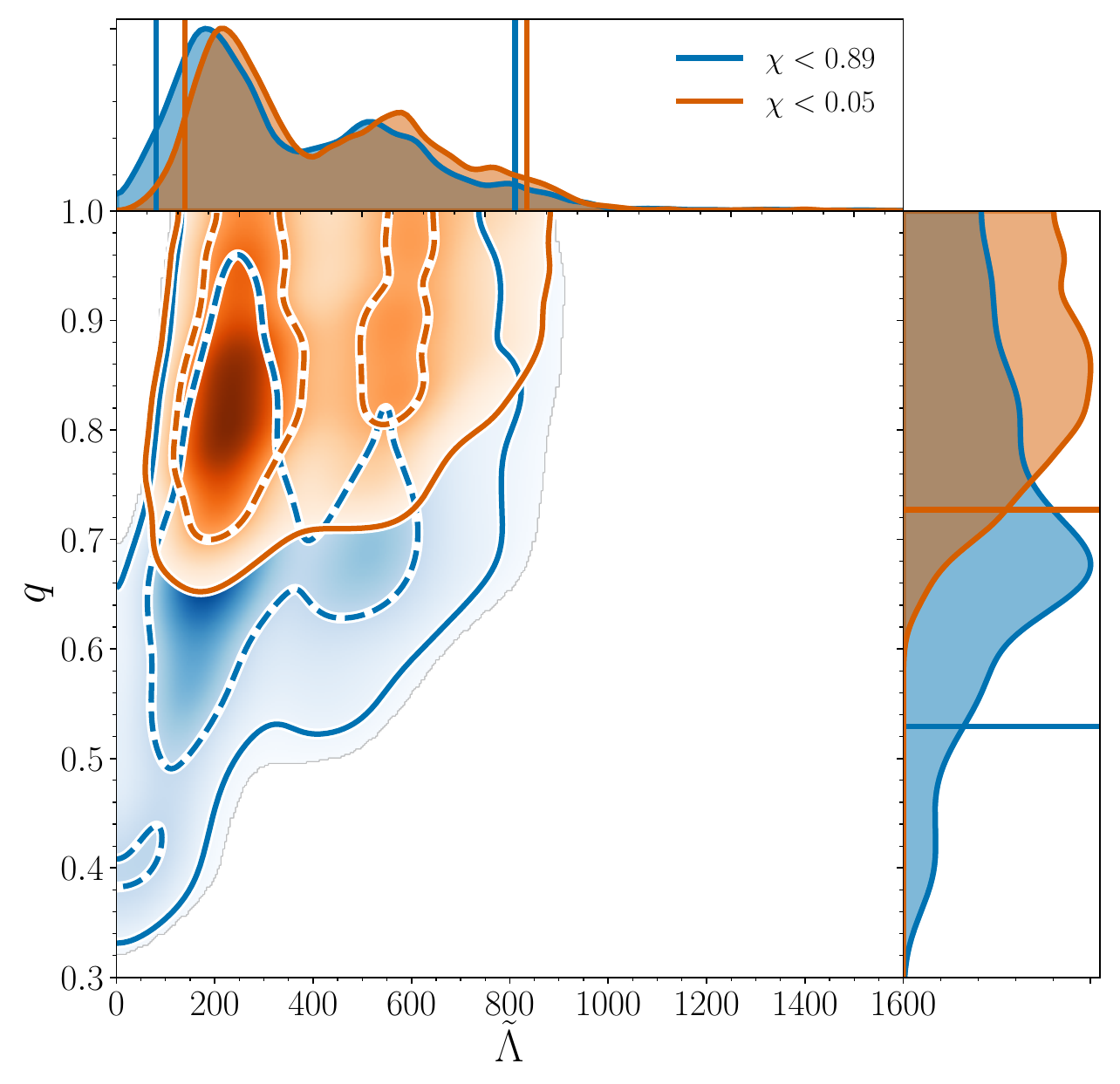}
\caption{Marginalized posterior for the mass ratio and $\tilde{\Lambda}$ of GW170817 showing the mild correlation between mass ratio and tides, reproduced from~\citet{LIGOScientific:2018hze}. }
\label{fig:GW170817mass-lambda}
\end{figure}

Despite the high PN term, $\tilde{\Lambda}$ is measurable at current sensitivity both due to the large prefactor in Eq.~\eqref{eq:Psitides} and the fact that nuclear equations of state predict $\tilde{\Lambda}\sim{\cal{O}}(10^2)$ in the relevant mass range.
Figure~\ref{fig:GW170817mass-lambda} shows the marginal posterior for the mass ratio $q$ and $\tilde{\Lambda}$ for GW170817 and for different assumptions about the neutron star spin~\cite{LIGOScientific:2018hze}.
The two parameters are moderately correlated ($\tilde{\Lambda}$ has a much stronger correlation with the time of coalescence $t_c$ that enters the phase at 4PN, see Eq.~\eqref{eq:GWphasePN}).
Regardless of spin, the GW170817 data place an upper limit on the tidal deformation of $\lesssim 700$ at the 90\% credible level~\cite{LIGOScientific:2018hze}.
Zero values of $\tilde{\Lambda}$ are disfavored, but not ruled out if spins are allowed to be large.
Higher-order tidal terms, and hence $\Lambda_1$ and $\Lambda_2$ individually, were not measured.
Consistent constraints are obtained with different waveform models, suggesting that they are robust against waveform systematics~\citet{LIGOScientific:2018hze}.

No competitive tidal constraints were achieved for GW190425 or any of the observed neutron star-black hole signals. 
Beyond their lower signal-to-noise ratio, this is also because the tidal imprint is much weaker for more massive neutron stars, $\Lambda ~\sim m^{-6}$ and more unequal-mass binaries, $\tilde{\Lambda} ~\sim q^{4}$.
Overall, among sub-$3\,M_{\odot}$ gravitational wave observations, only GW170817 resulted in informative constraints on tidal effects.

Translating the $\tilde{\Lambda}$ constraints of Fig.~\ref{fig:GW170817mass-lambda} to constraints on the neutron star radius requires additional steps, such as the use equation-of-state-independent relations~\cite{Yagi:2016bkt,Yagi:2015pkc,Chatziioannou:2018vzf,Carson:2019rjx,Zhao:2018nyf,De:2018uhw,Maselli:2013mva,Raithel:2018ncd,Annala:2017llu,Chatziioannou:2019yko} and generic equation-of-state models~\cite{Read:2008iy,OBoyle:2020qvf,Raithel:2016bux,Lindblom:2013kra,Lindblom2012,Lindblom:2010bb,Carney:2018sdv,Landry:2018prl,Essick:2019ldf,Landry:2020vaw,Alford:2013aca,Dietrich:2020efo,Steiner:2010fz,Tews:2018kmu}. 
Though these steps bring in their own uncertainties and model dependence~\cite{Greif:2018njt,Legred:2022pyp}, it is generally expected that tidal measurements achieved with gravitational wave observations constrain the nuclear equation of state for densities $\sim2-3 n_{\rm sat}$~\cite{Legred:2021hdx,Legred:2023als,Koehn:2024set}. 
Nonetheless, the $\tilde{\Lambda}$ upper limit for GW170817 implies that the radius of a $1.4\,M_{\odot}$ neutron star is $R_{1.4}\lesssim 13.5\,$km~\cite{LIGOScientific:2017vwq,LIGOScientific:2018hze,LIGOScientific:2018cki} at the 90\% level, see~\cite{Chatziioannou:2020pqz} for a review.

Though current constraints on GW170817 are robust~\cite{LIGOScientific:2018hze}, systematic uncertainties arising from incomplete knowledge of the two-body problem in general-relativity might impact constraints for future detections with a higher signal-to-noise ratio (SNR).
Numerical relativity simulation phase uncertainty grows rapidly in the very last few gravitational-wave cycles, when the stars come into contact. 
Numerical errors from the numerical relativity simulations alone dominate over statistical uncertainty in the measurement of tidal parameters from observations for SNRs $\gtrsim 80$~\citep{Gamba:2020wgg, Read:2023hkv}. 
Additional uncertainties arise for semi-analytic waveform models that, as previously discussed, start to deviate from numerical relativity data and from each other shortly before merger. 
In other words, while current models are adequate for GW170817 ($\mathrm{SNR} \simeq 32$), systematic uncertainties might dominate future events~\cite{Kunert:2021hgm}, particularly once next-generation detectors, such as Cosmic Explorer~\citep{Reitze:2019iox} and Einstein Telescope~\citep{Punturo:2010zz} come online. 
As such, the development of better simulations and waveform models is urgent. 
To this aim, new numerical schemes, such as discontinuous Galerkin methods~\citep{Radice:2011qr, Bugner:2015gqa, Hebert:2018xbk, Dunham:2020ugb, Tichy:2022hpa, Dumbser:2023see} are being explored.

\subsection{Merger and Post-merger}
\label{sec:BNSpostmerger}

After a tidally-accelerated late inspiral, the compact binary reaches gravitational wave frequencies $\gtrsim 1000\,$Hz and the merger stage.
Even though this stage is accompanied by large emission of gravitational energy, its high frequency places it in a region where the sensitivity of ground-based detectors rapidly deteriorates (the power spectral densities in Fig.~\ref{fig:GWspectrum} increase as $f^2$ above $1000\,$Hz).
No $\gtrsim 1000\,$Hz power has been unambiguously detected for any of the currently observed signals.
However, as we discuss below, this phase of the evolution is most consequential for the electromagnetic counter-parts to mergers. Moreover, the detection of the post-merger, kHz gravitational-wave signal for compact binaries is a key science goal for next-generation gravitational-wave experiments, such as Cosmic Explorer~\citep{Reitze:2019iox}, Einstein Telescope~\citep{Punturo:2010zz}, and NEMO \citep{Ackley:2020atn}.
The merger morphology and subsequent system evolution is qualitatively different for neutron star and mixed neutron star-black hole binaries, we therefore discuss each separately below.

\subsection{The outcome of neutron star binary mergers}
\label{sec:sim.outcome}

The outcome of a binary neutron star merger changes qualitatively depending on the parameters of the binary, with the total binary mass giving the leading-order effect~\citep{Shibata2016-lo, Shibata:2017xdx, Radice:2020ddv}. 
Because general-relativity has no intrinsic mass scale, it is the equation of state that provides one. 
This is typically taken to be the maximum (gravitational) mass of nonrotating neutron stars supported by the equation of state, $M_{\max}$. 
In other words, the merger outcome depends, to leading order, on $M_T/M_{\max}$, where $M_T = M_1 + M_2$ is the total binary mass at infinite separation.
\begin{itemize}
    \item If $M_T/M_{\max} \gtrsim 1.4$ then the merger results in ``prompt'' black hole formation~\citep{Shibata:2005ss, Hotokezaka:2011dh, Bauswein:2013jpa}.
    This scenario is the most likely for GW190425 owning to its high total mass~\cite{LIGOScientific:2020aai}.
    Gravitational wave emission is, in this case, characterized by the $\gtrsim 4\,$Hz ringdown radiation of the black hole which is prohibitively high in frequency for current detectors, but might be accessible with next-generation detectors \cite{Dhani:2023ijt}.
    The resulting gravitational wave sigal is depicted by the pink line in Fig.~\ref{fig:GWspectrum}.
    \item If $1.2 \lesssim M_T/M_{\max} \lesssim 1.4$, then the merger remnant is said to be a ``hypermassive neutron star'', because it cannot be supported by uniform rotation, unlike ``supramassive neutron stars'' which are discussed below.
    Hypermassive neutron stars are supported by \emph{differential} rotation. 
    They can collapse due to the loss of angular momentum to gravitational-waves in the early post-merger, when gravitational-wave losses are most intense, i.e., within ${\sim}20$\,ms (short-lived remnants)~\citep{Bernuzzi:2015opx, Zappa:2017xba}. 
    If this does not happen, the collapse can be triggered when turbulent viscosity erases the differential rotation. 
    This is expected to occur on a longer timescale of a few hundred milliseconds (long-lived remnants). 
    See~\citet{Radice:2020ddv} for a detailed discussion. 
    The mass threshold separating short- and long-lived hypermassive NSs is presently unknown.
    
    The typical gravitational wave signals produced by such remnants are depicted in green, orange, and purple in Fig.~\ref{fig:GWspectrum}. 
    Even though this scenario is the most likely for GW170817~\cite{Margalit:2017dij}, no signal was detected~\cite{LIGOScientific:2017fdd,LIGOScientific:2018hze}. 
    Numerical simulations suggest that the detector sensitivity was a factor of $\gtrsim 3$ too low and such a signal could be detected once LIGO achieves sensitivities $\gtrsim2-3$ over its design sensitivity~\cite{Torres-Rivas:2018svp}.
    \item If $1 \lesssim M_T/M_{\max} \lesssim 1.2$, then the merger remnant is called a ``supramassive neutron star''. Supramassive neutron stars are supported against gravity by the combined effects of pressure and centrifugal forces. Binaries in this mass range are expected to produce long-lived remnants that can only collapse after having lost angular momentum due to gravitational-waves and/or magnetized winds. The associated timescale for collapse is highly uncertain and could range from a few seconds to hours~\citep{Radice:2018xqa}. Shorter survival times are expected for the more massive binaries. In addition to the gravitational waves produced shortly after merger by the deformed massive neutron star remnant, very long-lived remnants might continue to emit low amplitude signals until they collapse. Such a signals could be targeted with continuous-waves techniques~\cite{Riles:2022wwz}. However, no detection was achieved for GW170817 or GW190425~\cite{LIGOScientific:2017fdd,LIGOScientific:2018urg,Grace:2024xty}.
    \item If $M_T/M_{\max} \lesssim 1$, then the merger results in the formation of a stable neutron star. This case appears to be unlikely, since the typical mass of neutron star binaries is $2.7\, M_\odot$, so the equation of state must be extremely stiff, or the binary mass extremely small for this case to occur. However, it is not completely ruled out \citep{Piro:2017zec}.
\end{itemize}

The threshold masses separating the different regimes are affected, to sub-leading order, by the mass ratio and spin of the remnants~\citep{Papenfort:2022ywx}, as well as other parameters of the EOS besides $M_{\max}$~\citep{Radice:2018xqa}. They are also only approximately known, because systematic exploration of the long-term outcome of neutron star binary mergers is beyond the capability of current simulation codes. 
One exception is the case of prompt collapse, which can be studied with relatively short simulations and is now well understood. 
We discuss it in detail below.

The ${\cal{O}}(10)\,$ms signal emitted by massive neutron stars remnants provides the best chance of postmerger detection due to its high total emitted energy. 
Even though the signal is complicated and likely depends on unresolved simulation physics, it exhibits certain morphological features whose origin and interpretation are considered robust~\citep{Zappa:2022rpd, Espino:2023dei}.
The main feature of the signal is that it is dominated by a single spectral peak, see Fig.~\ref{fig:GWspectrum}, whose frequency depends on the equation of state and system mass~\cite{Bauswein:2019ybt,Baiotti:2019sew,Bernuzzi:2020tgt}.
Subdominant features such as further spectral peaks have also been identified and interpreted in terms of the remnant dynamics~\cite{Bauswein:2015yca, Rezzolla:2016nxn}.

Measurement of the peak frequency from postmerger data combined with a measurement of the binary total mass from inspiral data would yield a ${\cal{O}}(10-100)\,$m radius measurement~\cite{Bose:2017jvk,Chatziioannou:2017ixj,Breschi:2023mdj,Criswell:2022ewn}.
Moreover, since the hypermassive remnant probes higher densities than the pre-merger neutron stars, comparison between the pre-merger and postmerger data would probe potential high-density phase transitions~\cite{Most:2018eaw,Bauswein:2018bma,Liebling:2020dhf,Prakash:2021wpz,Prakash:2023afe}.

Searches and inference for the postmerger signal is hampered by the signal's complicated morphology that, unlike the premerger signal, cannot be described from first principles and analytical calculations.
However, its short duration makes it ideal for morphologically-independent analyses~\cite{Clark:2014wua,Clark:2015zxa,Chatziioannou:2017ixj,Wijngaarden:2022sah,Criswell:2022ewn,Tringali:2023ray} or phenomenological models based on combinations of damped sinusoids~\cite{Tsang:2019esi,Breschi:2019srl,Easter:2020ifj,Soultanis:2021oia}.
Ultimately, detection of a postmerger signal would yield unprecedented constraints on the neutron star equation of state, c.f. Fig.~\ref{fig:GWspectrum}, where signals from different equations of state are more diverse postmerger than premerger.
However, the sharply decreasing high-frequency sensitivity of ground-based detectors makes detection of a postmerger signal with the current generation of detectors unlikely unless another exceptionally nearby signal is detected.

\subsubsection{Prompt black hole formation}
\label{sec:sim.prompt}

\begin{figure*}
    \includegraphics[width=\textwidth]{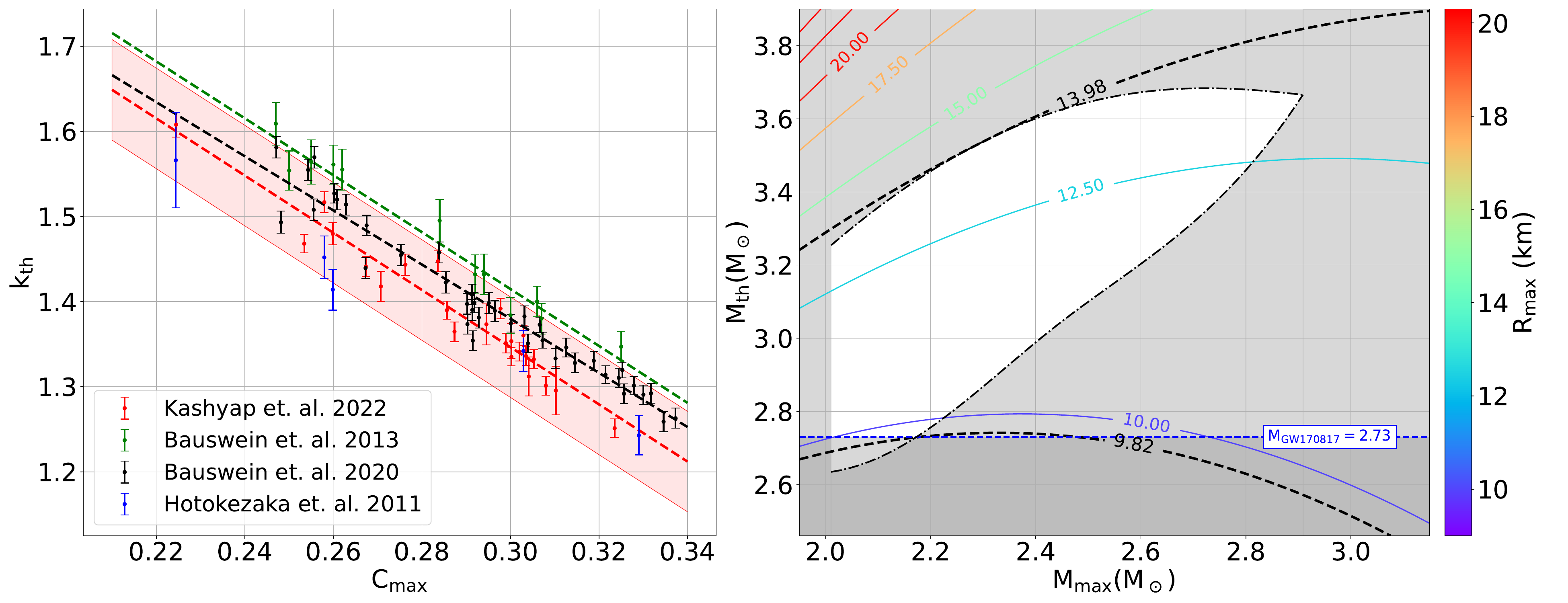}
    \caption{Threshold for prompt collapse $k_{\rm th}$ as a function of the compactness of the maximum mass NS (\emph{left panel}) and resulting constraints on $M_{\rm max}$ and $R_{\max}$. Adapted from Kashyap et al.~\cite{Kashyap:2021wzs}. } 
    \label{fig:Rmax_constraint}
\end{figure*}

The threshold for prompt collapse in binary neutron star mergers has been studied by several groups~\citep{Shibata:2005ss, Hotokezaka:2011dh, Bauswein:2013jpa, Zappa:2017xba, Bauswein:2017vtn, Koppel:2019pys, Bauswein:2020aag, Bauswein:2020xlt, Tootle:2021umi, Kashyap:2021wzs, Perego:2021mkd, Kolsch:2021lub, Papenfort:2022ywx, Cokluk:2023xio}. 
Unfortunately, different studies use inconsistent definitions for what is a ``prompt'' collapse. 
The most commonly adopted definition, which is also the one we endorse, is that the collapse is ``prompt'' if the stars do not exhibit any sign of a centrifugal bounce after merger. 
In this case, the maximum (rest-mass) density grows monotonically with time until black hole formation. 
This definition has two key advantages over alternatives. 
First, it is gauge and slicing independent, so it is well defined. 
Second, and most importantly, the bounce is related to the ejection of matter and the generation of an electromagnetic counterpart, Sec.~\ref{sec:sim.mma}. 
Indeed, for comparable mass-ratio binaries $q \gtrsim 0.75$, mass ejection occurs only as a result of the post-merger bounce of the remnant and of its subsequent evolution~\citep{Hotokezaka:2012ze, Bauswein:2013yna, Radice:2018pdn, Shibata:2019wef}. 
In absence of a bounce there is little to no ejecta. 
Since the main electromagnetic counterpart to binary neutron star mergers, the kilonova, is due to the radioactive decay of the expelled material, the absence of ejecta implies the absence of an electromagnetic counterpart. 
As such, whether or not prompt collapse has occurred for a particular binary can be determined observationally, when using this definition. 
For more asymmetric binaries, $q \lesssim 0.75$, the secondary star in the binary is tidally disrupted prior to merger~\citep{Dietrich:2016hky, Bernuzzi:2020txg}, potentially resulting in massive outflows even under prompt collapse.

Simulations have shown that, for equal mass binaries, prompt collapse occurs if~\citep{Bauswein:2013jpa, Bauswein:2017vtn, Kashyap:2021wzs}
\begin{equation}
    M > M_{\rm th} = k_{\rm th}\, M_{\max}\,,
\end{equation}
where $k_{\rm th} = 1.2{-}1.7$ depending on the equation of state. 
A typical value of $k_{\rm th}$, adopted here, is ${\sim}1.4$. Moreover, $k_{\rm th}$ strongly correlates with the compactness of the maximum mass nonrotating neutron star: $C_{\max} = G M_{\max}/ c^2 R_{\max}$ and is well fitted by
\begin{equation}\label{eq:prompt.kth}
    k_{\rm th} = a\, C_{\max} + b\,,
\end{equation}
with $a = -3.36 \pm 0.20$ and $b = 2.35 \pm 0.06$.\footnote{These relations were first discovered by~\citet{Bauswein:2013jpa}, but we report the more accurate values computed by~\citet{Kashyap:2021wzs}.} 
Figure~\ref{fig:Rmax_constraint} shows a compilation of $k_{\rm th}$ estimates from different works in the literature as a function of $C_{\max}$ (\emph{left panel}) and the resulting constraints on $R_{\max}$ (\emph{right panel}). 
Each point on the left panel corresponds to a different equation of state and/or to simulations by different groups. 
Although there is good qualitative agreement, there is a systematic offset between simulations employing approximate General Relativity~\citep{Bauswein:2013jpa, Bauswein:2020xlt} and those employing full-General Relativity~\citep{Hotokezaka:2011dh, Kashyap:2021wzs}. 
GW170817 had an electromagnetic counterpart, suggesting that $M_{\rm th} > M_{\rm GW170817} \simeq 2.74\, M_\odot$. 
Pulsar observations suggest that $M_{\max} > {2.01}\, M_\odot$, see Sec.~\ref{sec:radio.masses}. 
These lower limits on $M_{\rm th}$ and $M_{\max}$ and Eq.~\eqref{eq:prompt.kth} result in a constrain on the radius of the maximum-mass neutron star, $R_{\max} > {9.8}\,$km. 
Similarly, the coefficient $k_{\rm th}$ strongly correlates with the modified compactness of the $1.6\, M_\odot$ neutron star:
\begin{equation}\label{eq:prompt.kth.2}
    k_{\rm th} = a'\, C_{1.6}^\ast + b'\,, \quad
    C_{1.6}^\ast = \frac{G M_{\rm max}}{c^2 R_{1.6}}\,,
\end{equation}
from which it is possible to derive a lower limit on $R_{1.6}$ of $10.9\,$km \citep{Kashyap:2021wzs}.

Both constraints on $R_{\max}$ and $R_{1.6}$ neglect corrections on $M_{\rm th}$ due to the mass ratio. However, these are smaller than the systematic errors in Eqs.~\eqref{eq:prompt.kth} and \eqref{eq:prompt.kth.2}, when $q \gtrsim 0.75$, while $M_{\rm th}$ decreases with $q \lesssim 0.75$ \citep{Bauswein:2020xlt, Perego:2021mkd, Kolsch:2021lub}. As such, these lower limits are conservative.

The impact of mass ratio on the threshold for prompt collapse has been investigated by several groups~\citep{Bauswein:2020xlt, Perego:2021mkd, Kolsch:2021lub, Papenfort:2022ywx}. 
Although early studies proposed that $M_{\rm th}(q) / M_{\rm th}$ could be independent on the EOS~\citep{Bauswein:2020xlt, Tootle:2021umi}, a more systematic exploration by~\citet{Perego:2021mkd} showed that $M_{\rm th}(q)/M_{\rm th}$ depends on the equation of state. 
For example, while for some equations of state $M_{\rm th}(q)$ decreases monotonically with $q$, for others $M_{\rm th}(q)$ can initially grow as $q$ is reduced from $1$, before declining as $q$ is further reduced. 
\citet{Perego:2021mkd} suggested that the dynamics of prompt collapse binaries changes qualitatively when $q \lesssim 0.75$. 
For $0.75 \lesssim q \leq 1$ the dynamics is similar to the $q = 1$ case: as the stars are about to merge, a black hole is formed at their collisional interface. 
Shortly afterwards, both stars are swallowed by the horizon and little to no material is ejected. 
Within this range of $q$, $M_{\rm th}$ changes by less than 2\% compared to the $q = 1$ case. 
If $q \lesssim 0.75$, then the less massive component is tidally disrupted before merger and black hole formation is the result of accretion of matter onto the primary component. 
Such mergers are accompanied by significant mass ejection and are expected to be electromagnetically bright, despite the rapid black hole formation. 
In this regime, $M_{\rm th}(q)$ is always smaller than $M_{\rm th}$ and decreases rapidly with $q$ than for comparable mass ratio systems. 
However, is $M_{\rm th}(q)$ still confined to be within 10\% of the $q = 1$ value.

\citet{Perego:2021mkd} further reported that ${\rm d}M_{\rm th}/{\rm d}q$ is strongly correlated with the incompressibility parameter
\begin{equation}
    K_{\rm max} = 9\, \left. \frac{\partial p}{\partial n}\right|_{n = n_{\rm max}}\,,
\end{equation}
where $n_{\max}$ is the central density in the maximum-mass neutron star and the derivative is computed for $\beta$-equilibrated matter. 
In particular, as $K_{\max}$ increases, ${\rm d}M_{\rm th}/{\rm d}q$ decreases. 
At a fixed $q$, higher $K_{\max}$ then results in an increase of the threshold mass for prompt collapse, which is reasonable since $K_{\max}$ measures the pressure response of the primary star to accretion of material from the secondary. 
This correlation implies that a measurement of $M_{\rm th}(q)$ for two or more $q$'s would directly constrain $K_{\max}$.

\subsubsection{Delayed collapse: multimessenger constraints}
\label{sec:sim.mma}

The dynamics of binary neutron star systems close to merger has a profound impact on the properties of their electromagnetic counterparts. 
However, while there is a sharp change in the dynamics with prompt collapse, the mass of the ejecta, as well as the mass of the bound accretion disk left after merger, change smoothly with the parameters of the binaries. As a result, the kilonova signal can be used to disfavor ranges of parameters in a continuous fashion, yielding more stringent constraints on the equation of state.

\begin{figure}
    \centering
    \includegraphics[width=\columnwidth]{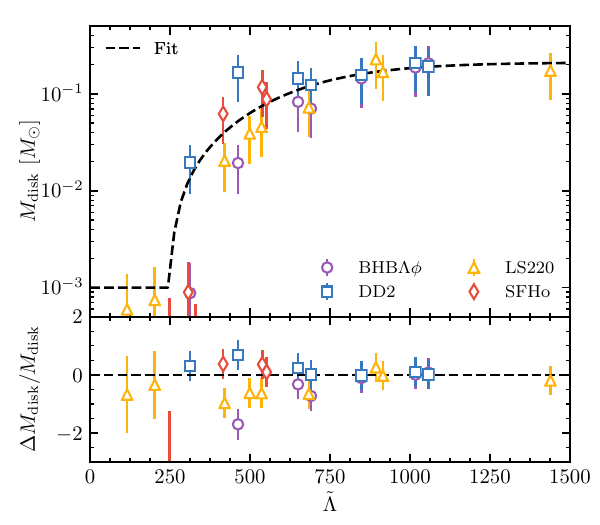}
    \caption{Bound debris formed in comparable mass-ratio binary mergers, as a function of the tidal parameter $\tilde\Lambda$. From~\citep{Radice:2018pdn}. }
    \label{fig:remnant_accretion_disk}
\end{figure}

\begin{figure}
    \centering
    \includegraphics[width=\columnwidth]{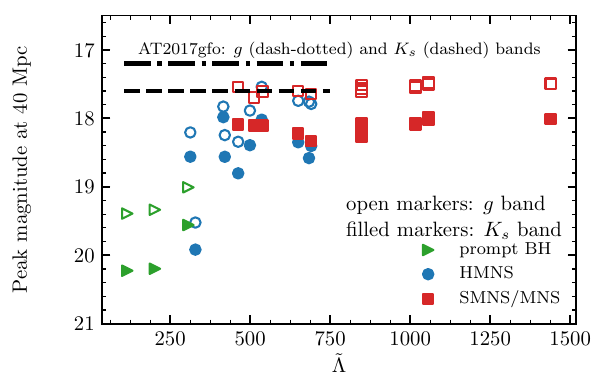}
    \caption{Predicted peak kilonova luminosity as a function of $\tilde\Lambda$ for a set of binary neutron star merger simulations. From~\citep{Radice:2020ddv}. }
    \label{fig:kilonova_vs_lambda}
\end{figure}

For example, Figure~\ref{fig:remnant_accretion_disk} shows the accretion disk mass measured for comparable mass ratio $q \gtrsim 0.85$ binary neutron star merger simulations as a function of the tidal parameter $\tilde\Lambda$ reported in~\citet{Radice:2018pdn}. 
Binaries with smaller $\tilde\Lambda$ are more compact and result in rapid formation of black holes with small ejecta mass and small accretion disks. 
As a result, their associated kilonova signal is expected to be up to two magnitude fainter and to evolve more rapidly, as shown in Figure~\ref{fig:kilonova_vs_lambda}. 
This picture changes qualitatively for $q \lesssim 0.75$, when the tidal disruption of the secondary neutron star can lead to the formation of massive disks, even in conjunction with prompt black hole formation. 
Overall, scenarios with very small tidal parameter and comparable mass ratios are disfavored by the presence of a kilonova in GW170817. 
This fact has been used to place more stringent constraints on the equation of satte with multimessenger data, as discussed below.

\citet{Radice:2017lry} showed that the sum of disk and ejecta masses drops sharply for binaries with $\tilde\Lambda \lesssim 400$ and proposed a constraint on the tidal deformability of GW170817 from below. \citet{Kiuchi:2019lls} warned against using the value of $\tilde\Lambda = 400$ as  a strict lower limit and showed examples of binaries with $\tilde\Lambda < 400$ and significant mass ejection. 
Indeed, multi-messenger analyses incorporating numerical relativity data and their systematic uncertainties find that the lower limit on $\tilde\Lambda$ is actually about $200$ at the 90\% confidence level~\cite{Radice:2018ozg}, corresponding to a radius constraint of $R_{1.4} \gtrsim 11~\,$km. 
The disk mass data presented in~\citet{Radice:2017lry} and~\citet{Radice:2018pdn} was later refitted by~\citet{Coughlin:2018fis} in terms of the threshold mass to prompt collapse and used in several works~\citep{Capano:2019eae, Dietrich:2020efo, Huth:2021bsp, Pang:2022rzc}. 
Fits using a larger set of simulations results, extending to much larger mass ratios and using simulations targeted to GW170817 have been presented in~\citet{Nedora:2020hxc, Nedora:2020qtd} and used in~\citet{Breschi:2019srl} for a more careful reanalysis of GW170817 and the associated kilonova AT2017gfo. 
Overall, results from different groups are in broad agreement and suggest that $11\, {\rm km} \lesssim R_{1.4} \lesssim 13\, {\rm km}$~\citep{Breschi:2019srl}. 
Combined constraints from GW170817 and NICER~\citep{Raaijmakers:2019dks, Raaijmakers:2021uju, Breschi:2024qlc} and heavy-ion collisions~\citep{Tsang:2023vhh} favor slightly larger $R_{1.4}$ and provide better constraints for the equation of state at densities beyond those probed in a $1.4\, M_\odot$ neutron star.

\subsubsection{Long-lived merger remnants}
\label{sec:sim.long_lived}

Neutron star binary systems with sufficiently low total mass can result in the formation of ``long-lived remnants'', which avoid black hole formation for a timescale that is long (possibly infinite for absolutely stable remnants) compared to the postmerger gravitational wave emission timescale of ${\sim}10{-}20\,$ms. 
Such remnants might collapse as a result of angular momentum loss to a magnetized wind and/or to residual gravitational wave emission. 
The formation of a long-lived remnant is expected to be relatively common~\citep{Piro:2017zec, Radice:2018xqa}, since there are low-mass neutron star binary candidates in our galaxy that would produce a long-lived remnant given the current lower limit on $M_{\max}$. 
Long-lived remnants have further been invoked to explain X-ray tails in short gamma-ray bursts~\citep{Dai:1998bb, Dai:1998hm, Zhang:2000wx, Dai:2006hj, Metzger:2007cd, Rowlinson:2013ue, Murase:2017snw, Dimple:2023wvs}. 
The formation of a long-lived remnant is a possible explanation for a radio flare following GRB 201006A~\citep{Rowlinson:2023eip}, although such an interpretation is debated~\citep{Sarin:2024uhe}. 
On the other hand, because long-lived remnants have a rotational energy of ${\sim}10^{53}\,$erg~\citep{Margalit:2017dij}, if such remnants are formed and if their energy is radiated in the form of a wind, their formation should be accompanied by an extremely bright kilonova counterpart~\citep{Gao:2016uwi, Sarin:2022wby, Wang:2023qww} and the formation of a loud radio remnant~\citep{Nakar:2011cw, Gao:2013rd}. Neither signature has been observed~\citep{Schroeder:2020qbf, Ghosh:2022aks, Eddins:2022rlw}. 

\begin{figure*}
\centering
\includegraphics[width=\textwidth]{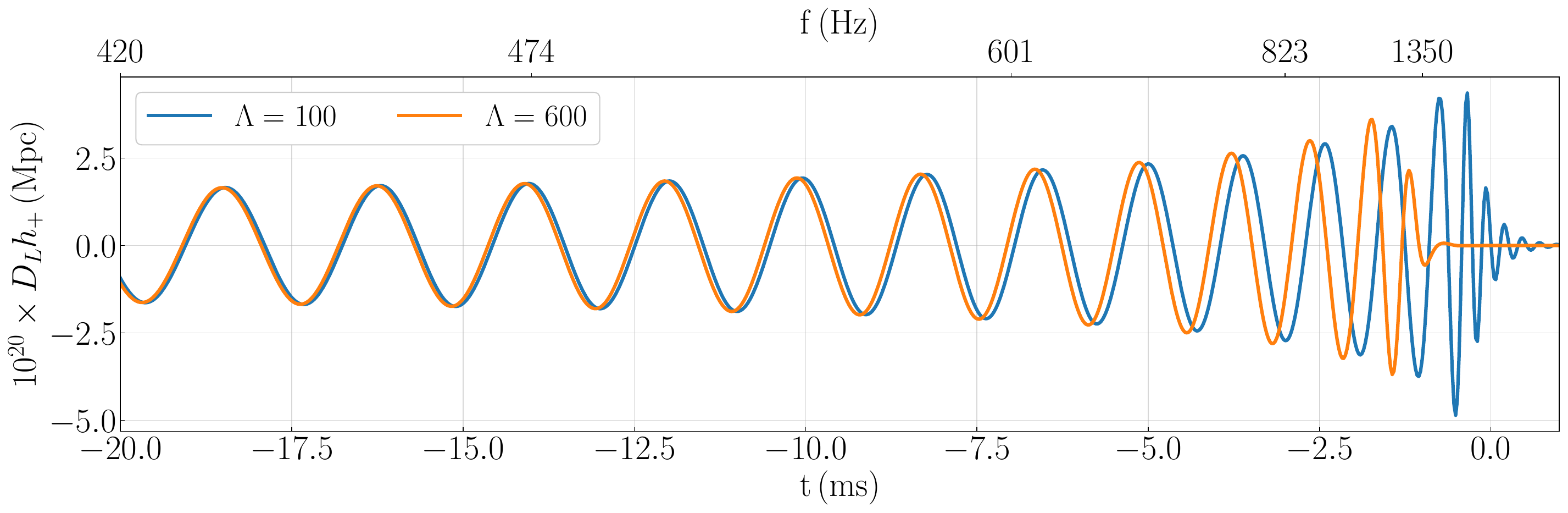}
\caption{Final inspiral and merger stages of a neutron star-black hole binary for different values of the tidal deformability of the neutron star in the time domain. 
We plot the waveform using {\tt IMRPhenomNSBH}~\cite{Thompson:2020nei} for a nonspinning $1.4\,M_{\odot}$ neutron star and a nonspinning $3\,M_{\odot}$ black hole. 
The upper x-axis denotes the corresponding frequency for the $\Lambda=100$ binary. 
Tidal deformation speeds up the binary evolution.
The more deformable neutron star (orange) disrupts before merger, resulting in a rapid signal attenuation.
The less deformable neutron star (blue) plunges into the black hole, leading to a ringdown-like signal. }
\label{fig:GWNSBHwithtides}
\end{figure*}

The implications for the equation of state are unclear and it is not excluded that even remnants believed to be ``long-lived'' might collapse due to effects that are currently not modeled in the simulations~\citep{Radice:2018xqa, Beniamini:2021tpy, Margalit:2022rde, Radice:2023zlw}, or that most of the rotational energy is lost to gravitational waves~\citep{Ai:2019rre, Yuan:2023fqk} and/or fails to thermalize the merger ejecta~\citep{Wang:2023yqy, Ai:2024zcn}. 
In the case of GW170817, the absence of a signature of a long-lived remnant has been broadly interpreted as placing an upper limit on the maximum mass of NSs of $2.17{-}2.3\,M_\odot$, with more recent works generally favoring the more conservative upper limit in the light of the current uncertainties~\citep{Margalit:2017dij, Shibata:2017xdx, Rezzolla:2017aly, Ruiz:2017due, Margalit:2022rde}. 
On the other hand, if long-lived remnants are assumed to be required to produce gamma-ray burst with late-time activity, then about 1/4 of the mergers need to form long-lived remnants. 
Given our current knowledge of the mass distribution of neutron star binary mergere, this would translate in a lower limit on the maximum mass of neutron stars of ${\sim}2.1\, M_\odot$~\citep{Lasky:2013yaa, Ai:2019rre}. 

Overall, a better theoretical understanding of the fate of such remnants and of the associated electromagnetic emission is needed to make such constraints robust.

\subsection{Neutron star-black hole binaries}

Two qualitatively different merger scenarios are possible for mixed neutron star-black hole binaries depending on the system properties, depicted in Fig.~\ref{fig:GWNSBHwithtides}.
In the first scenario (orange), the neutron star is severely deformed by the tidal field of the black hole to the point of complete disruption~\cite{Lackey:2013axa}.
The gravitational wave signal then effectively turns off rapidly as the system is no longer comprised of two compact objects and the overall orbital quadrupole moment (and its derivatives) are small.
Disruption therefore strongly affects the gravitational wave amplitude~\cite{Lackey:2013axa,Pannarale:2013uoa,Pannarale:2015jka,Thompson:2020nei,Matas:2020wab}.
In the second scenario (blue), the neutron star is more mildly deformed and plunges through the black hole horizon. 
The resulting merger gravitational wave signal is then similar to that of a binary black hole as the remnant black hole rings down to a steady state~\cite{Foucart:2013psa}.

Which scenario is realized depends on the system properties~\cite{Lackey:2013axa,Foucart:2013psa,Foucart:2018lhe,Foucart:2018rjc}. 
Neutron star disruption becomes more likely for systems with a high black hole mass (as the curvature and hence the tidal field strength is inversely proportional to the mass), large black hole spin aligned with the orbital angular momentum (as the black hole horizon is smaller for larger spins and thus the neutron star can continue orbiting closer to the black hole before merger), and a neutron star that is less compact (as then it is more deformable).
Detection of a disrupting signal would place strong constraints on the above system properties.
Indeed since premerger tidal deformation~\cite{Kumar:2016zlj} is in general expected to be small for neutron star-black hole binaries as $\tilde{\Lambda}\sim q^4$, the merger conditions and the frequency at which the neutron star might disrupt~\cite{Pannarale:2015jia} are complementary probes of the system properties and the equation of state.

None of the currently detected neutron star-black hole signals are informative about either scenario~\cite{LIGOScientific:2021qlt,LIGOScientific:2024elc} due to the high disruption frequency.
Disruption is also in general expected to be more rare~\cite{Foucart:2018rjc}, especially since observationally most black holes in binaries have low spins, masses above $\sim5\,M_{\odot}$~\cite{KAGRA:2021duu} and GW170817 places an upper limit on the neutron star compactness~\cite{LIGOScientific:2018hze}. 
The best candidate for disruption and an electromagnetic counterpart is GW230529~\cite{LIGOScientific:2024elc} due to its low primary mass, however even in this case no direct observational evidence exists in the gravitational wave signal.

\section{Electromagnetic observations}

We will now present a review of recent pulsar mass measurements obtained with radio observations. The primary focus will be on measurements of extremely massive neutron stars obtained via the relativistic Shapiro delay \cite{Demorest10,Fonseca21}. Shapiro delay observations of a larger number of somewhat less massive neutron stars through pulsar timing arrays and other efforts will also be discussed. We will detail the relatively limited number of systematic uncertainties associated with this technique and discuss perceived trends in mass measurements (increasing or decreasing) as their uncertainties improve over time. We will briefly touch on recent extremely-high-mass results obtained from multiwavelength studies of ``spider'' pulsars, some of which have been assisted by radio timing campaigns. Lastly, we will discuss how upcoming and planned radio facilities will improve our efforts to more precisely measure more neutron star masses. 

In this section we will also describe constraints on the equation of state from the (predominantly) thermal emission of isolated neutron stars, soft-X-ray transients, and quiescent low-mass X-ray binaries. 
Relativistic effects imprint information about neutron star masses and radii in the emissions that we measure.
By modeling these effects we can recover information on neutron-star structure.
We will review constraints arising from spectral modeling of neutron stars exhibiting crust cooling or photospheric radius expansion bursts, and those being delivered by pulse profile modeling, a technique being pioneered by NICER. 
This section will describe recent observations, their connections to the equation of state, and the extent to which systematic uncertainties may impact those connections.  

\subsection{Radiation processes}
\label{sec:radiation}

Neutron stars emit across the electromagnetic spectrum, from low frequency radio \citep{LOTAAS} to very high energy gamma-ray \citep{Hess23}.  But there are two classes of electromagnetic emission that are particularly important for measurements that let us constrain the EOS.  

The beamed non-thermal magnetospheric emission that leads many neutron stars to manifest as radio or gamma-ray pulsars \citep{Philippov22} - due to a misalignment between the radiation beam and the rotational axis - enables precision measurement of the spin of many neutron stars. Long-term monitoring of spin and orbital evolution in relativistic binaries permits the measurement of neutron star masses, discussed in more detail in Sec~\ref{sec:radio.masses}.  

The other involves thermal emission from the surface, typically (due to the temperatures involved) in the X-ray. There are various physical processes that can heat the neutron star surface to X-ray emitting temperatures: some internal (latent heat from birth or internal magnetic field decay, thermonuclear bursting due to unstable nuclear burning in accreted oceans); and some external (accretion, magnetospheric return current heating).  Heating may be uniform or non-uniform: when it is non-uniform, and X-ray emitting hotpots are misaligned with the rotational axis, the resulting emission will be pulsed at the spin rate of the neutron star.   

As photons emitted from the surface layers propagate through the neutron star space-time towards the observer, relativistic effects encode information about mass and radius\footnote{The radius being measured is the photosphere: how this relates to the radius as defined in the TOV equations and any subsequent EOS analysis will need to be considered.} in the spectrum and timing properties of the radiation. A rotating neutron star is not perfectly spherical, and the spacetime is not Schwarzschild.  While ray-tracing in a numerically-computed spacetime \citep{Stergioulas95,Vincent18} is possible, there exist approximations that are computationally faster and sufficiently accurate for most current problems.   For slowly-rotating stars the key effects to be modelled are gravitational redshift and light deflection, which leads to more of the stellar surface being visible \citep[see e.g.][]{Nattila17}.  As rotation increases one must also consider Doppler boosting, aberration, time delays, the effects of rotationally-induced stellar oblateness, and perhaps the mass quadrupole moment, with the ray-tracing being more complex if surface emission is not uniform.  There exists a large body of work developing computationally efficient space-time approximations that take these effects into account \citep{Pechenick83,Strohmayer92,Miller98,Braje00,Beloborodov02,Poutanen03,Cadeau07,Morsink07,Baubock12,Psaltis14b,AlGendy14,Baubock15,Nattila18,Bogdanov19b,Silva21,Oliva21}.  The use of relativistic ray-tracing models to extract mass and radius from X-ray data is discussed in more detail in Secs.~\ref{sec:xrayspec} and~\ref{sec:ppm}.

\subsection{Radio observations: masses}
\label{sec:radio.masses}

Pulsars are a rapidly rotating, highly magnetized subset of the neutron star population whose lighthouse-like beams are perceived as pulsations given a favorable orientation with the Earth's line of sight. 
Roughly $\sim$1/10 of these objects are millisecond pulsars that have been spun up to exceptional ($\sim$ms) rotational rates through the accretion of matter from a companion star. 
Millisecond pulsars are ideal targets for pulsar timing, a process by which a pulsar's clock-like nature is exploited to create a model that will predict a pulse time of arrival accounting for every rotation over long time spans while maintaining phase coherence. 
High-precision ($\sim$$\mu$s) pulsar timing facilitates a comparison between expected and measured times of arrival, indicating if the pulses have been influenced by some unaccounted-for phenomenon. 
Millsecond pulsar timing has been used to great effect in efforts to test general relativity, detect low-frequency (nHz) gravitational waves, and constrain the neutron star equation of state. 

The power of pulsar timing in directly constraining the neutron star interior equation of state is straightforward. 
Because each equation of state dictates a maximum neutron star mass, a well-constrained mass measurement for a sufficiently massive pulsar will rule out equations of state that predict a lower maximum mass. 
The troublesome aspect of this approach is the difficulty with which precise mass measurements for pulsars can be obtained; however, in the subset of systems that permit such measurements, radio pulsar timing offers a relatively model-independent (assuming the validity of general relativity) and direct method for obtaining neutron star masses. 

\begin{figure}
\centering
\includegraphics[width=0.45\textwidth]{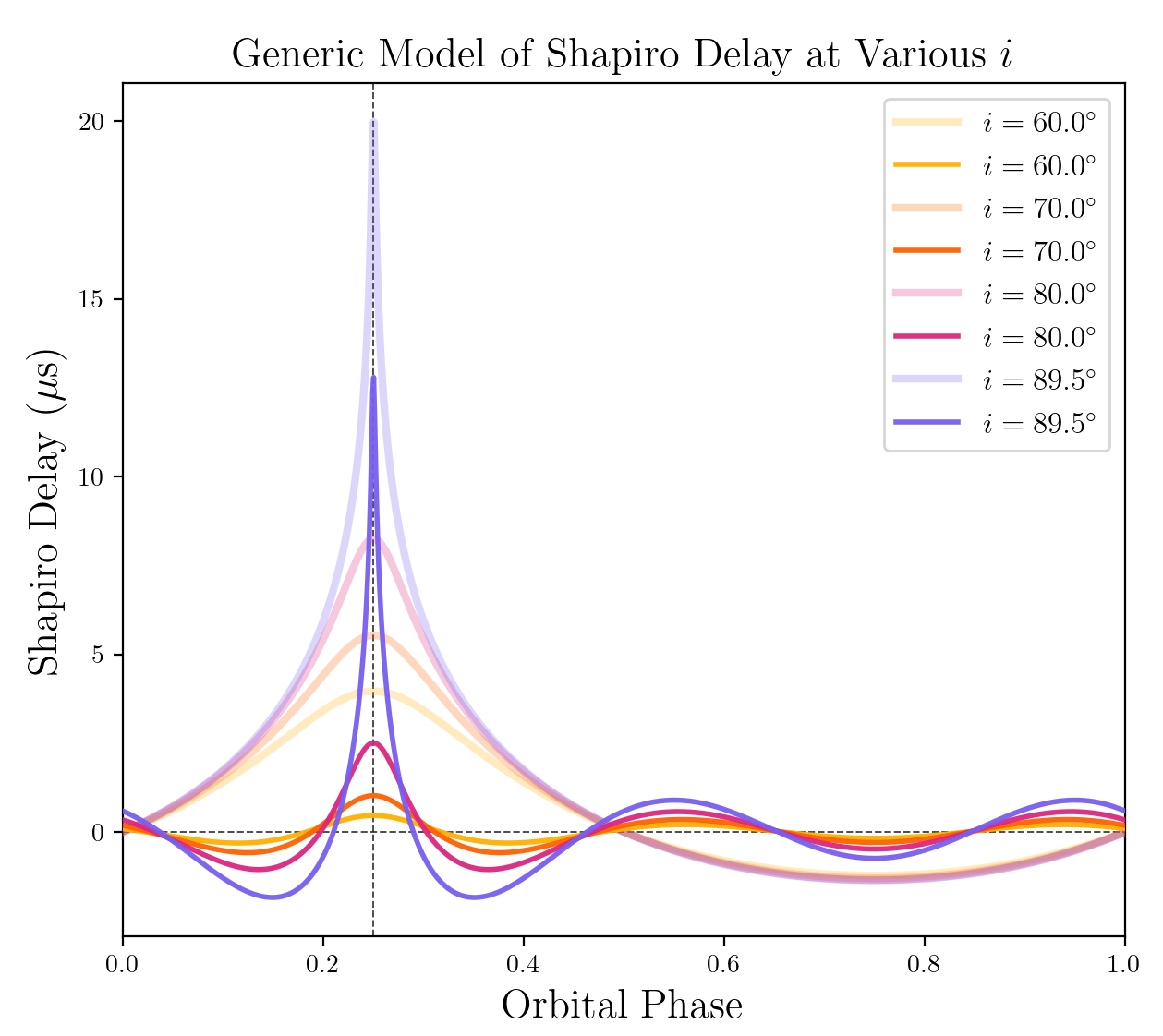}
\caption{Pulse arrival delay induced by the the Shapiro time delay. Translucent curves represent the absolute Shapiro delay, while opaque curves indicate the measurable Shapiro delay (the amount of detectable Shapiro delay when simultaneously fitting for all orbital parameters).}
\label{fig:SD}
\end{figure}

For binary systems, an adequate pulsar timing model requires five Keplerian orbital parameters: the projected semimajor axis $x\equiv a\,\textnormal{sin}(i)/c$, longitude of periastron $\omega$, time of periastron passage $T_\textnormal{0}$, orbital period $P_\textnormal{b}$, and orbital eccentricity $e$. As these parameters are measured to remarkably high precision, all but the pulsar and companion masses ($m_{\textnormal{p}}$ and $m_{\textnormal{c}}$, respectively) nearly complete the Keplerian mass function:
\begin{equation}
f(m_{\textnormal{p}}, m_{\textnormal{c}}) = \frac{4\pi^2}{G}\frac{(a\,\textnormal{sin}i)^3}{P_b^2} = \frac{(m_c \,\textnormal{sin}i)^3}{(m_p + m_c)^2}\,,
\end{equation}
leaving only the component masses unknown. In a relatively small number of observed Millisecond pulsar binaries, one is able to measure post-Keplerian general relativistic parameters: the rate of periastron advance $\dot{\omega}$, Einstein delay $\gamma$, orbital period decay $\dot{P_{\textnormal{b}}}$, and the Shapiro delay ``range'' and ``shape'' parameters $r$ and $s$:
\begin{align}
r &= T_{\odot} m_\textnormal{c}\,,
s &= \textnormal{sin}(i) = x\left(\frac{P_\textnormal{b}}{2\pi}\right)^{-2/3}T_{\odot}^{-1/3}M_T^{2/3}m_\textnormal{c}^{-1}\,,
\end{align}
where $M_T$ is the total mass and $T_{\odot}$ = $GM_{\odot}/c^3$ $\approx$ 4.9\,$\mu$s. 
The post-Keplerian parameters provide additional measurements that depend on the orbital parameters and masses, so the observation of two post-Keplerian effects will break the degeneracy and isolate $m_{\textnormal{p}}$ and $m_{\textnormal{c}}$ (see e.g.~\citet{Stairs03}, \citet{Freire10} for more details). 
Any possibility of constraining post-Keplerian parameters depends on the system in question. For example, high-$e$ orbits are favorable for $\dot{\omega}$ and $\gamma$, long timing baselines are required for $\dot{P_{\textnormal{b}}}$, and nearly edge-on orbits with sin($i$) $\approx$ 1 are required for Shapiro $r$ and $s$ (see Figure~\ref{fig:SD}).

Measuring the relativistic Shapiro delay in highly-inclined Millisecond pulsar binaries, which appears as a delay in pulse times of arrival as the pulsar passes behind its companion during superior conjunction, has been one of the most fruitful ways to constrain pulsar masses, and in turn, the equation of state. \citet{Demorest10} presented the first evidence for a $\sim$\,2\,M$_{\odot}$ neutron star via pulsar timing measurements of J1614$-$2230, measuring its mass to be 1.97 $\pm$ 0.04\,M$_{\odot}$ and challenging a number of softer equations of state. 
Further timing of this millisecond pulsar led to an improved constraint of 1.928 $\pm$ 0.017\,M$_{\odot}$~\citep{Fonseca16}. 
More recently, but in a similar fashion, \citet{Cromartie20} measured the mass of the millisecond pulsar J0740+6620 to be 2.14$^{+0.10}_{-0.09}$\,M$_{\odot}$ (68.3\% credibility interval). Adding data from the Canadian Hydrogen Intensity Mapping Experiment (CHIME) to the updated Green Bank Telescope data set, \citet{Fonseca21} further constrained the mass to be 2.08 $\pm$ 0.07\,M$_{\odot}$. 
An independent radio-derived constraint on the mass of J0740+6620 facilitated improved constraints on its radius (see~\ref{sec:xrayspec}).

Caution should be exercised in cases where the Shapiro delay cannot be precisely constrained, as is frequently the case in non-edge-on ($i\lessapprox85$ degrees) orbits where the induced timing delay is comparable to the achievable timing precision. 
If $r$ and $s$ cannot be measured to high precision, unmodeled delays induced by e.g.~astrometry, intrinsic red noise, or chromatic red noise from interstellar medium dispersion can negatively impact the accuracy and precision of pulsar mass constraints. 
This is not a concern for most near-edge-on systems, especially if the pulsar has been timed for a duration sufficiently long to constrain slowly-varying parameters. 
For example, in the case of \citet{Fonseca21}, it was shown that varying the dispersion measure modeling scheme changed the measured pulsar mass by only $\sim1\%$. However, in cases such as the gamma-ray bright millisecond Pulsar J1231$-$1411, with a measured inclination angle closer to 80 degrees, choices in modeling the interstellar medium and solar wind dispersion measure, astrometry, and intrinsic red noise can have a more deleterious impact (Cromartie et al.~2024, \emph{in prep.}). 

Observations of systems where other post-Keplerian parameters can be measured have also yielded profound results. For example, \citet{Barr24} significantly measured $\dot{\omega}$ in an eccentric globular cluster millisecond pulsar J0514$-$4002E using the MeerKAT telescope, implying the total system mass to be 3.8870 $\pm$ 0.0045\,M$_{\odot}$. Combining upper limits on $r$, $s$, and $\gamma$ (none of which were significantly detected) with the mass function yields a pulsar mass $<$ 1.79\,M$_{\odot}$ and a companion mass $>$ 2.09\,M$_{\odot}$ (95\% credible interval). 
Because the companion was not detected in optical or near-ultraviolet Hubble Space Telescope observations, it may be a black hole or massive neutron star existing in the mass gap between the maximum pulsar mass and minimum black hole mass. 

Several massive millisecond pulsars have been measured using techniques that target $m_\textnormal{c}$ through optical spectroscopy and photometry. \citet{Antoniadis13} combined radio pulsar timing observations with radial velocity measurements derived via phase-resolved optical spectroscopy to constrain the mass of J0348+0432 to be 2.01 $\pm$ 0.04\,M$_{\odot}$. 
More recently, \citet{Romani22} measured the mass of the highly accelerated ``black widow'' millisecond pulsar J0952$-$0607 to be 2.35 $\pm$ 0.17\,M$_{\odot}$ (68\% credible interval). Although this binary system does not suffer from significant pulsar heating or high Roche lobe filling, determination of the companion's mass is dependent on the model used for light curve fitting. The faint nature of the white dwarf companion implies that this measurement will improve as larger-class optical telescopes come online. 

Pulsar Timing Array experiments aim to detect nHz gravitational waves of astrophysical or cosmological origin by searching for quadrupolar correlations in pulse time of arrival deviations. Six Pulsar Timing Array collaborations are currently active: the European Pulsar Timing Array (EPTA), the Indian Pulsar Timing Array (InPTA), the MeerKAT Pulsar Timing Array (MPTA), the North American Nanohertz Observatory for Gravitational Waves (NANOGrav), the Parkes Pulsar Timing array (PPTA), and the Chinese Pulsar Timing Array (CPTA). 
Recently, Pulsar timing Arrays have produced evidence for a stochastic background of nHz gravitational waves, likely produced by merging supermassive binary black holes \citep{Agazie23, EPTAGW, PPTAGW, CPTAGW}.

Because sensitivity to gravitational waves increases as more pulsars are timed to better precision over longer periods of time, the resulting data sets are expansive and ripe for synergistic science pursuits. 
Both the second EPTA and third PPTA data releases, along with the NANOGrav 15-year data set, contained newly constrained pulsar masses from measurements of post-Keplerian parameters \citep{EPTADR2, PPTADR3, Agazie23data}. The high-mass millisecond pulsars J1614$-$2230 and J0740+6620 are both included in the NANOGrav data set. The Pulsar Timing Array's high-cadence timing data, when combined with targeted observations over superior conjunction, made it possible to constrain these pulsars' masses; without a ``hint'' of Shapiro delay emerging in NANOGrav data, it is possible that these sources would have never been timed over sufficiently long periods to meaningfully constrain their masses. Of the 68 millisecond pulsars included in the NANOGrav 15-year data release \citep{Agazie23data}, 50 are in binaries. Seventeen of these systems show significant constraints on companion mass and inclination via the relativistic Shapiro delay. 

Some concern has been expressed over the apparent tendency of Shapiro-derived mass measurements to decrease with time; however, this is not supported by an analysis of trends in mass constraints as they improve. For example, the median mass for seven of the millisecond pulsars in the NANOGrav 15-year data release decreased, while seven either did not significantly change or increased. The perceived trend in decreasing masses may be a function of selection bias, as one (reasonably) will focus more on minute measurement improvements in the highest-mass neutron stars. 
Critically, the reported credible intervals for masses of millisecond pulsars such as J1614$-$2230 and J0740+6620 have stayed self-consistent with the addition of data. In the case of J0740+6620, the 3$\sigma$ lower limit on the measured mass is unchanged between \citet{Cromartie20} and \citet{Fonseca21}. Because the mass distribution of neutron stars (not just the maximum mass) impacts theories of binary evolution and neutron star formation, the value of constraining \emph{any} neutron star mass significantly, even if $m_\textnormal{p} \ll$ 2\,M$_{\odot}$, should not be overlooked. 

\subsection{X-ray spectroscopy 
}
\label{sec:xrayspec}

One method of determining mass and radius is to fit the X-ray spectrum emitted by the compact object to an ``atmosphere'' model which assumes that the surface is emitting X-rays 
uniformly.~\cite{Zavlin96,Rutledge99,Heinke06,Webb07}. Atmosphere models must choose the composition of the atmosphere (typically either H or He), the magnitude of the magnetic field, and may take into account relativistic effects on the emitted radiation as described in Sec.~\ref{sec:radiation}. This technique has been applied most extensively to quiescent Low Mass X-ray binaries, where the surface has been heated by accretion which has then stopped. The accretion process generates a significant X-ray luminosity which obscures the surface, but when the accretion subsides, the surface is visible. The accretion is also believed to bury the magnetic field, possibly simplifying the analysis.

Mass and radius constraints have also been obtained from 
thermonuclear bursters, using the emission during an X-ray burst that is triggered by unstable thermonuclear burning in the accreted ocean. Bursts which are particularly energetic may lift the photosphere off of the neutron star surface~\cite{Paczynski83,Ebisuzaki83}, named ``Photospheric Radius Expansion'' bursts. Photospheric Radius Expansion bursts create a unique opportunity to measure the neutron star radius (pioneered in \cite{Ozel06}).
Mass and radius constraints from quiescent Low Mass X-ray Binaries and Photospheric Radius Expansion X-ray bursts was used to predict the neutron star tidal deformability \citep{Steiner15un} as was measured in GW 170817 two years later~\citep{LIGOScientific:2017vwq}.

Spectral modelling of these sources is complicated by a number of factors.  Ideally the distance must be known: while distances for quiescent Low Mass X-ray Binaries in globular clusters may be known to better than 10\% this is not always the case for the bursters.  
There are uncertainties in atmospheric modelling and spectral corrections, and the possible role of residual or ongoing accretion.  Pile-up in detectors may be an issue. 
Rotational corrections could be significant even if the rotation rate cannot be measured, and surface emission could be non-uniform even if it is not detectable (for example if a hot spot were centered on the rotational pole).  
This has led to a great deal of debate in the literature about these methods \citep[see for example][]{MillerASSL21,OzelFreire16,Suleimanov20,Baubock15,Elshamouty16}, and the latest modelling attempts to address these issues. 

Recent modelling of X-ray bursts fits bursting atmosphere models directly to spectra during the cooling tails of bursts, avoiding some of the earlier issues \citep{Nattila17}, and rotational corrections are now also being included \citep{Suleimanov20}.  
The latest modelling of quiescent Low Mass X-ray Binaries takes into account uncertainties in the distance, atmosphere composition, and the possible presence of undetected asymmetries in the emission \citep{Steiner18}.   
The more recent analyses also directly couple in information from current EOS models \citep[see e.g.][]{Baillot19}. \citet{AlMamun21ce} attempted to detect systematic uncertainties in quiescent Low Mass X-ray Binary and X-ray burst constraints on neutron star masses and radii via a Bayesian inference with observations from GW 170817 and found that no additional systematic uncertainties were required to explain the data.
Nevertheless, gravitational wave observations will likely supercede these methods of obtaining information on neutron star masses and radii in the near future, as the GW observations have smaller systematic uncertainties. This will mean that, in the future, better knowledge of the mass-radius curve from GW observations will translate into a better understanding of the nature of X-ray emission from quiescent Low Mass X-ray Binary and X-ray bursting sources.

\subsection{X-ray pulse profile modeling}
\label{sec:ppm}

Pulse Profile Modeling, 
just like the techniques described in Sec.~\ref{sec:xrayspec}, exploits the fact that General and Special Relativity
leaves mass- and radius-dependent imprints on thermal X-rays emitted from the neutron star surface.
However it involves rotational-phase resolved spectral modelling applied to stars whose X-ray emission varies as the star rotates, due to non-uniformities - often called {\it hot spots} - in the surface temperature distribution. The emission from these sources is as a consequence {\it pulsed} at the spin frequency of the neutron star.  

The properties of the pulsations - the flux and spectrum as a function of rotational phase - are determined in part by the various effects described in Sec.~\ref{sec:radiation}. Higher compactness (mass/radius), for example, suppresses amplitude; while Doppler boosting (which depends primarily on surface velocity and hence radius) increases harmonic content. These complementary effects enable us - in principle - to disentangle mass and radius.  
Since effects like Doppler boosting are stronger for more rapid spin, Pulse Profile Modeling is most typically discussed in the context of rapidly spinning neutron stars: rotation-powered millisecond pulsars; accretion-powered millisecond X-ray pulsars \citep{Patruno21}; and thermonuclear burst oscillation 
sources \citep{Watts12}. 

\begin{figure*}
\centering
\includegraphics[width= \textwidth]{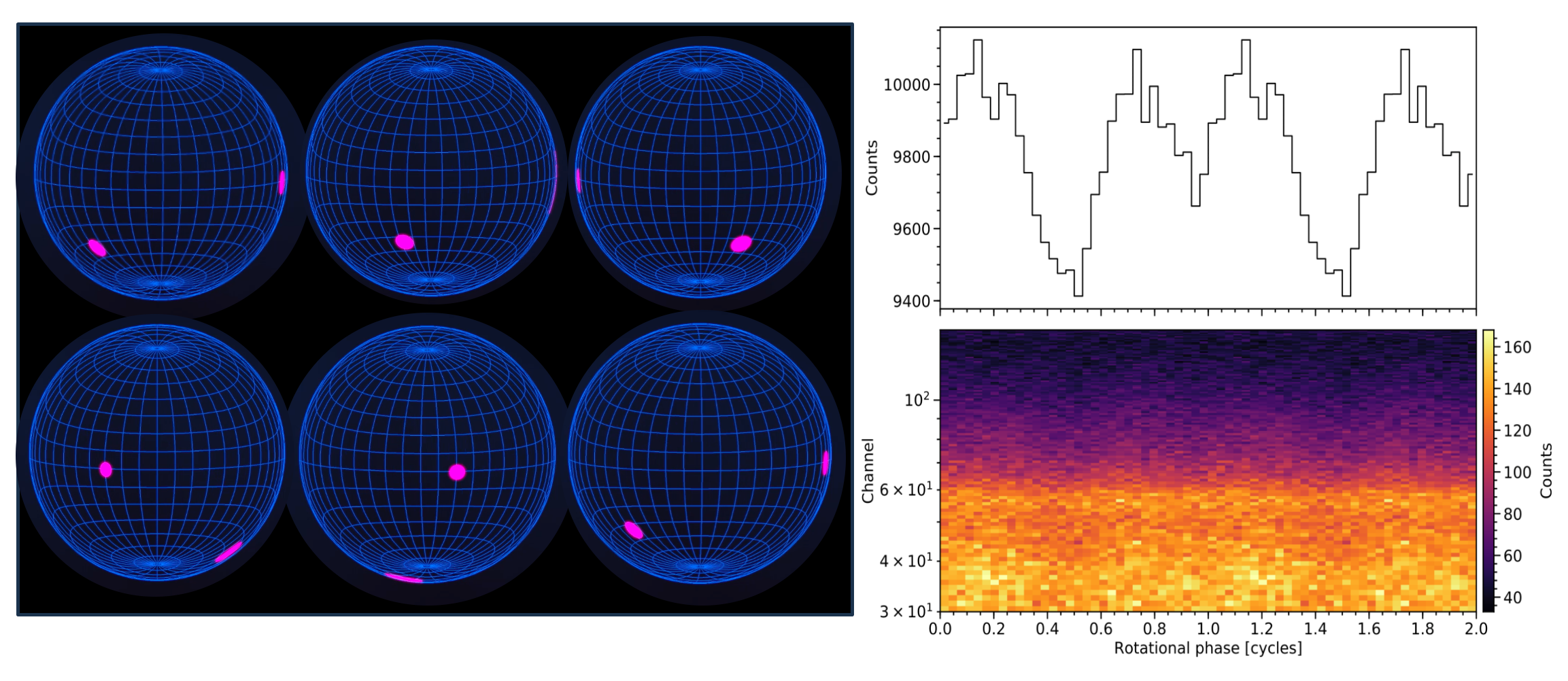}
\caption{The formation of a pulse profile.  Left: snapshots at various points of the star's spin cycle as a star with two hot spots rotates (credit: NASA/Sharon Morsink); the effects of gravitational lensing on spot visibility are clearly visible. Right: the resulting bolometric (summed over energy-channels) and energy-resolved pulse profile as recorded by the NICER telescope, from \citet{Riley21}. Counts denotes number of photons recorded. }
\label{fig:ppm}
\end{figure*}

As photons arrive at the X-ray telescope
two things must be recorded: the time of arrival, and the energy channel.  The former can be mapped to a rotational phase bin using a spin ephemeris for the neutron star, while the latter depends on the incident photon energy and the instrumental response matrix. 
Over time we build up a {\it pulse profile}, as shown in Fig.~\ref{fig:ppm}.  
Good time and energy resolution are essential, with constraints improving as the number of photons in the profile increases \citep{Lo13,Psaltis14a}.  

To determine the mass and the radius we need a physical model of how the pulse profile is formed.  The main ingredient is the relativistic ray-tracing model itself:  the ability to propagate photons from the emitting surface to the detector through a space-time defined by the mass, radius and spin. Here we build on decades of work building up a good understanding of the space-time of rapidly-rotating neutron stars (Sec.~\ref{sec:radiation}):  current modeling relies primarily on the Oblate Schwarzschild space-time model \citep{Morsink07}. This includes the most important General and Special Relativistic effects plus modifications due to rotation-induced stellar oblateness, making use of an approximation for the oblate shape of the neutron star that is largely EOS-independent \citep{AlGendy14}. 
Other factors which affect the pulse profile and which must be modeled include the atmospheric beaming function, the instrument response, source distance, observer inclination, interstellar absorption, and the properties (size, shape, location, temperature) of the hot spots. With these in place one can simulate pulse profiles for comparison to data.

Early efforts to model pulse profiles, using data sets from RXTE and XMM-Newton, involved $\chi$-squared fitting for pulse profiles with a limited number of energy bands \citep[see for example][]{Poutanen03,Bhattacharyya05,Bogdanov13}.   
A step change has been enabled by NICER \citep{nicer}, an instrument specifically designed to collect large high quality pulse profile data sets for millisecond pulsars.  
The NICER millisecond pulsars are also radio pulsars that can be timed very precisely.  This provides both the spin ephemeris that is form the pulse profile (assigning photons to phase bins) but can also 
- if the pulsar is in a binary, sometimes provide priors on mass, distance and inclination (see Sec.~\ref{sec:radio.masses}).  
The data sets built up by NICER (over repeated observations, since the millisecond pulsars are very stable and the timing solutions good enough to phase-connect over long periods of time) are large enough that Pulse Profile Modeling can be done using the full energy resolution of the instrument. We also now fully sample
the posterior distributions for all of the model parameters including the mass and radius \citep{Bogdanov19b,Bogdanov21}.  

\begin{figure*}
\centering
\includegraphics[width=0.9\textwidth]{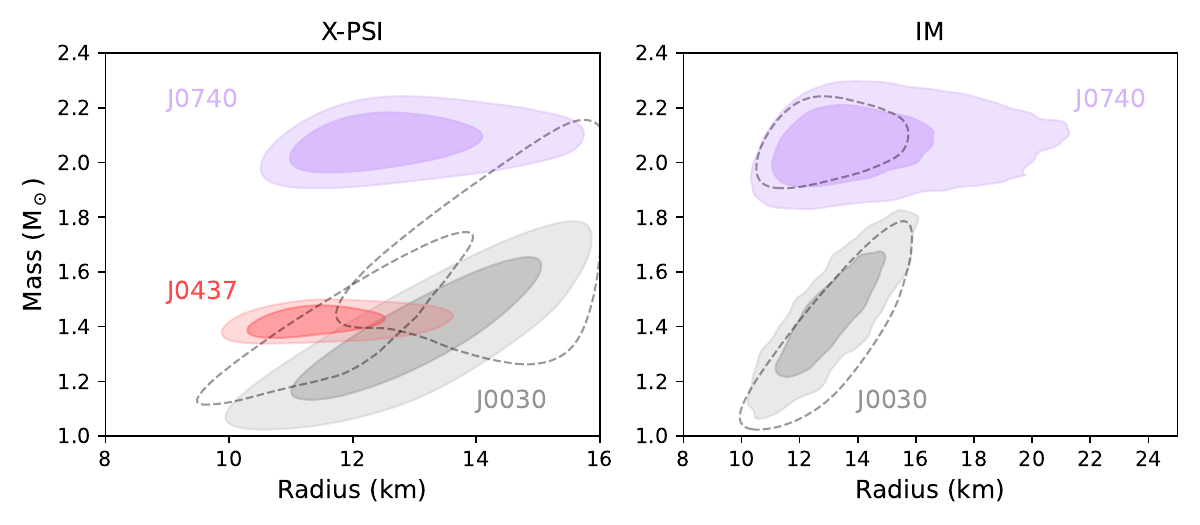}
\caption{Credible regions for mass-radius (68\% and 95\% regions where shaded) reported from Pulse Profile Modeling analyses of NICER data.  Left:  results derived using the X-PSI \citep{Riley23} pipeline, for PSR J0740+6620 \citep{Salmi24}, PSR J0437+4715 \citep{Choudhury24} and PSR J0030+0451 \citep{Vinciguerra24}. The shaded regions for PSR J0030+0451 are for the preferred model when analysing only NICER data; the black dashed lines show the 95\% credible regions for two solutions (different models) identified in a joint analysis of NICER and XMM-Newton data for this source (see text). Right:  results derived using the Illinois-Maryland (IM) pipeline, for PSR J0740+6620 \citep{Miller21} and PSR J0030+0451 \citep{Miller19} (black dashed lines show the corresponding 95\% X-PSI contours, for comparison). The upper bounds on the x-axes reflect the different radius priors used by the different pipelines (and this should be borne in mind when comparing the results).
}
\label{fig:nicerresults}
\end{figure*}

NICER was launched in 2017 and since that time has been building up pulse profile data sets for seven X-ray bright millisecond pulsars (the faintest, PSR J0740+6620, was added to the target list after launch following the measurement of its mass by \citealt{Cromartie20}).  The inferred masses and radii reported to date are summarized in Fig.~\ref{fig:nicerresults}.

One of the more complex aspects of the NICER modelling is how to describe the properties of the hot spots. 
For the millisecond pulsars, the X-rays are thought to be thermal emission due to return current heating of the magnetic poles. 
The expected shape of the heated regions depends on the physics of the magnetosphere and the field geometry, and various shapes are possible \citep[see for example][]{Harding01,Lockhart19} including spots, rings and arcs.  
To date, studies have used parametrized models that encompass a very wide range of possibilities, constructed for example from overlapping circles or ovals.  
Associated with the inferred mass and radius there is also an inferred hot spot geometry that is sensitive to this modeling choice. 

The first source for which results were reported, PSR J0030+0451, is a 205\,Hz isolated millisecond pulsar at a distance of 329\,pc.  
Since it is isolated, there is no mass measurement from radio timing; a prior range of 1-3\,M$_\odot$ was assumed. Analysis of NICER data from July 2017 to December 2018 \citep[$\sim$ 1.9\,Ms in total,][]{Bogdanov19a} by two independent pipelines has yielded very similar results.   
\citet{Riley19}, using the X-PSI code \citep{Riley23}, obtained a mass of $1.34^{+0.15}_{-0.16}$\,M$_\odot$ and a radius of $12.71^{+1.14}_{-1.19}$\,km (68\% credible intervals), while \citet{Miller19} found a mass and radius of $1.44^{+0.15}_{-0.14}$\,M$_\odot$ and $13.02^{+1.24}_{-1.06}$\,km respectively. 
The inferred hot spot properties were very similar despite different parametrizations; a small spot and a long extended emitting region in the same hemisphere, thus pointing to a complex magnetic field geometry \citep{Bilous19}.  

Subsequent simulations \citep{Vinciguerra23} and a more detailed analysis of this data set using the X-PSI pipeline \citep{Vinciguerra24} paint a more complex picture. 
Analysing only the NICER data set with the same hot spot models gives results that are highly consistent with the earlier results of \citet{Riley19}.  
However the likelihood surface is multimodal, which becomes important once background constraints are included via joint modelling with XMM-Newton data for this source (see below). Subject to some caveats, the analysis finds inferred masses and radii $\sim$ [1.4\,M$_\odot$, 11.5\,km] or $\sim$ [1.7\,M$_\odot$, 14.5\,km], 
depending on the assumed spot model (credible intervals are similar to the analysis without background constraints). 
The two hot spots need no longer be in the same hemisphere, and may have temperature gradients rather than being elongated.   
This source also appears to be sensitive to different choices regarding the atmosphere model \cite{Salmi23}. Ongoing studies using larger data sets may resolve some of these ambiguities, but in the mean time EOS studies should consider the impact of the possible multimodality when using mass-radius posteriors from this source \citep[see e.g.][]{Christian24,Rutherford24}. 

The second source for which NICER pulse profile modeling analysis has been done is the 2.1\,M$_\odot$ millisecond pulsar PSR J0740+6620 \citep[for the mass measurements from radio timing, see][and Sec.~\ref{sec:radio.masses}]{Cromartie20,Fonseca21}.   
This source was the first for which background constraints were included in the analysis, via joint modelling with XMM-Newton data. 
{\it Background}, in this context, means emission from anything other than the hot spots, astrophysical or instrumental, and is assumed to be unpulsed (a phase-invariant spectral component in pulse profile).  
Pulsed emission must come from the hot spots, but they can also contribute an unpulsed component, for example if a spot overlaps the rotational pole, or the star is highly compact such that the spots stay in view even when rotated to the far side of the star.  
Putting limits on the NICER background is challenging \citep[see e.g.][]{3C50}; for XMM-Newton data sets it is easier to get a good estimate of background.  
Modelling the high quality spectral-timing NICER data simultaneously with even phase-averaged XMM-Newton data, with a background estimate for the latter, puts indirect constraints on the NICER background. 

Using data from September 2018 to April 2020, \citep[1.6\,Ms in total,][]{Wolff21}, \citet{Riley21} reported a radius of $12.39^{+1.30}_{-0.98}$\,km, while \citet{Miller21} found $13.7^{+2.6}_{-1.5}$\,km.
The inferred mass is essentially unchanged from the radio prior, at $2.08\pm 0.07$\,M$_\odot$. 
The reasons behind the differences in the inferred radii from the two pipelines are discussed in depth in the papers, but the main factors are differences in priors, assumptions about NICER/XMM-Newton cross-calibration uncertainty, treatment of the XMM-Newton background information, and sampler choice.  
However both analyses disfavour lower radii.  
Subsequent analysis using NICER background models \citep{Salmi22} delivered consistent results compared to the joint analysis with XMM-Newton. 
The hot spot properties for this source appear to be simpler than for PSR J0030+0451 although once again the two hot spots are not antipodal, pointing to a magnetic field that is not a simple centered dipole.  
The results for this source are less sensitive to variations in atmosphere model, perhaps due to the highly constrained geometry, mass and distance. More recent analysis using NICER data up to April 2022 has produced consistent but slightly tighter results on radius \citep{Salmi24,Dittmann24}.  The fact that the credible intervals for PSR J0740+6620 are similar to those of PSR J0030+0451, despite the good mass prior on the former, reflects the difference in data set quality: PSR J0740+6620 is further away, at $\sim$ 1.2 kpc, and hence has lower  signal to noise. 

NICER's closest and brightest target is the 174\,Hz millisecond pulsar PSR J0437-4715.  
This source is in a binary with a favorable inclination for measurement of the Shapiro delay, enabling a tight mass measurement (used as a prior in the Pulse Profile Modeling) of $1.418 \pm 0.044$ \,M$_\odot$.   
This source must be observed off-axis, due to the presence of a bright Active Galactic Nucleus in NICER's field of view.  
Despite this, the AGN still contributes to the NICER background, and this must therefore be taken into account when modelling the source.  
Using NICER data from July 2017 to July 2021, and NICER background estimates, \citet{Choudhury24} reported an inferred mass of $1.418 \pm 0.037$\,M$_\odot$ and a radius of $11.36^{+0.95}_{-0.63}$\,km. 
Once again, the hot spot geometry, for the models studied, was inconsistent with a simple centered dipole magnetic field.


\section{Future theoretical and experimental/observational challenges and opportunities}

In this final section we will give a summary of future challenges and directions about the constrain of the nuclear EOS, higher density, and future neutron star observations.

Observations of neutron stars will provide exciting, high-precision constraints on dense matter in the next decades.
To analyse these data, it is crucial that constraints from  nuclear theory and experiment keep up with this. 
For theoretical calculations of the EOS, at low densities uncertainties can be improved by pushing nuclear interactions from chiral EFT to higher orders.
In addition, it is important to study open issues in chiral EFT, such as the dependence of results on the regulator scale, the power-counting scheme, or new approaches to constructing chiral interactions~\citep{Tews:2020hgp}.
Similarly, constraints at asymptotically high densities can be obtained from perturbative QCD~\citep{Kurkela:2014vha,Gorda:2022jvk,Somasundaram:2022ztm,Komoltsev:2023zor}.
Improvements in these calculations might enable them to provide valuable constraints on the EOS~\cite{Komoltsev:2023zor,Koehn:2024set}.
In addition to these approaches, unfortunately there exist no systematic order-by-order approaches at intermediate densities. 
At these densities, we will only be able to compare different models  of dense matter, some of which including phase transitions to forms of quark matter, and to understand compare predictions to astrophysical data.

In addition to these modeling advances, it will be key that all theoretical predictions provide meaningful uncertainties.
Recently, several approaches have been explored~\cite{Epelbaum:2014,Drischler:2020hwi,Keller:2022crb}, some of which used Gaussian Processes to provide a statistical interpretation of the error bands. 
The shape of the assumed uncertainty has an impact on the results~\cite{Koehn:2024set} and additional work needs to investigate this in greater detail.
Ideally, uncertainties should be propagated directly from distributions of low-energy couplings to desired observables, instead of estimating uncertainties {\it a posteriori}. 

Finally, future experimental constraints on the EOS will provide valuable opportunities.
The recent measurement of the neutron-skin thickness by PREX~\cite{Adhikari_2022} provided constraints on the symmetry energy, albeit with large uncertainties.  
A similar future experiment in Mainz might improve on the current situation, and provide valuable constraints.
Similarly, heavy-ion collision experiments provide and excellent way of constraining the EOS at intermediate densities~\cite{Huth:2021bsp,Tsang:2023vhh}.
Analyses of these experiments suffer from larger systematic uncertainties but improved theoretical analysis tools, such as transport models, will enable HIC experiments to bridge the gap between constraints from nuclear physics and astrophysics.

\begin{figure}
\centering
\includegraphics[width=0.5\textwidth]{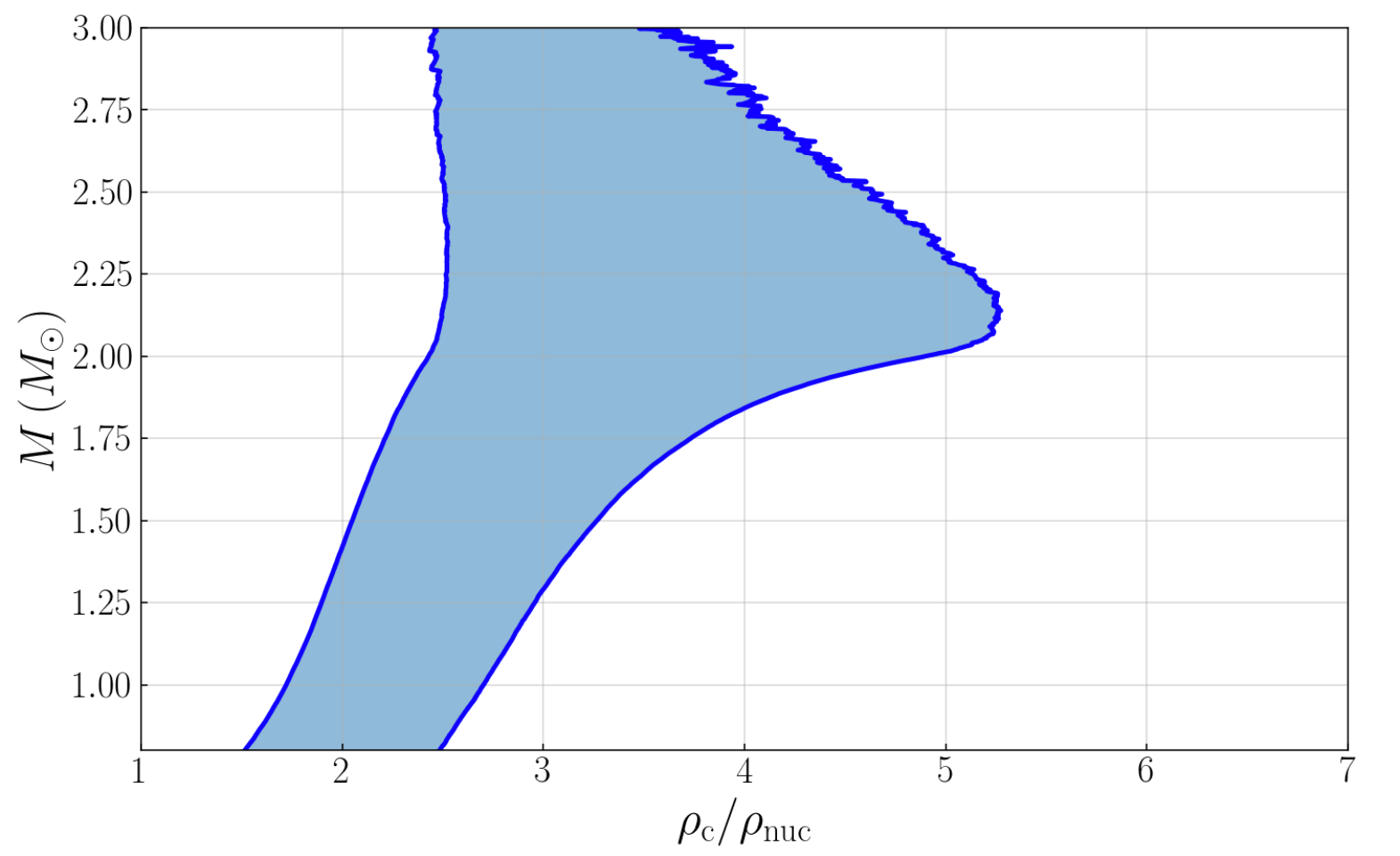}
\caption{
Mass-central density posterior for neutron stars including gravitational wave, X-ray, and radio data.
}
\label{fig:Mrhoc}
\end{figure}

Beyond the current fourth observing run of the ground-based gravitational wave detector network~\cite{KAGRA:2013rdx}, the next fifth observing run is expected around 2027 with ${\cal{O}}(1-10)$ anticipated detections of binaries containing neutron stars.
Further proposed detectors include Voyager~\cite{LIGO:2020xsf}, A\#~\cite{Gupta:2023lga}, NEMO~\cite{Ackley:2020atn}, and the third-generation detectors Cosmic Explorer~\cite{Reitze:2019iox} and Einstein Telescope~\cite{Punturo2010} that will be sensitive to edge of the observable universe.

The third generation detectors are expected to observe tens of thousands of signals containing neutron stars~\cite{Sachdev:2020bkk,Borhanian:2022czq,Johnson:2024foj}, including binaries originating from cosmological distances~\cite{Haster:2020sdh}.
This wealth of information can lead to ${\cal{O}}(10-100)\,$m radius measurements~\cite{Chatziioannou:2021tdi,Rose:2023uui,Iacovelli:2023nbv,Puecher:2023twf,Huxford:2023qne} from gravitational wave data alone and enable novel studies that are inaccessible to current-generation detectors.
For example, louder signals allow us to place constraints on the tidal properties of increasingly more massive neutron stars~\cite{Chen:2020fzm} and gain access to the postmerger emission and the complementary information it includes~\cite{Torres-Rivas:2018svp,Breschi:2023mdj,Prakash:2023afe}. 
Detections across the mass spectrum will also enable a detailed exploration of the equation of state across masses and the possibility of deviations from the hadronic expectation of almost-constant radius that could signal a phase transition~\cite{Chatziioannou:2015uea,Han:2018mtj,Chatziioannou:2019yko,Pang:2020ilf,Essick:2023fso}

With an increased measurement precision, either from a few very loud sources or by combining thousands of quieter sources, a number of challenges emerge.
Inferring the neutron star masses and tidal parameter hinges on accurate waveform models that can coherently track the signal phase evolution through the late inspiral stages and merger.
Currently-available waveform models are robust for GW170817, but systematics in modeling both the point-particle and the tidal sector need to be continuously addressed as the sensitivity improves~\cite{Dudi:2018jzn,Samajdar:2019ulq,Samajdar:2018dcx,Kunert:2021hgm,Chatziioannou:2021tdi,Nagar:2023zxh,Purrer:2019jcp,Gamba:2020wgg}.
Moreover, neutron star binaries will be observed for minutes to hours and multiple sources will coincide in the detector data stream~\cite{Samajdar:2021egv,Pizzati:2021apa,Relton:2021cax}.
Data analysis techniques that simultaneously analyze multiple signals (and other data components such as the detector noise) will need to be developed~\cite{Pizzati:2021apa,Alvey:2023naa,Janquart:2022nyz}.
The time-frequency density of signals is not expected to be high enough to lead to Gaussian confusion noise~\cite{Johnson:2024foj}, suggesting that techniques being developed to address much worse confusion noise can be adapted~\cite{Littenberg:2023xpl}.

Though only upper limits have been placed on the gravitational-wave emission of isolated pulsars, those upper limits are reaching the level of $10^{-8}$ in ellipticity~\cite{LIGOScientific:2021hvc} and are  becoming competitive with proposed theoretical upper limits, see~\citet{Riles:2022wwz} for a comprehensive review.

Direct mass constraints from radio observations will improve with the development of new instrumentation. Future facilities such as the ngVLA, SKA, and DSA-2000 will provide opportunities for more frequent and more precise timing observations, both of which are critical for constraining post-Keplerian orbital parameters. New facilities will also allow for deeper searches for new millisecond pulsars during which many high mass neutron stars might be discovered. New ``ultra-wideband'' receivers such as those developed for the Parkes and Green Bank radio telescopes will also improve timing precision. As Pulsar Timing Array data sets grow, slowly-varying orbital parameters will become increasingly well measured, meaning post-Keplerian effects can be measured more precisely.

NICER continues to operate and build up data for Pulse Profile modeling on the brightest rotation-powered millisecond X-ray pulsars: this will lead to improved constraints on the sources for which have already been published, and analysis will also be completed for several new sources.  One challenge that is emerging is how to handle multi-modality and how to select between two good but different solutions. One option may be to rely more heavily on the sources with mass priors from radio timing, and to seek consistent solutions for the other sources. The large-area X-ray telescope Athena, due for launch in the 2030s, will enable us to study fainter rotation-powered millisecond pulsars, including several more with mass priors from radio timing. Background for a mission like Athena will be much lower than for NICER, which will also help to improve constraints. The largest challenge for pulse profile modeling is probably the surface pattern uncertainty: better coupling to magnetosphere and multi-wavelength pulsar emission models may help to reduce this \cite{Chen20,Kalapotharakos21}. 

Pulse profile modeling is also being extended to other classes of rapidly-rotating neutron star with non-uniform surface emission: thermonuclear burst oscillation sources \citep{Lo13,Miller15,Kini23,Kini24a,Kini24b} and accretion-powered millisecond pulsars, where data from X-ray polarimetry missions such as IXPE can help to reduce uncertainties relating to geometry.  \citep{Salmi18,Salmi21}. While these sources tend to spin faster than the rotation-powered millisecond pulsars being studied by NICER (enhancing the relativistic effects that pulse profile modeling exploits), there are complicating factors such as hot spot variability  and additional sources of X-ray emission such as the accretion disk that need to be included in the model. Large broad-band spectral-timing X-ray telescopes, such as the proposed mission STROBE-X \cite{strobex}, have pulse profile modeling of these sources as a primary science goal.  One benefit of accreting neutron stars is that one can apply different mass-radius inference techniques to the same star (e.g. there are stars that have both accretion-powered pulsations and thermonuclear burst oscillations), which will help to uncover systematic errors. And with a sufficiently large population of sources, one can compare the results from pulse profile modeling for the accreting stars to those from the rotation-powered millisecond pulsars, and the EOS constraints from pulse profile modeling to those derived from gravitational waves. All of these things will further reduce systematic errors.

\section*{Acknowledgements}
\noindent
We thank P.~Ziomek for providing Fig.~\ref{fig:binary} and R.~Kashyap for providing Fig.~\ref{fig:Rmax_constraint}.
K.C. acknowledges support from the Department of Energy under award number DE-SC0023101 and the Sloan Foundation.  A.L.W. acknowledges support from ERC Consolidator Grant No.~865768 AEONS and NWO ENW-XL grant OCENW.XL21.XL21.038 {\it Probing the phase diagram of Quantum Chromodynamics}. 
D.R. acknowledges funding from the U.S. Department of Energy, Office of Science, Division of Nuclear Physics under Award Number(s) DE-SC0021177, DE-SC0024388, from the National Science Foundation under Grants No.~PHY-2011725, PHY-2020275, PHY-2116686, and AST-2108467, and from the Sloan Foundation.
The work of S.G. and I.T. is supported by the U.S. Department of Energy, Office of Nuclear Physics, under contract No. DE-AC52-06NA25396, and by the Office of Advanced Scientific Computing Research, Scientific Discovery through Advanced Computing (SciDAC) NUCLEI program. S.G. is also supported by the Network for Neutrinos, Nuclear Astrophysics and Symmetries (N3AS), through the National Science Foundation
Physics Frontier Center Award No. PHY-2020275.

\bibliographystyle{apsrmp4-1}
\bibliography{biblio}

\end{document}